\documentclass[12pt,preprint]{aastex}

\newcommand{\tsf}{$t_{\rm SF}$}
\newcommand{\OP}{$\Omega_{\rm P}$}
\newcommand{\Rcr}{$R_{\rm CR}$}

\newcommand{\co}{$^{12}$CO(1--0)}
\newcommand{\Ha}{H$\alpha$}
\newcommand{\HI}{H{\sc i}}
\newcommand{\HII}{H{\sc ii}}

\newcommand{\kms}{km s$^{-1}$}
\newcommand{\kmskpc}{km s$^{-1}$ kpc$^{-1}$}

\shorttitle{Star Formation Timescale and Pattern Speed}
\shortauthors{Egusa et al.}

\begin{document}

\title{Determining Star Formation Timescale and Pattern Speed
in Nearby Spiral Galaxies}

\author{Fumi Egusa\altaffilmark{1,2}, Kotaro Kohno\altaffilmark{1}, Yoshiaki Sofue\altaffilmark{3},
Hiroyuki Nakanishi\altaffilmark{3}, \\and Shinya Komugi\altaffilmark{1}}
\email{fegusa@astro.caltech.edu}


\altaffiltext{1}{Institute of Astronomy, the University of Tokyo, Mitaka, Tokyo 181-0015, Japan}
\altaffiltext{2}{Current address: California Institute of Technology, MC 249-17, Pasadena, CA 91125, USA}
\altaffiltext{3}{Faculty of Science, Kagoshima university, 1-21-35 Korimoto, Kagoshima 890-0065, Japan}

\begin{abstract}
 We present a revised method for simultaneous determination of the pattern speed (\OP)
and star formation timescale (\tsf) of spiral galaxies, which is originally proposed in our previous work.
 As this method utilizes offsets between molecular and young-stellar arms, 
we refer to it as the ``Offset Method".
 Details of the method, its application, and 
results for CO and \Ha ~images of 13 nearby spiral galaxies are described here.
 CO data are from our observations with the Nobeyama Millimeter Array for 2 galaxies, 
and from the BIMA SONG for the rest.
 Out of 13 galaxies, we were able to derive \OP ~and \tsf ~for 5 galaxies.
 We categorize them as ``C" galaxies as their offsets are clear.
 Our findings from these galaxies are as follows.
(1) The corotation radius calculated by the derived \OP ~is close to the edge of the CO data, 
and is about half of the optical radius for 3 galaxies.
(2) The derived \tsf ~is roughly consistent with the free-fall time of typical molecular clouds, 
which indicates that the gravitational instability is the dominant mechanism 
triggering star formation in spiral arms.
(3) The \tsf ~is found to be almost independent of 
surface density of molecular gas, metallicity, or spiral arm strengths.
 The number of ``C'' galaxies and the quality of CO data, however, are not enough to confirm these relationships.
 We also find that 2 other galaxies show no offsets between CO and \Ha, although 
their arms are clearly traced, and categorize them as ``N" galaxies.
 The presence of a bar could account for this feature, since these 2 galaxies are both barred.
 With one galaxy excluded from our analysis due to its poor rotation curve, 
offsets of the remaining 5 galaxies are found to be ambiguous.
 Either their dependence on the rotational frequency cannot be explained by our picture, or 
the number or quality of data is not sufficient for the analysis. 
 We categorize them as ``A" galaxies.
 The possible reasons for this ambiguity are 
(1) the density wave is weaker, and/or (2) observational resolution and sensitivity are not enough to detect the spiral arms 
and their offsets clearly.
 The former is supported by our finding that the arm strengths of ``A" galaxies 
are slightly weaker than that of ``C" galaxies.
\end{abstract}




\section{Introduction}
\subsection{Pattern Speed\label{sec.intro_Op}}
 Ever since the spiral density wave theory was proposed by \citet{LS64} to solve 
the winding problem of spiral arms, it has been known as the 
most successful theory to explain observational features of spiral galaxies.
 The pattern speed (\OP), defined as the angular rotational velocity of 
a spiral pattern or underlying gravitational potential, is 
one of fundamental parameters in the density wave theory, 
since it determines the existence and location of kinematical resonances.

 The one that has been extensively discussed is the ``Corotation Resonance" (CR), 
described as 
\begin{equation}
\Omega (r=R_{\rm CR}) =\Omega_{\rm P}, 
\end{equation}
where $\Omega$ is the angular rotational velocity of materials.
 This equation means that at the CR, the rotational speed of materials 
and the pattern are the same.
 The radius $R_{\rm CR}$, which satisfies this equation, 
is called the ``Corotation Radius".
 As the rotation speed of gas 
in the pattern-rest frame becomes small around the CR, 
a galactic shock \citep{Fuji68, Robe69} 
would not occur. 
 Since galactic-scale star formation is thought to be enhanced 
by this shock, star formation should be less efficient around the CR. 
 As the relative velocity of materials viewed from the pattern 
changes its sign at the CR, 
the direction of streaming motions due to the galactic shock 
should be different between the inside and outside of the CR. 
 These observable features have been used to locate the CR 
as described later in this section. 

 Another important resonance is the ``Lindblad Resonance" (LR), 
which is defined by
\begin{equation}
\Omega (r=R_{\rm LR}) = \Omega_{\rm P} \pm \frac{\kappa}{m},
\label{eq.LR}
\end{equation}
where $m$ is any integer but a frequency of the strongest modes or 
the number of arms is mostly used, and $\kappa$ is the epicyclic frequency, 
which is expressed as 
\begin{equation}
\kappa = 2\Omega\sqrt{1+\frac{r}{2\Omega}
\frac{{\rm d}\Omega}{{\rm d}r}}.
\label{eq.kappa}
\end{equation}
 With an assumption that spiral arms are tightly winding, 
the dispersion relation for the stellar disk is different from that of a gaseous disk \citep{GD}, 
so that gases and stars are theoretically predicted to show different behaviors 
at the resonances.

 As described above, the pattern speed not only locates the resonances, but also 
influences the star formation activities. 
 Therefore, its determination 
is very important to the study of galaxies.
 It cannot be determined directly from observations, however, 
since the pattern structure is not a material structure but a density wave. 

 Several methods have been proposed for its determination.
 The earlier and common technique is to locate specific resonances 
on radii where some properties of arms change. 
 \citet{RRS75} adopted \OP ~for 24 galaxies 
so that the corotation radius is 
nearly coincident with the radius where arms of \HII ~regions disappear, 
or the radial extent of the ``easily visible" disk.
 \citet{EEM92} used $B$-band images to locate 
5 resonances for 18 galaxies. 
 Such techniques are, however, subject to uncertainties  regarding the sensitivity and/or resolution of 
the imaging observations. 
 In addition, the tracers used must be carefully 
considered, since stars and gases can behave differently at resonances 
and their distributions are not identical.
 
 \citet{CB90} derived the star formation efficiency (SFE) 
in the arm and interarm region
for NGC 628 and NGC 3992 from \Ha ~and \HI ~data. 
 The arm-to-interarm ratio of the SFE is expected to be larger than 1 
as the star formation is thought to be enhanced in arm regions, but they 
found that its value drops to almost unity at a certain radius. 
 Since star formation at spiral arms are thought to be less efficient 
around the CR, they concluded that this radius is the corotation radius. 
 It is difficult to derive the SFE values in interarm regions, however,  
since there are fewer stars and gases than in arm regions, 
and thus sensitive observations are required. 

 For our Galaxy, a time periodicity of the star formation history of about 0.5 Gyr 
has been found at the solar neighborhood \citep{Her00,dlFM204}.
 Assuming that this periodicity is due to the passage of a spiral potential 
and that the potential has a two-armed pattern as suggested by \citet{Dri00}, 
the pattern speed can be calculated as 
$\Omega_{\rm P}=\Omega(r=R_\odot)-\pi/(0.5~{\rm Gyr})=21$ \kmskpc.
 A consistent result from numerical simulations of the stellar and gaseous response to 
the spiral potential are presented by \citet{Mar04}.
 These results for the Galaxy also indicate a strong relationship between the star 
formation and the spiral potential.

 \citet{Canz93} showed that the residual velocity field, 
obtained by subtracting the axisymmetric component from 
the observed velocity field, should be different 
inside and outside the corotation radius, and thus 
a change in direction of the non-axisymmetric component of the velocity field 
could be used to locate the CR. 
 \citet{Semp95} and \citet{Canz97} applied this method to \HI ~and \Ha ~data of NGC 4321, 
respectively, and they obtained a comparable location of the CR.
 Since this method requires precise kinematical information 
over the entire disk, a velocity field 
covering the whole galaxy with high velocity and spatial resolution is essential.

 \citet{TW84} presented a method which did not use any morphological 
locations of resonances, but used  
the continuity equation for a surface brightness 
of galaxies. 
 This method is called the ``Tremaine-Weinberg method" (hereafter TW method).
 As it is based on the continuity equation, 
it has been applied mostly to early type galaxies \citep[e.g.,][]{MK95, GKM99, DW04}, 
whose star formation is less active than that of late type galaxies.
 In recent studies, however, \HI ~and CO data have come to be used 
for the application to gas-dominated, late type galaxies \citep[e.g.,][]{West98, Zim04}. 
 Meanwhile, \citet{Deb03} showed by $N$-body simulations that 
the results of the TW method is sensitive to the uncertainty in adopted position angle (P.A.) of outer disk.
 \citet{RW04} applied the TW method to CO data of nearby galaxies with varying P.A.s 
and confirmed that the uncertainty in P.A.\ made the error of the resultant \OP ~larger.
 They also showed that the best-fitted \OP ~only to the
 inner bar was larger than that of the remaining disk, 
even though the galaxy in interest was not classified as SB galaxy.
 \citet{Mei08} used a modified method which explicitly allowed a radial variation of \OP, and  
found multiple spiral patterns for the nearby spiral galaxy M 51 (NGC 5194) with outwardly decreasing \OP.
 The dependence on P.A.s, however, was found to be still large.

 The results of numerical simulations also show 
a large dependence of \OP ~on kinematics and spiral structure \citep[e.g.,][]{Wada98}, 
so that they can be used to derive the best value of \OP ~by comparison with observations. 
 \citet{Oey03} adopted an evolutional model of the \HII ~luminosity 
function to draw isochrones of massive stars, and 
fitted these isochrones to an observed distribution of \HII ~regions to derive \OP.
 These approaches, of course, depend on their modelings, 
and often have difficulty estimating the accuracy of 
the derived values and the effect of other parameters on the results.

\subsection{Star Formation Timescale\label{sec.intro_tsf}}
 Since all the physical processes from molecular clouds to young stars 
are not yet clearly understood, especially for massive objects, 
it has been difficult to derive a timescale for star formation from a theoretical point of view. 
 A number of numerical simulations regarding the GMC formation and successive star formation 
have been performed \citep[e.g.,][]{V-S07,Wada08} and have inferred on the timescale, 
but the full processes are not traced 
since the dynamic range of the size scale and the number of processes are too large to be included in 
calculations.
 Instead, it is common to estimate ages of \HII ~regions from equivalent widths of emission 
lines from ionized gas, such as \Ha ~and H$\beta$, on the basis of population synthesis models, 
which do not require any information about the parental molecular clouds. 

 \citet{Lei99} compiled physical processes and parameters for star formation such as the initial mass function, 
star formation history, and metallicity, to 
provide the package {\sc Starburst99} to calculate observable properties such as colors, 
spectra, equivalent widths, and luminosity.
 \citet{Bast05} estimated the equivalent width of \Ha ~emissions and 
derived ages of star cluster complexes in NGC 5194 as 5--8 Myr 
using this package assuming solar metallicity. 
 Even if twice the solar metallicity is assumed, the derived ages do not change largely.
 They also applied the GALEV SSP models \citep{GalevSSP} to optical colors of individual star clusters
and found the ages of the majority of the clusters to be 4--10 Myr.
 With spectroscopic observations of the Br$\gamma$ line, 
\citet{Gros06} derived ages of $K$-band knots, presumably sites for massive star formation, 
to be about 7--10 Myr for the nearby spiral galaxy, NGC 2997.
 They also found that these knots were located 
slightly upstream from the smooth peak of $K$-band flux, regarded as the potential minimum of the spiral arm.

 We should note that ages derived from models depend on parameters used in the model, 
and that the exact age is often difficult to determine due to degeneracies between 
internal extinction, mass, and metallicity.
 In addition, this age only represents the timescale from a newly formed star, and thus 
has no information about the molecular cloud collapse in contrast with the timescale 
we take into account in this paper.

 As described above, it is in principle difficult to determine the pattern speed 
and star formation timescale from observations, although they are both important parameters 
for understanding kinematics and star formation activities in spiral galaxies.
 In this paper, we propose a method to determine both parameters simultaneously.
 We refer to this method as the ``Offset Method" as it uses offsets between  CO and \Ha ~arms.
 This method is a revised version of what we have presented in a previous paper \citep{egu04}.
 To apply this method, we have obtained CO, \Ha, and rotation curve data from 
our observations and the literature.

 We describe the basic idea and characteristics of the method in \S \ref{sec.method}, 
and processes of the application in \S \ref{sec.appl}, 
followed by results in \S \ref{sec.dis_indivi}, which includes a categorization of the sample, 
discussions for individual galaxies, and properties of each category.
 We discuss derived parameters and the validity of the offset method in \S \ref{sec.dis_param}, 
and conclude this paper by giving a summary in \S \ref{sec.end}.

\section{Offset Method\label{sec.method}}
 This ``Offset Method" is proposed for simultaneous determination of  
the pattern speed and star formation timescale by the use of offsets observed in spiral galaxies. 
 As the basics of this method have been presented by our previous paper \citep{egu04}, 
we repeat important concepts and explain several modifications here.

\subsection{Basic Idea and Formulation\label{sec.idea}}
 For simplicity, 
we assume that the spiral pattern is rigid, and that 
materials in the disk rotate in pure circular orbits. 
 The former assumption means that 
the spiral structure is sustained by the density wave, 
and that the pattern speed (\OP) is constant. 
The velocity fields of most disk galaxies 
show a spider diagram pattern, 
which is a velocity field that results from pure circular rotations. 
 The streaming motion and the velocity dispersion 
generates some non-circular motions, 
but they are typically about 10--50 \kms ~\citep[e.g.,][]{AW96, CB97, Kuno97}, which 
is small compared to the circular rotational velocity of around 200 \kms. 
 These observational results support the latter assumption.
 We should note that we do not consider nor include any bar structures 
in our analysis, 
since the pattern speed of the bar would be different from 
that of spiral \citep[e.g.,][]{Wada98} and 
particles within the bar potential move in elliptical orbits 
with higher eccentricity.

 We define the star formation timescale, \tsf, 
as the {\it average} time for the massive star formation
from molecular clouds in spiral arms, which are agglomerated or compressed by the spiral structure. 
 If the physical processes of star formation at spiral arms 
do not vary extremely with radius, 
this timescale can be regarded as a constant parameter 
representing a typical value of the entire disk.
 We should note again that this timescale \tsf ~would be different from 
ages of \HII ~regions mentioned in \S \ref{sec.intro_tsf}, 
since \tsf ~includes the time needed for clouds to be collapsed.

\begin{figure*}
\begin{center}
\includegraphics[width=\linewidth]{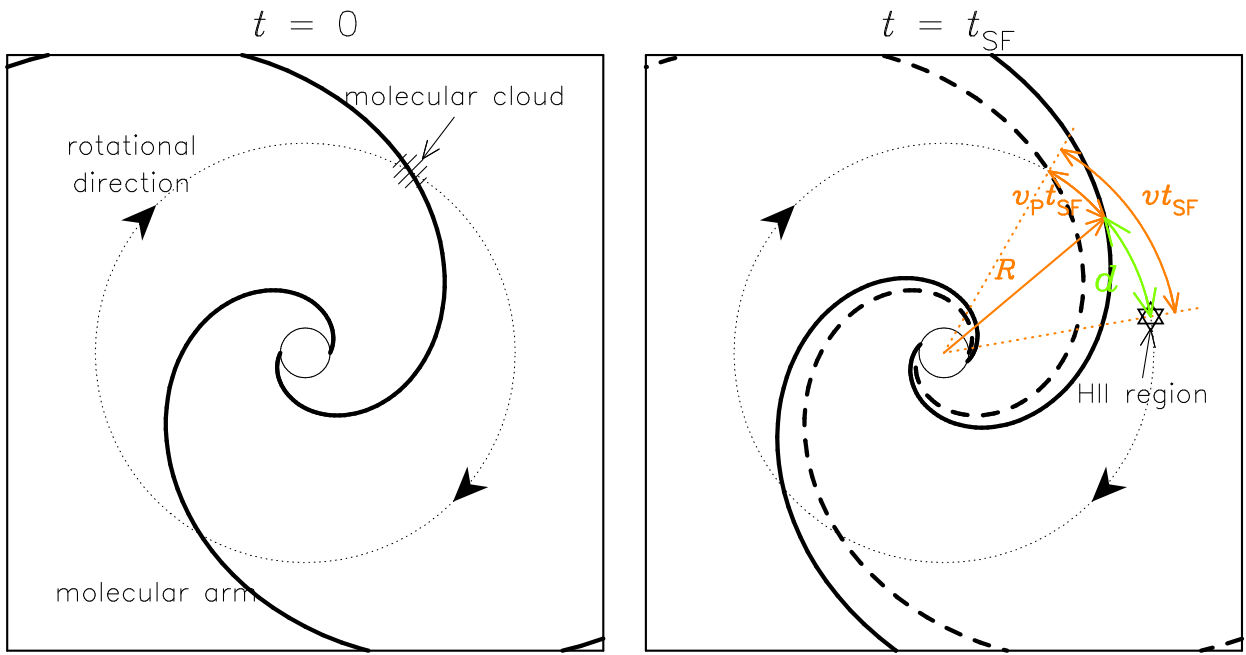}
\end{center}
\caption{Basic idea of our method. 
 If we observe a face-on spiral galaxy at $t=0$ (the left panel), 
the same galaxy will be observed  
as the right panel at \tsf.
 The thick solid lines are molecular arms at time $t$ of each panel.
 The thick dashed lines in the right panel (\tsf) 
show the position of molecular arms in the left panel ($t=0$).
 The offset distance between the massive stars and molecular arm is $d$, 
expressed in equation (\ref{eq.d}).}
\label{fig.draw}
\end{figure*}

 Figure \ref{fig.draw} illustrates our concept regarding the relationship between the spiral structure and 
star formation processes inside the CR, 
where $\Omega > \Omega_{\rm P}$.
 If we observe a face-on spiral galaxy at $t=0$ (the left panel), 
at $t=t_{\rm SF}$ the same galaxy will be observed  
as in the right panel 
and the offset distance between the arm of young stars and molecular clouds, 
$d$, can be written as
\begin{equation}
 d=\Big(\frac{v}{\rm km~s^{-1}}\Big)
  \Big(\frac{t_{\rm SF}}{\rm s}\Big)
  -\Big(\frac{v_{\rm P}}{\rm km~s^{-1}}\Big)
  \Big(\frac{t_{\rm SF}}{{\rm s}}\Big)
  \quad{\rm [km]},
\label{eq.d}
\end{equation}
where $v$ is the velocity of materials, 
and $v_{\rm P}$ is the velocity of the pattern.
 If the outside of the CR is taken into account, 
massive stellar arms will be seen on the concave side of 
molecular arms, and $d$ will be negative.

 Dividing both sides of equation (\ref{eq.d}) by radius $r$ [kpc],  
we obtain
\begin{eqnarray}
\theta = \Big[
   \Big(\frac{\Omega}{\rm km~s^{-1}~kpc^{-1}}\Big)
   -\Big(\frac{\Omega_{\rm P}}{\rm km~s^{-1}~kpc^{-1}}\Big)
\Big]
   \times\Big(\frac{t_{\rm SF}}{\rm s}\Big)
   \quad{\rm [km~kpc^{-1}]},
\label{eq.ofst}
\end{eqnarray}
where $\Omega\equiv v/r$, 
$\Omega_{\rm P}\equiv v_{\rm P}/r$, and
$\theta$ is the azimuthal offset.
 This equation shows the relation  
between two observables, $\Omega$ and $\theta$, 
and we can rewrite it as
\begin{equation}
 \theta = 0.586~\Big[
   \Big(\frac{\Omega}{\rm km~s^{-1}~kpc^{-1}}\Big)
   -\Big(\frac{\Omega_{\rm P}}{\rm km~s^{-1}~kpc^{-1}}\Big)
\Big]
   \times\Big(\frac{t_{\rm SF}}{\rm 10^7~yr}\Big)
   \quad{\rm [degree]}.
\label{eq.linear}
\end{equation}
 Assuming \tsf ~to be constant over the spiral disk,
$\theta$ is a linear function of $\Omega$,
since \OP ~is also assumed to be constant. 
 Therefore, by plotting $\theta$ against $\Omega$
and fitting them with a line, both $\Omega_{\rm P}$ and 
$t_{\rm SF}$ can be simultaneously determined as a horizontal-axis intercept 
and gradient of the fitted line, respectively.

 The fitting is performed by the least $\chi^2$ method and each data point 
is weighted by errors in both axes, $\Delta \theta$ and $\Delta \Omega$.
 The inclusion of the uncertainty in $\Omega$ and the use of the least $\chi^2$ method 
for the fitting are major methodological revisions from what we presented in the previous paper.

\subsection{Uniqueness and Merits}
 One major advantage of the offset method is 
its independency from the tight-winding approximation for spiral structure and 
any models for cloud collapse and population synthesis.
 The results thus will provide constraints or parameters to theories and models 
of both galactic dynamics and star formation on the basis of observational data.

 Another uniqueness is that it gives \OP ~and \tsf ~simultaneously, 
as described in the previous section.
 Measuring offsets in a wide range of radii makes this possible.
 In addition, statistical uncertainties of both \OP ~and \tsf ~are determinable by the fitting.
 We should emphasize that this is the first method proposed that derives \tsf ~observationally with 
quantitative estimates of statistical errors.

\subsection{Requirements for Application}
 Application of the offset method, requires tracing 
 both molecular and young-stellar arms at a wide range of radii. 
 Nearby galaxies with small inclination angle satisfy 
this requirement, while distant and highly-inclined galaxies do not.
 Grand-design galaxies are usually preferable to see spiral structures. 
 We do not pay attention to whether a galaxy has a bar in its center or not, 
since we put importance on spiral structures only and exclude bar regions from the analysis, if any.
 Images used in the analysis should have sufficient quality to detect and resolve 
spiral arms.
 For our application, CO and \Ha ~images are used to trace molecular gas and young stars, respectively.

 In addition to the image maps, we need information about the rotational velocity. 
 As the line-of-sight velocities of an almost face-on galaxy 
do not give velocities parallel to the disk, 
the inclination angle must not be close to zero. 
 The velocity data are available from spectroscopic observations 
in the optical and radio wavelengths.

 In short, we need nearby spiral galaxies which are 
moderately inclined and show clear CO and \Ha ~spiral structures.
 To satisfy these requirements, we observed two nearby spiral galaxies, 
NGC 4254 and NGC 6181, in the \co ~line with the Nobeyama Millimeter Array (NMA).
 For NGC 4254, we have combined the single dish data from \citet{Kuno07} with the NMA data 
to recover the missing flux.
 Details of the observations and data reduction will be presented in another paper. 
 In addition to our data, we selected 11 galaxies from BIMA SONG \citep{BIMA2}.
 The whole sample of 13 nearby spiral galaxies is described in \S \ref{sec.sample}.

\section{Data Analysis\label{sec.appl}}
\subsection{Property of Sample Galaxies\label{sec.sample}}
 In order to apply the offset method, 
we selected 13 nearby spiral galaxies. 
 The selection criteria are that the galaxy shows a molecular spiral structure in CO data, and that 
its inclination angle is small but not zero.
 As we pay attention to spiral structures of molecular gas and \HII ~regions in the analysis, 
we do not take into account the bar type nor arm class (AC) from \citet{EE87} for the selection.
 For these sample galaxies, we obtained \Ha ~images and rotation curve data from the literature.
 Most of the images were obtained via the NASA/IPAC Extragalactic Database (NED).

 The general properties of our sample are listed in Table \ref{sample.tb}.
 Out of the 13 galaxies, 6 are classified as SA (no bar), another 6 are 
classified as SAB (weakly barred), and 1 is classified as SB (strongly barred).
 Regarding the AC, 12 are grand-design galaxies (AC=9 or 12), 
and 1 is a flocculent galaxy (NGC 4736, AC=3).

\begin{figure*}[h!]
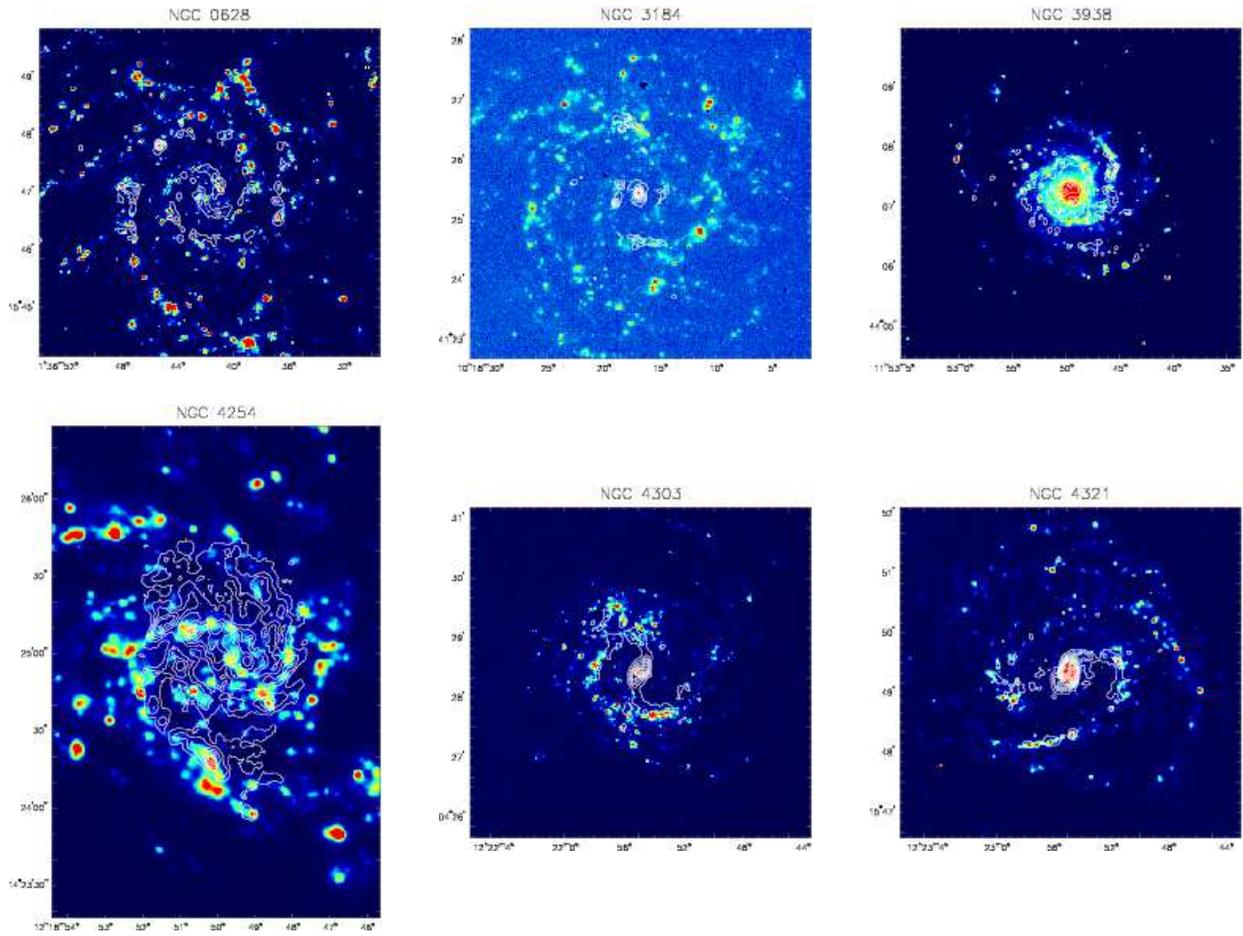

\begin{center}
\begin{minipage}{0.3\linewidth}
\begin{center} 
\includegraphics[width=\linewidth]{f2a.eps2}
\end{center}
\end{minipage}
\hspace{0.03\linewidth}
\begin{minipage}{0.3\linewidth}
\begin{center} 
\includegraphics[width=\linewidth]{f2b.eps2}
\end{center}
\end{minipage}
\hspace{0.03\linewidth}
\begin{minipage}{0.3\linewidth}
\begin{center} 
\includegraphics[width=\linewidth]{f2c.eps2}
\end{center}
\end{minipage}\\
\vspace{12pt}
\begin{minipage}{0.3\linewidth}
\begin{center} 
\includegraphics[width=\linewidth]{f2d.eps2}
\end{center}
\end{minipage}
\hspace{0.03\linewidth}
\begin{minipage}{0.3\linewidth}
\begin{center} 
\includegraphics[width=\linewidth]{f2e.eps2}
\end{center}
\end{minipage}
\hspace{0.03\linewidth}
\begin{minipage}{0.3\linewidth}
\begin{center} 
\includegraphics[width=\linewidth]{f2f.eps2}
\end{center}
\end{minipage}
\caption{CO contours on an \Ha ~image of sample galaxies.}
\label{COHA.fig}
\end{center}
\end{figure*}

 We show CO contours overlaid on an \Ha ~image for the sample galaxies in Figure \ref{COHA.fig}.
 Properties of the data are listed in Table \ref{sample.data.tb}.
 The spatial resolution of CO data is about $3''$--$7''$, and typically 500 pc at the galaxy's distance.
 As it is comparable to the typical value of offsets and arm widths, the CO data quality should be 
sufficient to resolve offsets for most of the sample galaxies.
 The \Ha ~image seeing is typically about $2''$, which is smaller than that of CO.

\subsection{Rotation Curve{\label{sec.RC}}}
 In general, rotation curves (RCs) are obtained from spectroscopy, 
such as optical slit observations and radio mapping observations.
 With the CO data cube, we made velocity field maps and 
applied the task {\tt GAL} in the {\tt AIPS} software in order to derive the RCs. 

 This task fits the tilted-ring model to the velocity field, 
we simultaneously derive the kinematical center, systemic velocity, position angle 
of major axis (P.A.) and inclination of galactic disk ($i$), 
in addition to the rotational velocity ($V_{\rm rot}$).
 Initial guesses for the fitting are the observing center for kinematical center, values from 
BIMA SONG, \citet{RC99}, or RC3 \citep{RC3} for P.A., $i$, and systemic velocity.
 For the BIMA data, however, inclination angles were derived to be about 90 degree 
for almost all galaxies, although the sample was selected to be nearly face-on. 
 This discrepancy is presumably due to the clumpiness of CO distributions, and we use 
values of $i$ from the literatures (BIMA SONG, \citet{RC99}, or \citet{SINGSHa}).
 Derived parameters from this procedure are listed in Table \ref{param.tb}, 
with adopted values for P.A. and $i$.

 Fixing all the parameters except for $V_{\rm rot}$, 
the rotation curve with statistical errors can be calculated from the velocity field. 
 We could not derive meaningful kinematical parameters for 
3 galaxies (NGC 3184, NGC 4535, and NGC 4736) out of 13 galaxies, however.
 For NGC 3184 and NGC 4736, the kinematical center could not be determined.
 For NGC 4535, the kinematical center and the systemic velocity could be determined, but 
the rotational velocities were not properly derived at some radii. 
 This failure could be attributed to that CO data for NGC 4736 are rather noisy and 
those for the other 2 galaxies do not contain single-dish data, 
so that the flux from extended components are resolved out.
 For the remaining 10 galaxies we derive rotational velocities, but some of them  
are limited in the range of radii or have larger errors in the outer regions. 

 Given that RCs from CO are not sufficient for all the galaxies,  
we searched the literature for RCs with a wider range of radii. 
 \citet{RC99} compiled data from optical (\Ha ~and [N{\sc ii}] line) 
and radio (CO and \HI ~line) observations to present central-to-outer rotation curves 
for nearby spiral galaxies. 
 \citet{SINGSHa} observed a part of the SINGS galaxies in the \Ha ~line via Fabry-Perot 
spectrometry and obtained kinematical information and rotation curves 
in the same way as in this study.

\begin{figure*}
\begin{center}
\includegraphics[width=0.3\linewidth]{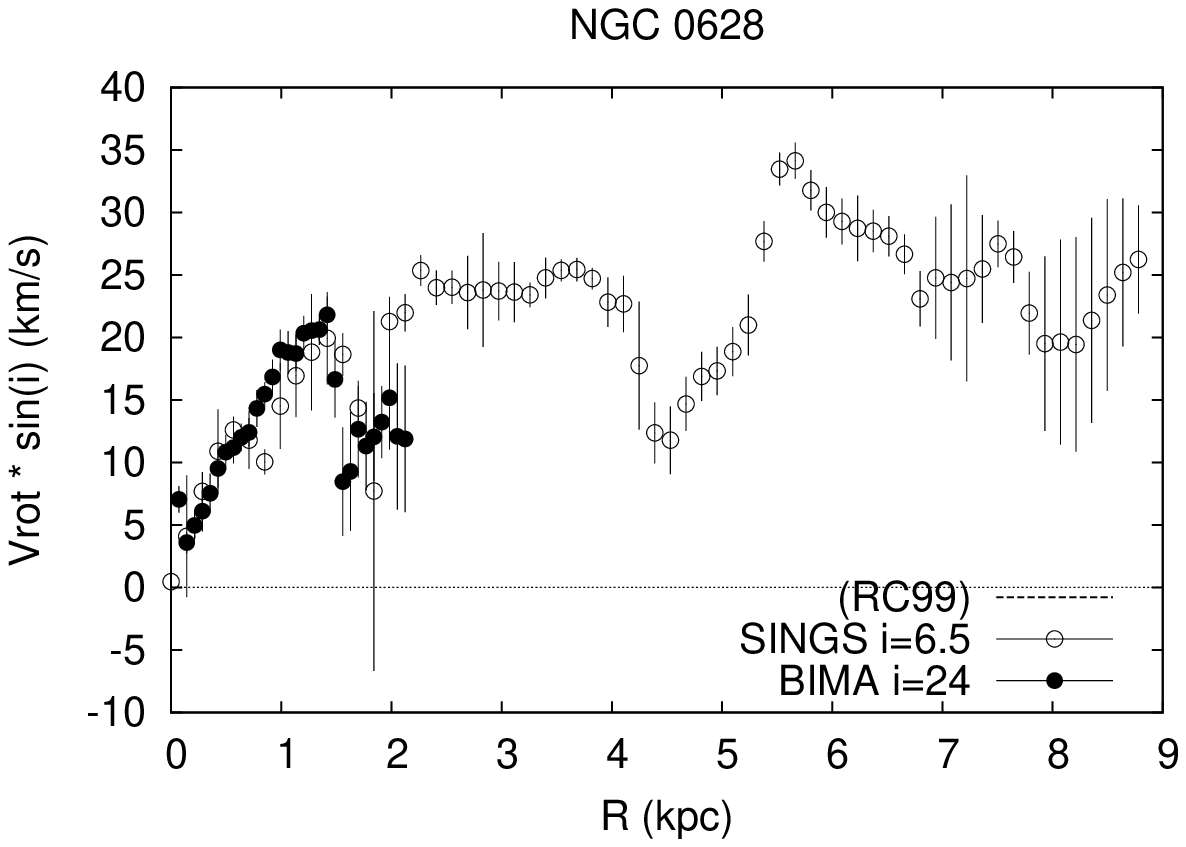}
\hspace{0.03\linewidth}
\includegraphics[width=0.3\linewidth]{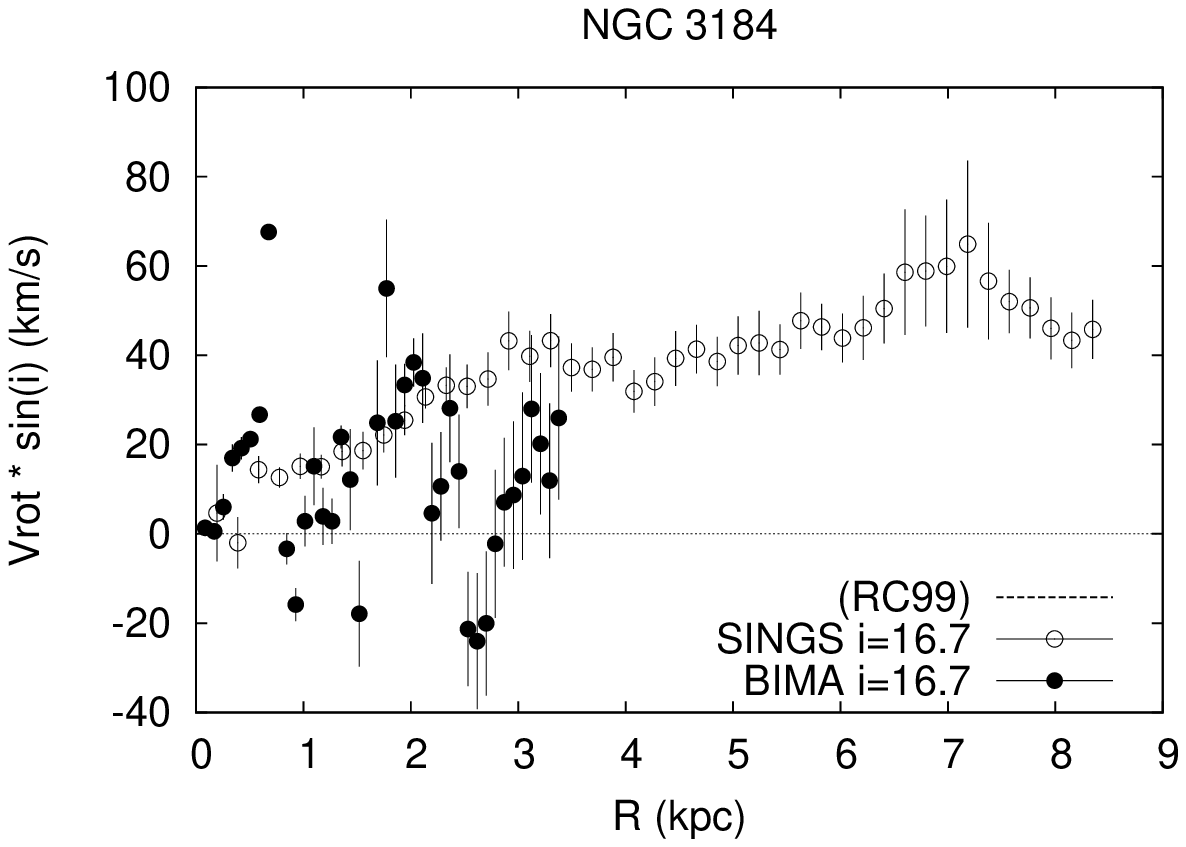}
\hspace{0.03\linewidth}
\includegraphics[width=0.3\linewidth]{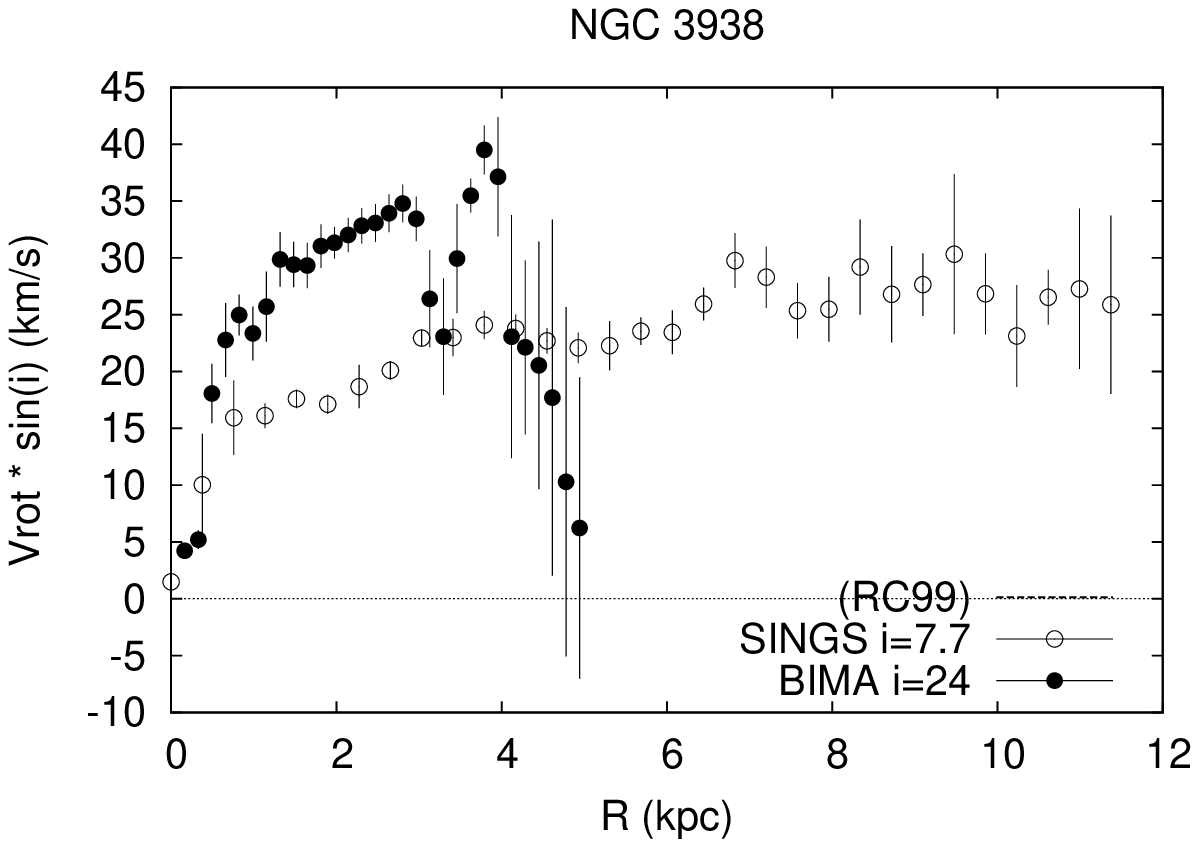}\\
\vspace{12pt}
\includegraphics[width=0.3\linewidth]{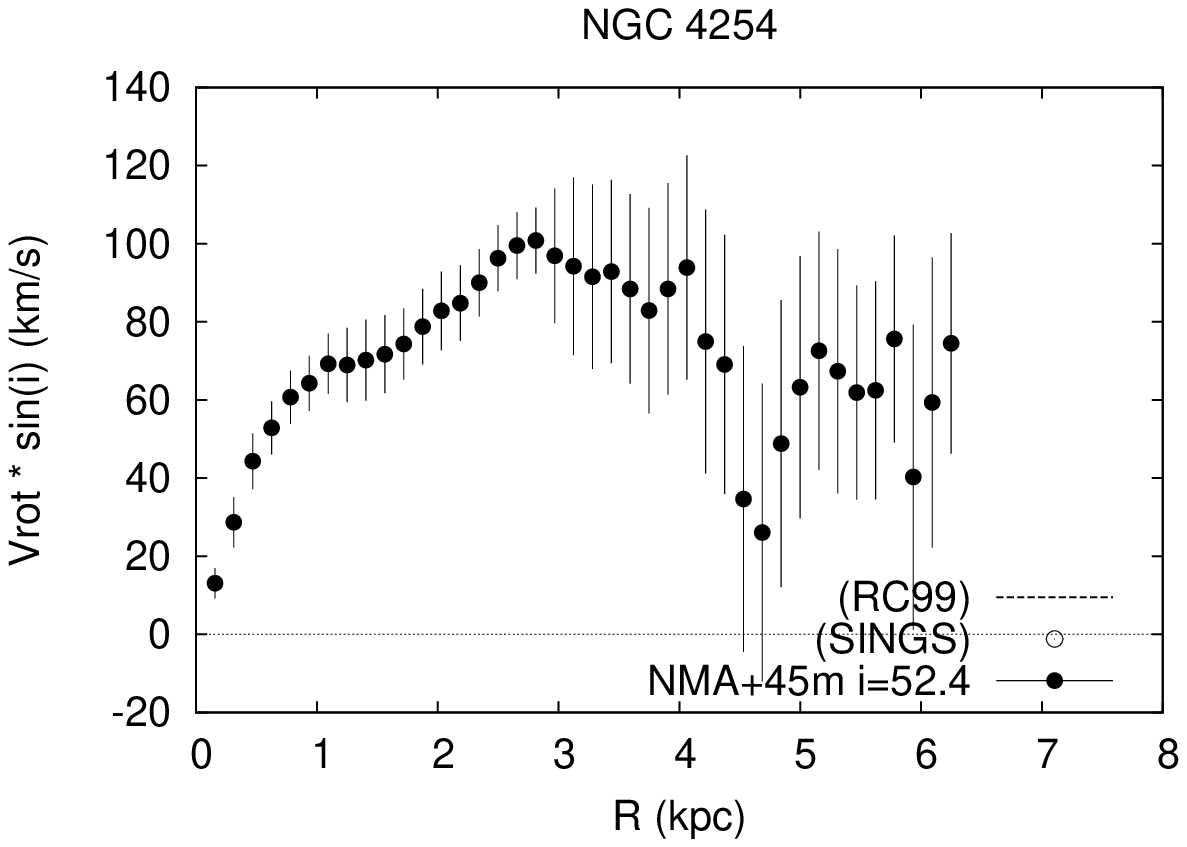}
\hspace{0.03\linewidth}
\includegraphics[width=0.3\linewidth]{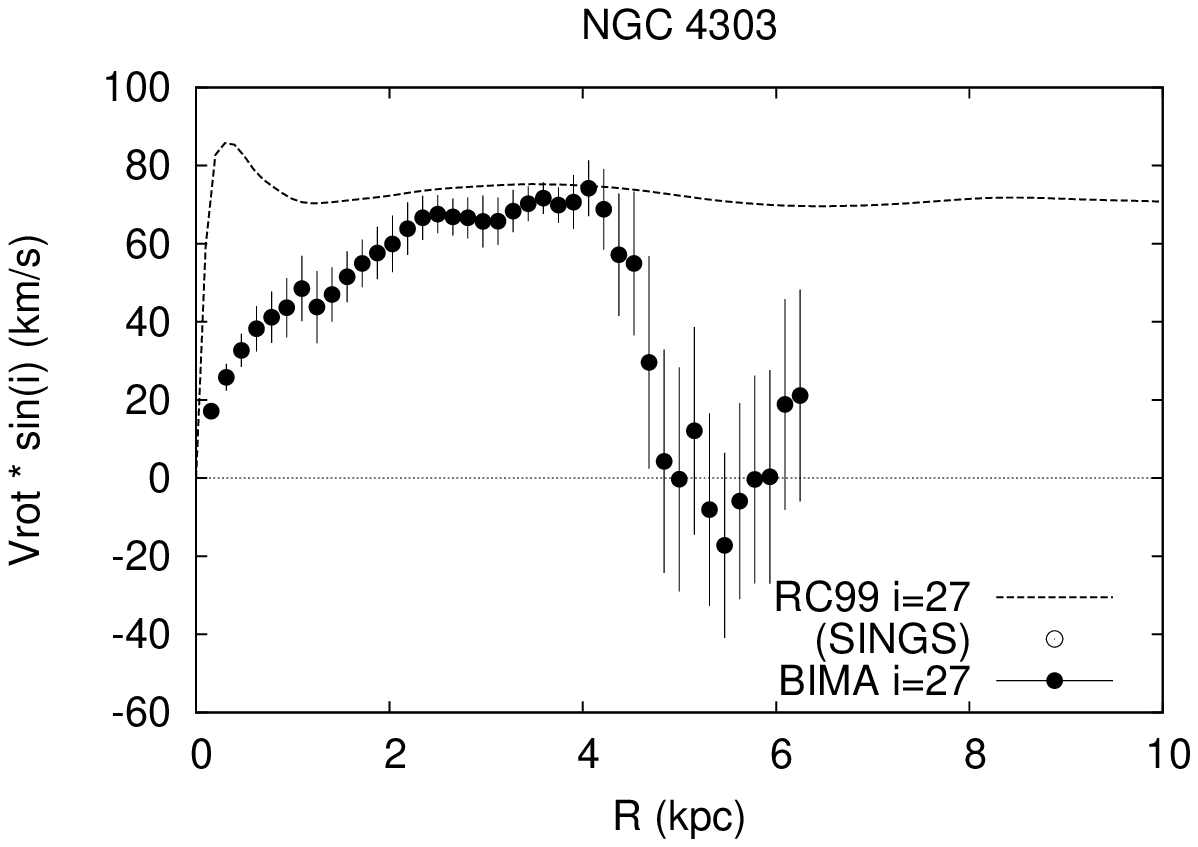}
\hspace{0.03\linewidth}
\includegraphics[width=0.3\linewidth]{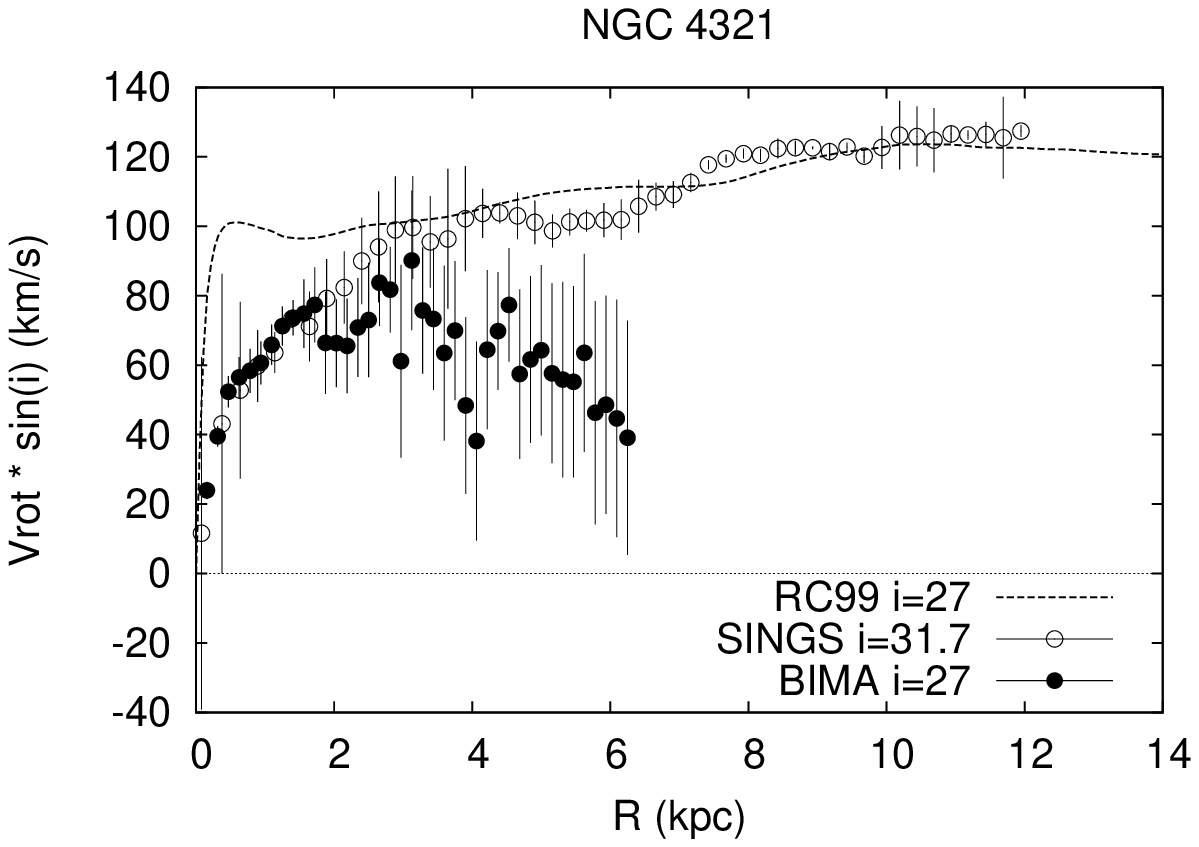}\\
\vspace{12pt}
\includegraphics[width=0.3\linewidth]{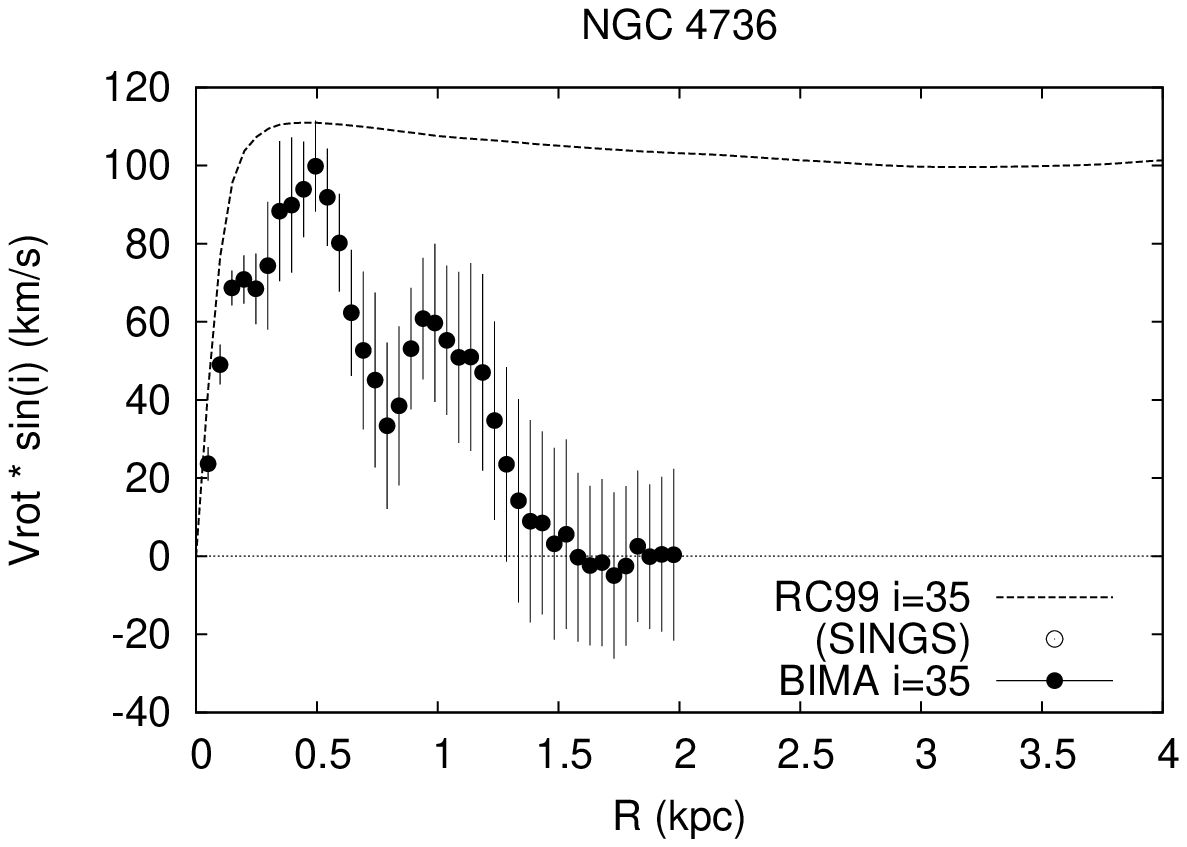}
\hspace{0.03\linewidth}
\includegraphics[width=0.3\linewidth]{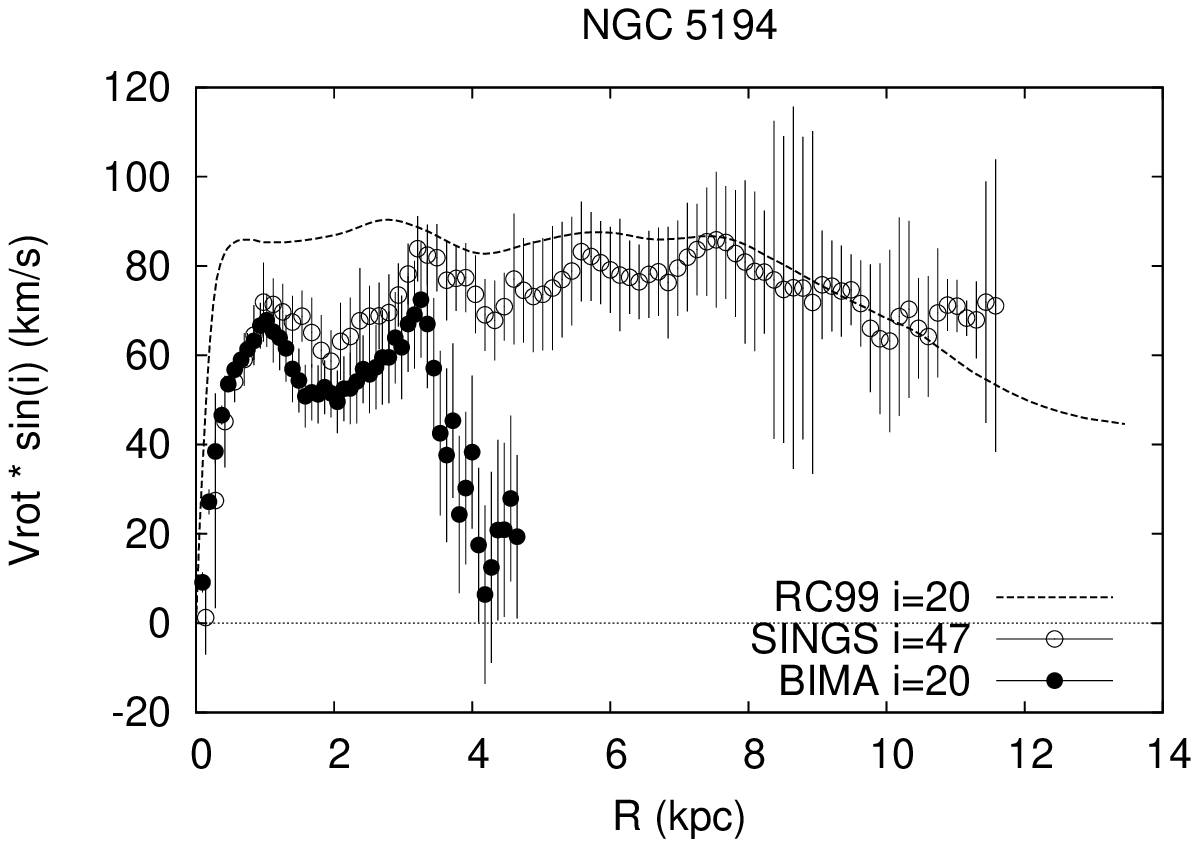}
\hspace{0.03\linewidth}
\includegraphics[width=0.3\linewidth]{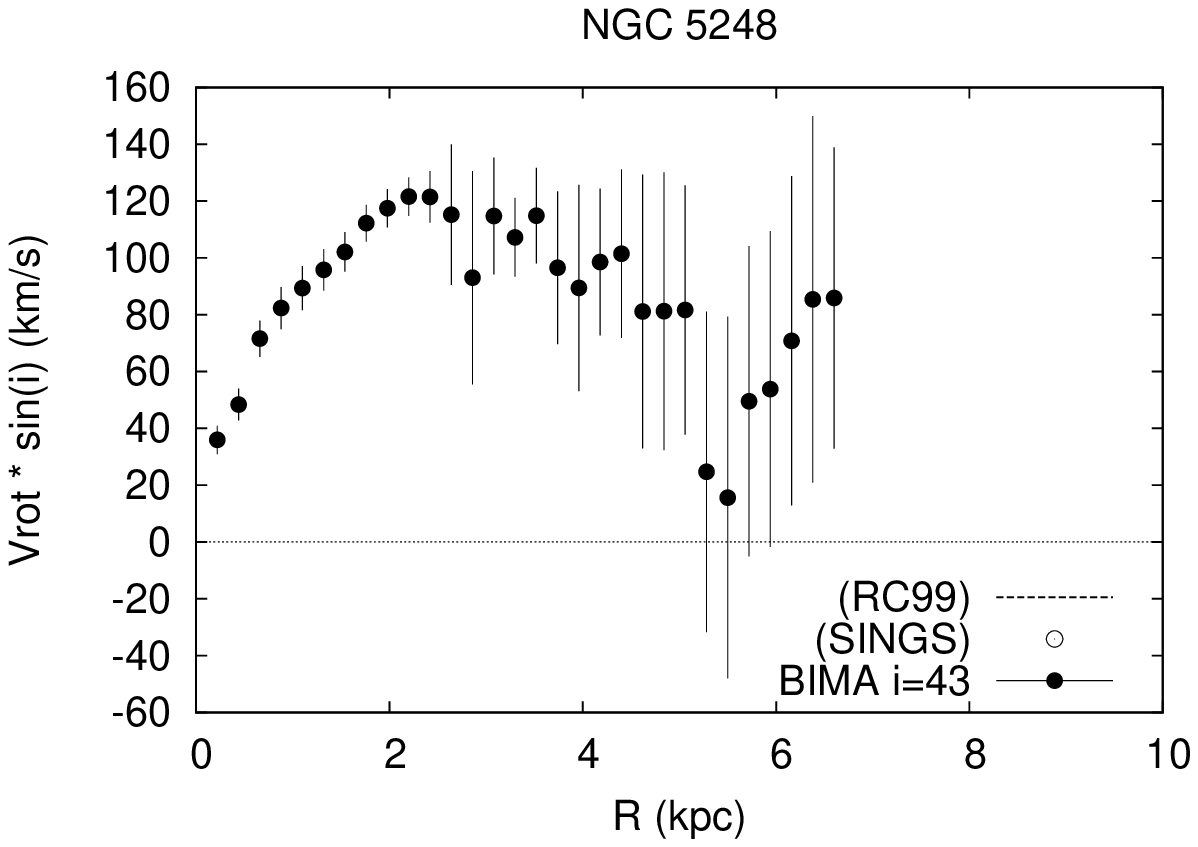}\\
\vspace{12pt}
\includegraphics[width=0.3\linewidth]{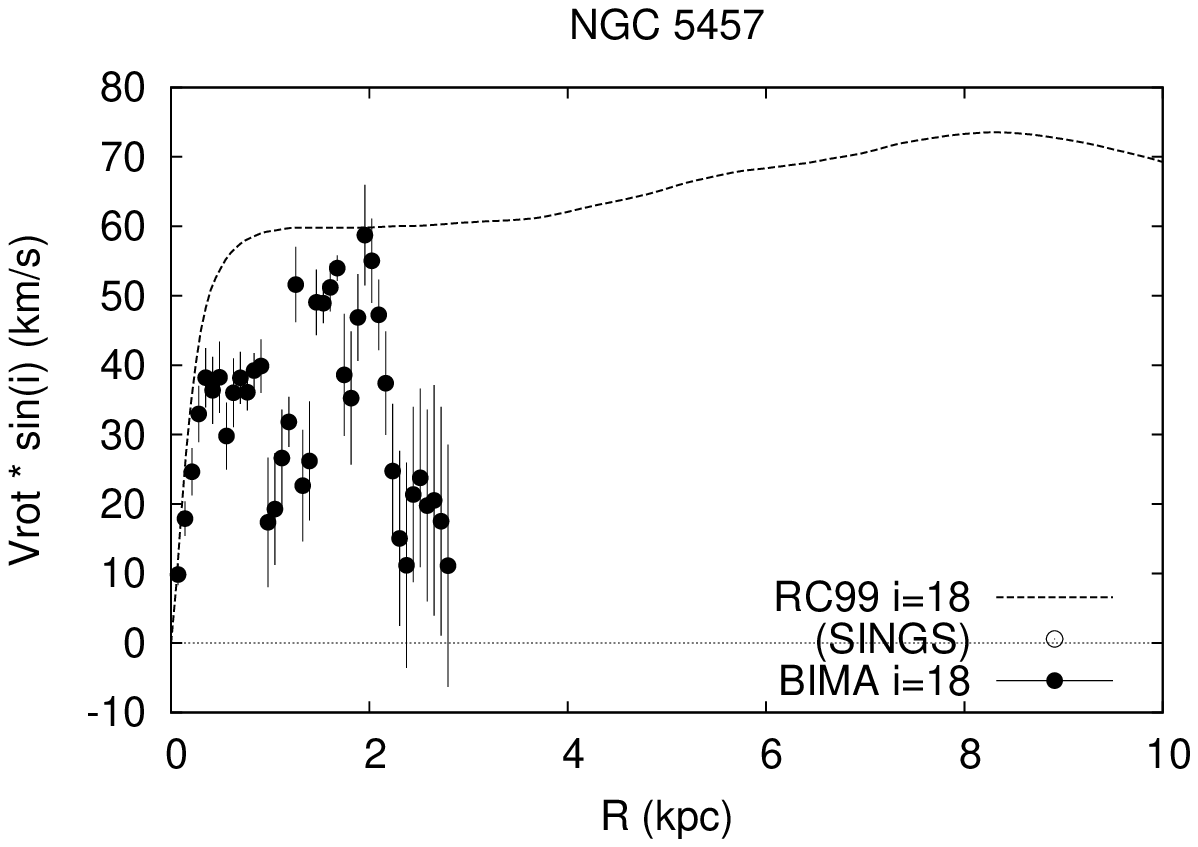}
\hspace{0.03\linewidth}
\includegraphics[width=0.3\linewidth]{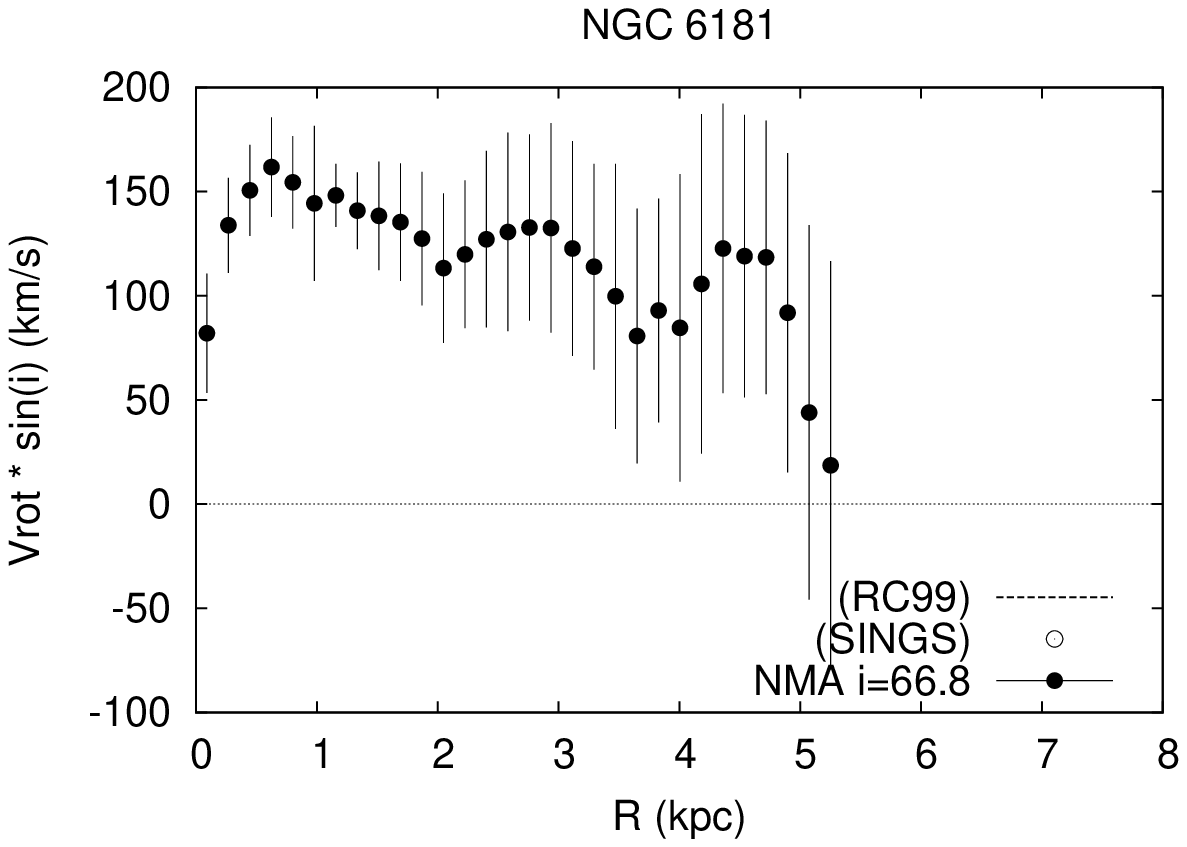}
\hspace{0.03\linewidth}
\includegraphics[width=0.3\linewidth]{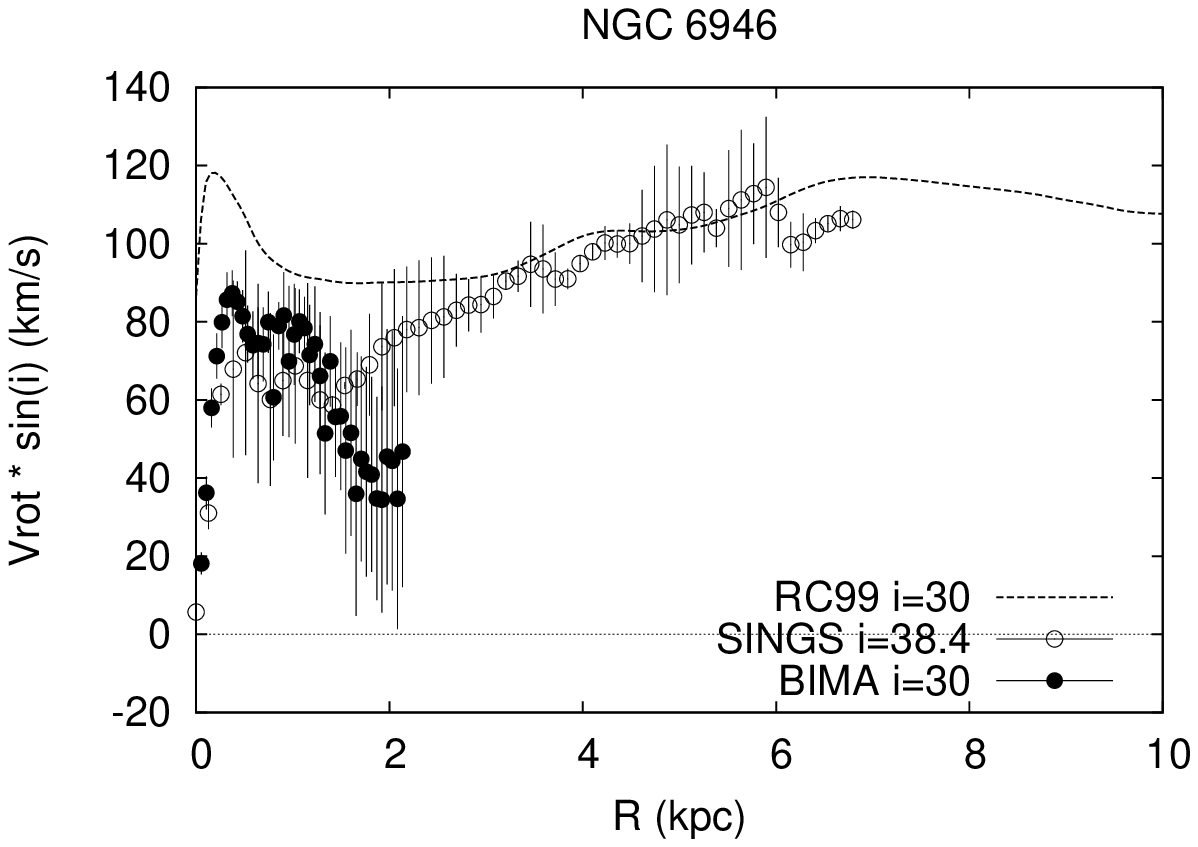}
\caption{Available rotation curves from \citet{RC99}, \citet{SINGSHa}, 
and {\tt GAL} applied to CO velocity field of \citet{BIMA2} or from our observations, labelled as ``RC99", 
``SINGS", ``BIMA", and ``NMA", respectively. 
Labels in parentheses mean that their data are not available.}
\label{RC3.fig}
\end{center}
\end{figure*}

 In Figure \ref{RC3.fig}, available rotational velocities multiplied by the sine of 
the assumed inclination angle ($V_{\rm rot}\sin i$) are plotted against radius.
 In the central regions, 
rotational velocities from \citet{RC99} are larger than those from the others, while in outer regions 
\citet{RC99} and \citet{SINGSHa} give almost consistent results.
 This difference is principally due to that 
\citet{RC99} traces terminal velocities in a position-velocity diagram, while \citet{SINGSHa} and this study trace 
intensity-weighted velocities, and that these two velocities are known to be often substantially different
in the inner regions.
 RCs from CO data generally follow those from \citet{SINGSHa} 
but drop around the outer edge of the CO map.

 From these available RCs, we selected which to use 
considering errors and radial coverage, in the following analysis.
 In principle, we try to avoid the use of CO data, since they are not reliable in the outer regions.
 Although not all the data are consistent each other, 
most of them are in good agreement where we measure the offsets.
 For some galaxies, however, a poor RC makes the application of offset method difficult, 
indicating the uncertainty of the results could be underestimated. 
 We could not obtain a satisfactory RC for NGC 4535, 
so we decide to exclude this galaxy from the following analysis.
 Available RC data and the choice of RC are summarized in Table \ref{sample.data.tb}.

\subsection{Phase Diagram and Offsets}
 The image analysis was started by checking the coordinates of CO and \Ha ~data.
 If the coordinates of \Ha ~images are incorrect or not available, foreground stars 
seen in the $R$-band image with the same field of view as \Ha ~were used to derive the coordinates.
 The Aladin Sky Atlas was used to obtain 
the coordinates of foreground stars and the task {\tt ccmap} of {\tt IRAF} was used to calculate 
and register the image coordinates.

\begin{figure*}
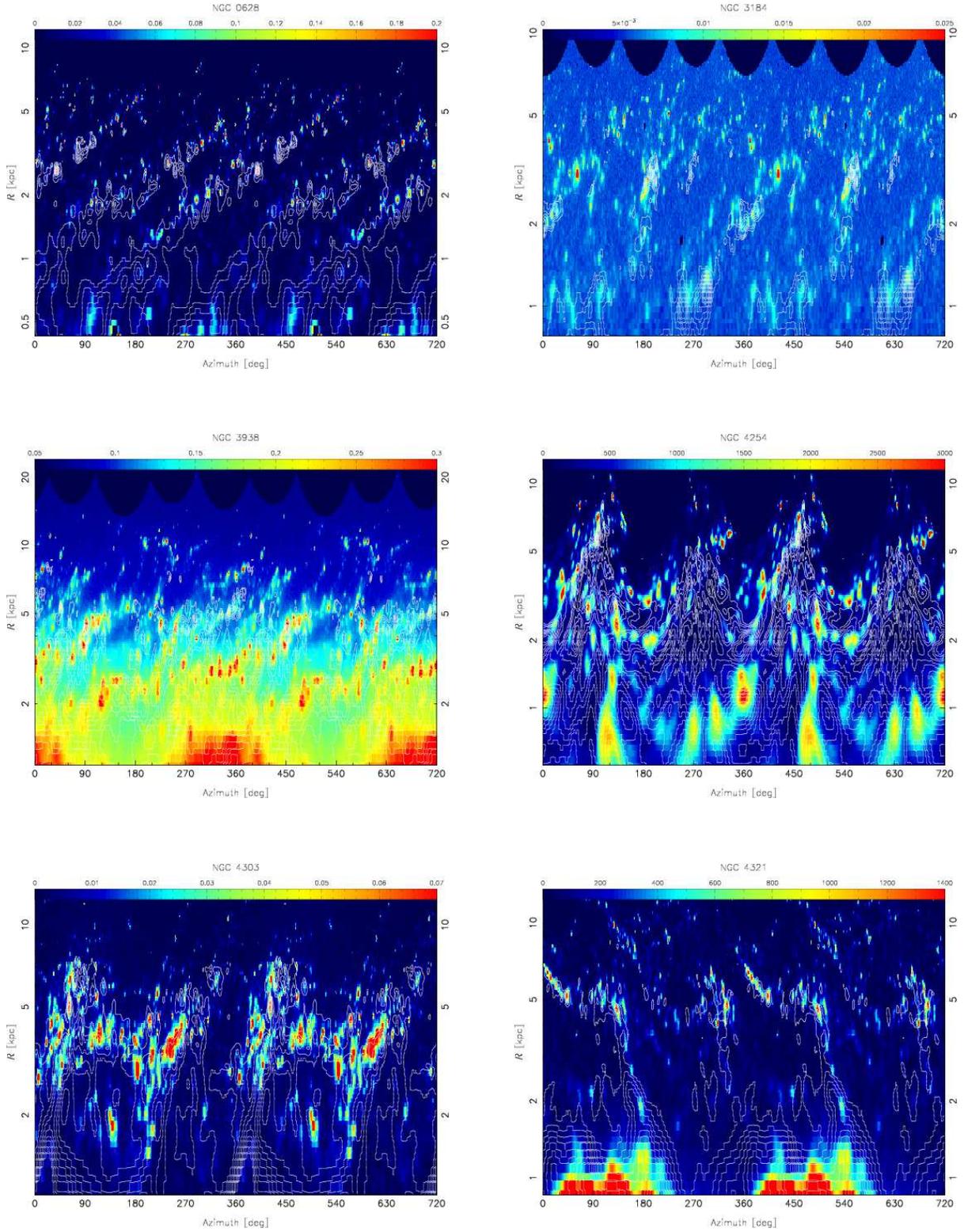

\begin{center}
\includegraphics[width=0.45\linewidth]{f4a.eps2}
\hspace{0.05\linewidth}
\includegraphics[width=0.45\linewidth]{f4b.eps2}\\
\vspace{0.05\textheight}
\includegraphics[width=0.45\linewidth]{f4c.eps2}
\hspace{0.05\linewidth}
\includegraphics[width=0.45\linewidth]{f4d.eps2}\\
\vspace{0.05\textheight}
\includegraphics[width=0.45\linewidth]{f4e.eps2}
\hspace{0.05\linewidth}
\includegraphics[width=0.45\linewidth]{f4f.eps2}
\caption{Phase diagram of each galaxy with CO contours on an \Ha ~image. 
The ordinate is the radius in logarithmic scale and the 
abscissa is the azimuthal angle and shown for two periods, or $0-720$ degree, to delineate spiral arms.}
\label{phase2.fig}
\end{center}
\end{figure*}

 Projected images of galaxies in the (RA, Dec) coordinate were transformed 
into deprojected images in the polar coordinate 
(radius $r$, azimuth $\phi$), by the use of the task {\tt pgeom} of {\tt AIPS} with 
the dynamical center derived above and adopted values of P.A. and $i$ listed in Table \ref{param.tb}. 
 For NGC 3184 and NGC 4736, whose dynamical centers could not be derived, 
the center of the CO map, which is almost the same as the NED position, was adopted.
 In Figure \ref{phase2.fig}, we show CO contours on an \Ha ~image in the ($\phi$, $\log r$) coordinates 
for 12 galaxies whose RC are available.
 Images in this coordinate, called the phase diagram, are used to recognize spiral arms, since 
spiral arms are often expressed as $\ln r=\tan (i_{\rm pitch}) \phi$, 
where $i_{\rm pitch}$ is the pitch angle of arms, and appear as a line in this diagram.

 To measure the offset, we first derived the average flux density in $r\pm\frac{1}{2}\Delta r$ for each azimuthal angle.
 The step in radius $\Delta r$ was set to a value ranging from one-third to half of the CO beamsize. 
 Then, peaks of CO and \Ha ~flux belonging to a spiral arm were searched,  
and the offset was measured as the azimuthal separation of CO and \Ha ~peaks.

\subsection{The $\Omega-\theta$ Plot and Fitting}
\begin{figure*}
\begin{center}
\begin{minipage}{0.3\linewidth}
\begin{center} 
\includegraphics[width=\linewidth]{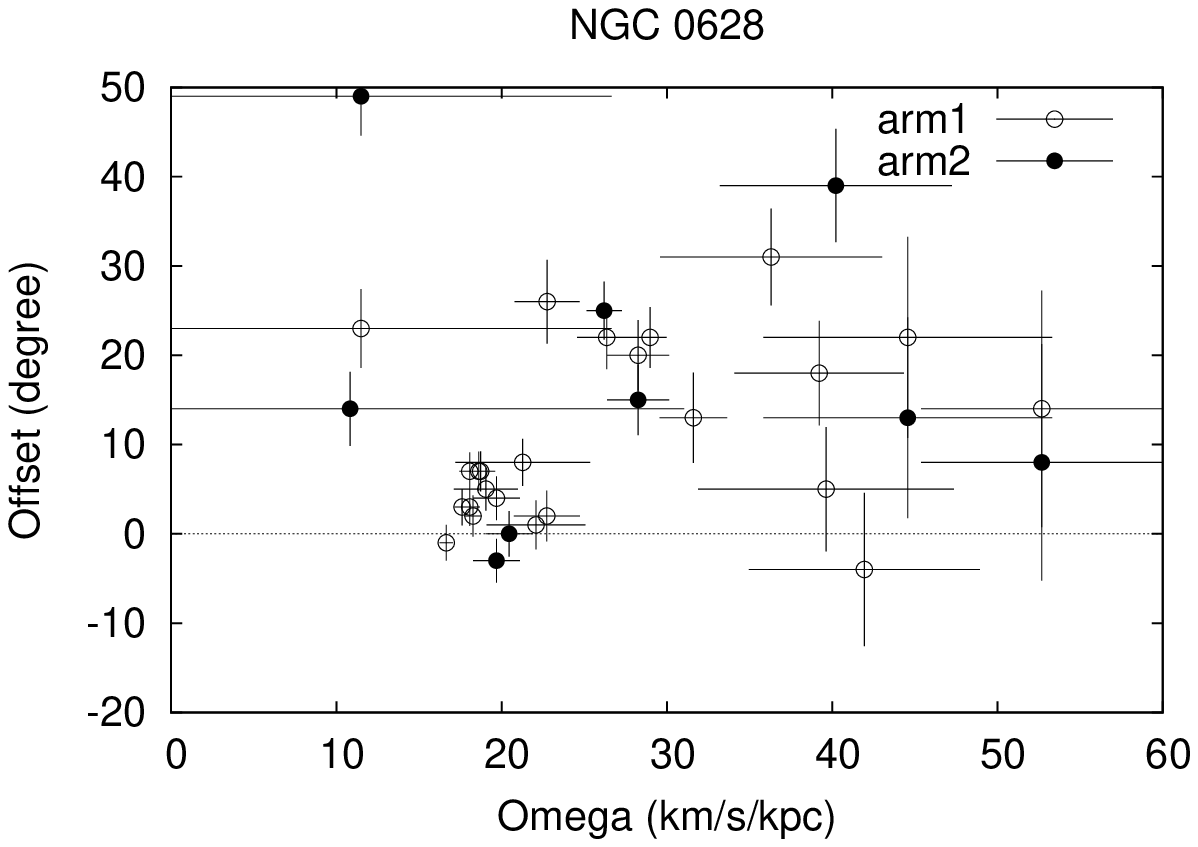}
\end{center}
\end{minipage}
\hspace{0.03\linewidth}
\begin{minipage}{0.3\linewidth}
\begin{center} 
\includegraphics[width=\linewidth]{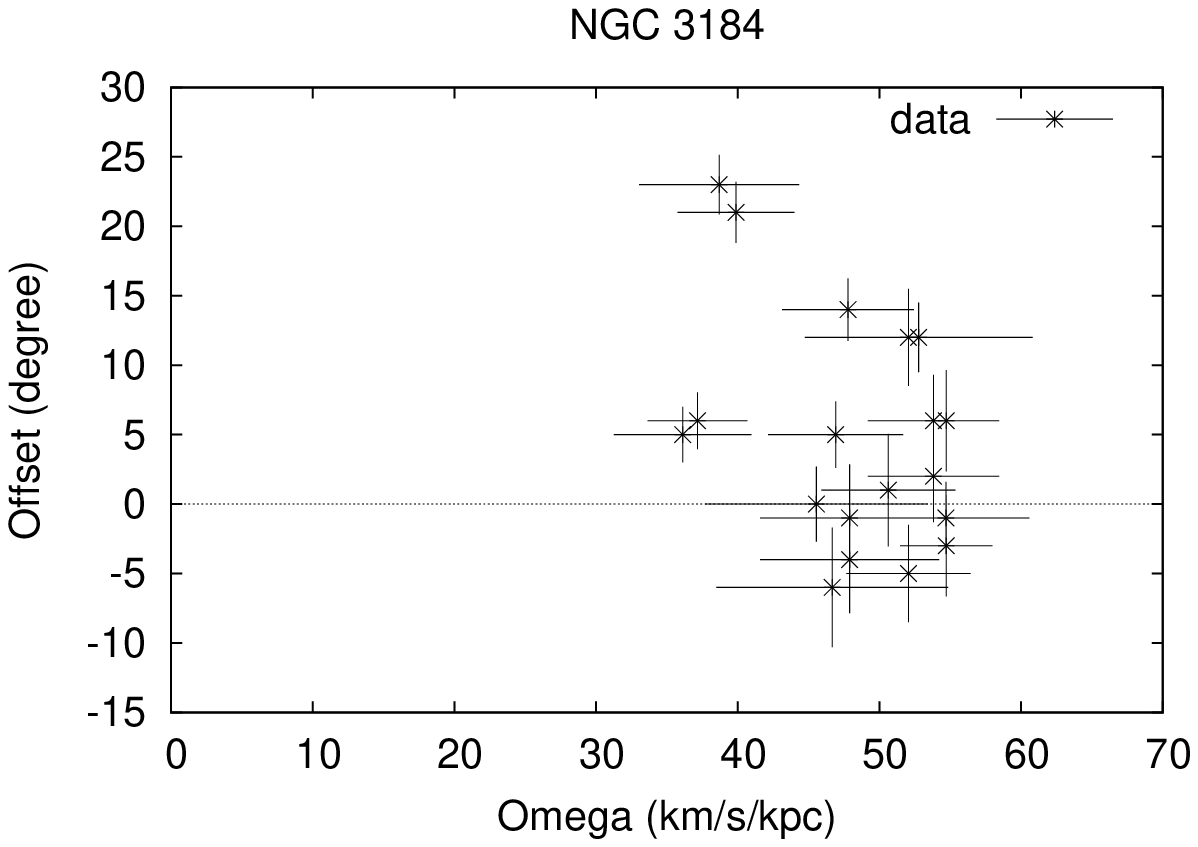}
\end{center}
\end{minipage}
\hspace{0.03\linewidth}
\begin{minipage}{0.3\linewidth}
\begin{center} 
\includegraphics[width=\linewidth]{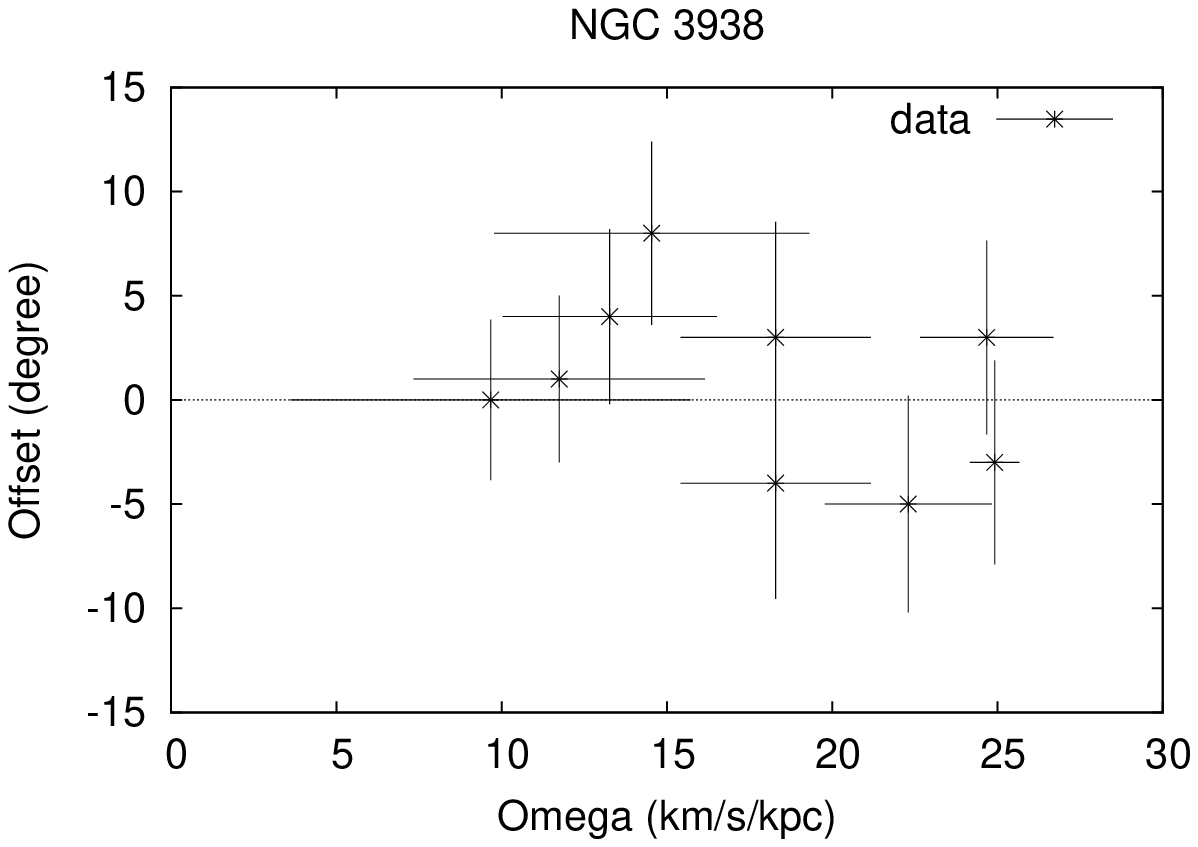}
\end{center}
\end{minipage}\\
\vspace{12pt}
\begin{minipage}{0.3\linewidth}
\begin{center} 
\includegraphics[width=\linewidth]{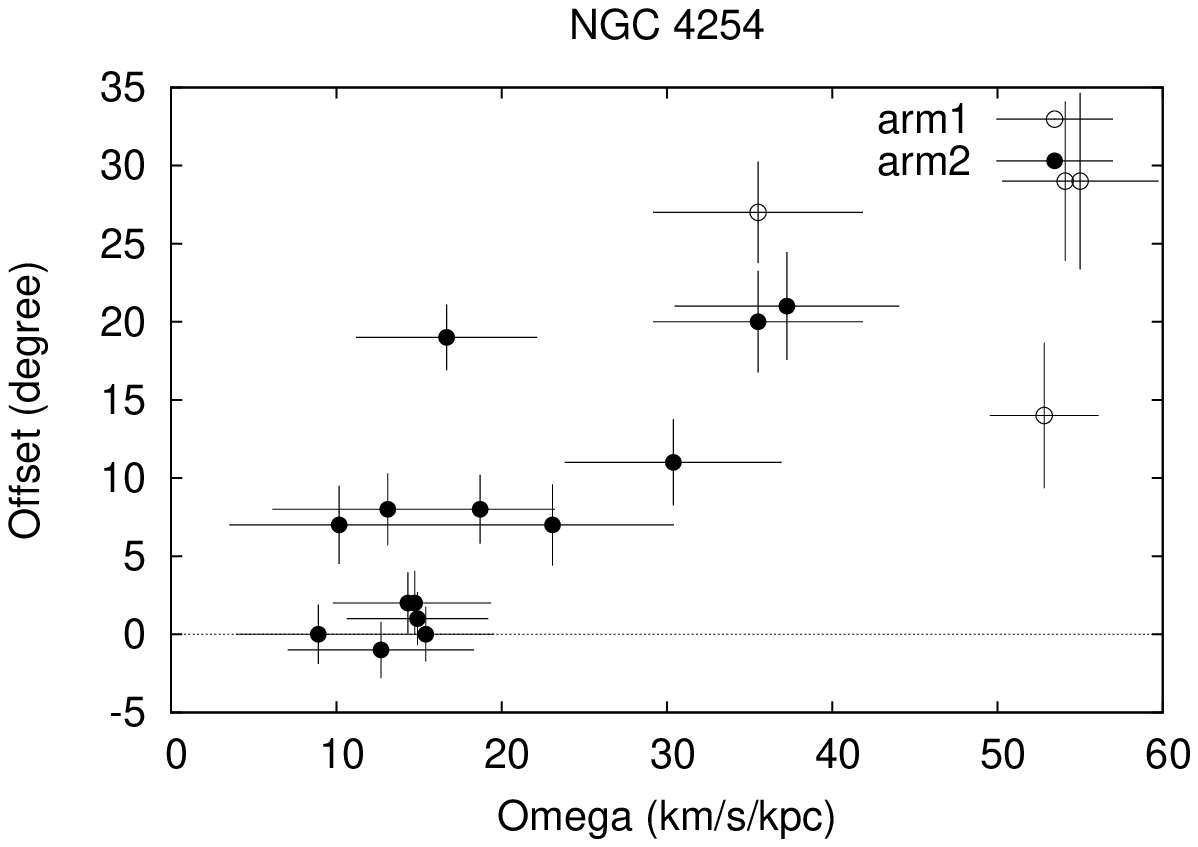}
\end{center}
\end{minipage}
\hspace{0.03\linewidth}
\begin{minipage}{0.3\linewidth}
\begin{center} 
\includegraphics[width=\linewidth]{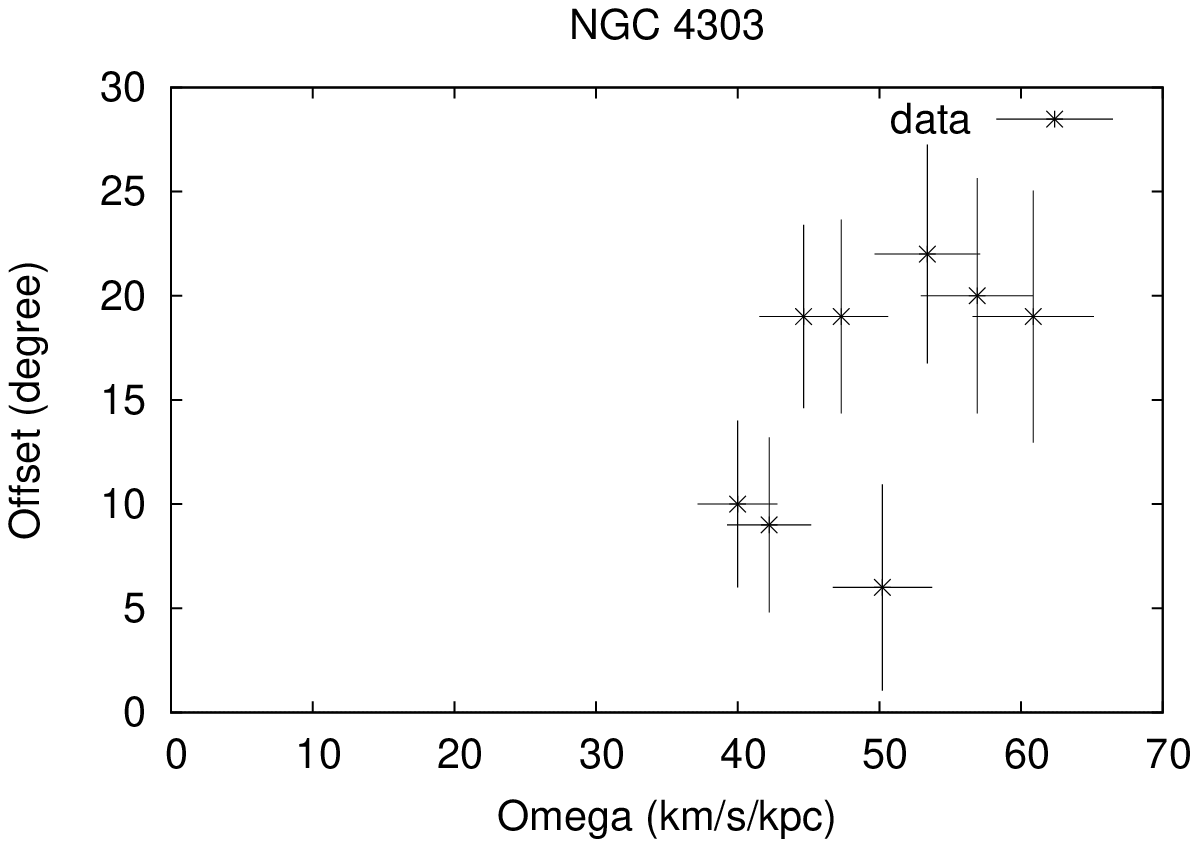}
\end{center}
\end{minipage}
\hspace{0.03\linewidth}
\begin{minipage}{0.3\linewidth}
\begin{center} 
\includegraphics[width=\linewidth]{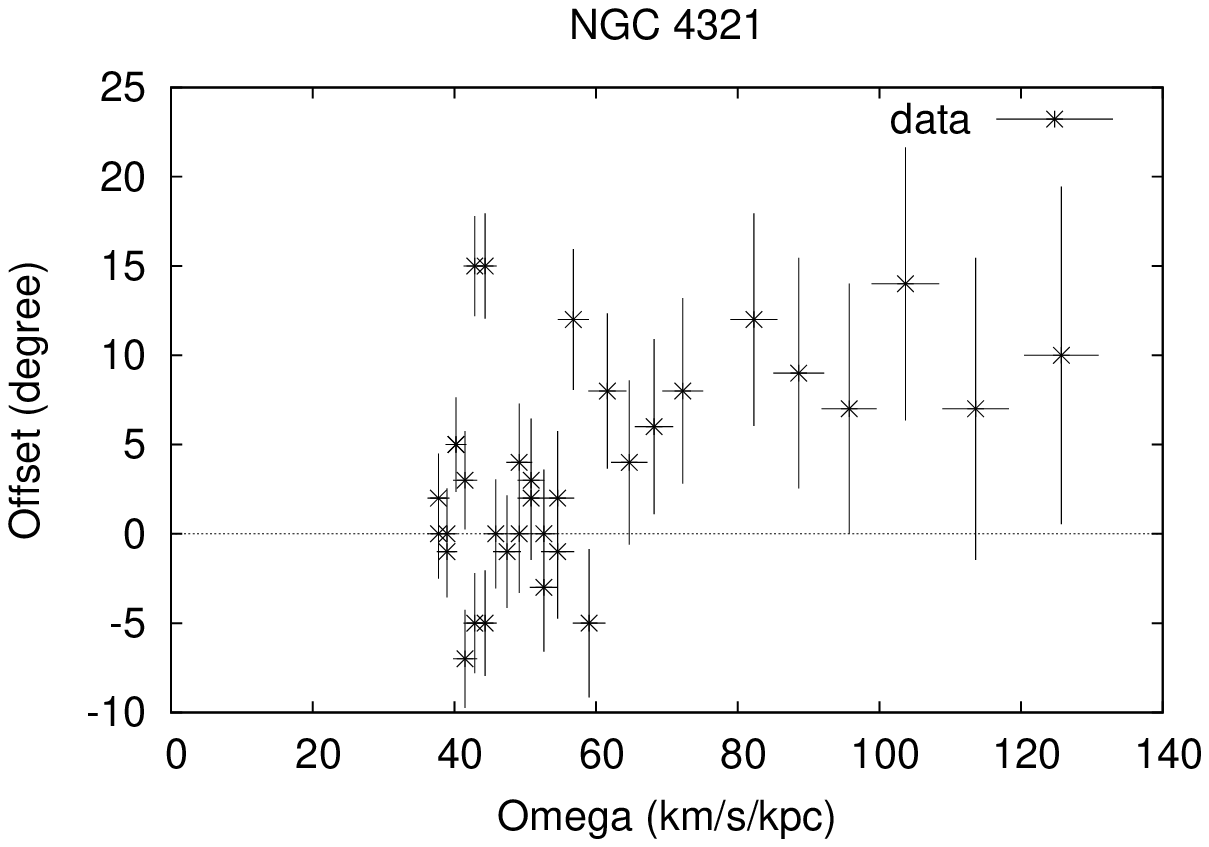}
\end{center}
\end{minipage}\\
\vspace{12pt}
\begin{minipage}{0.3\linewidth}
\begin{center} 
\includegraphics[width=\linewidth]{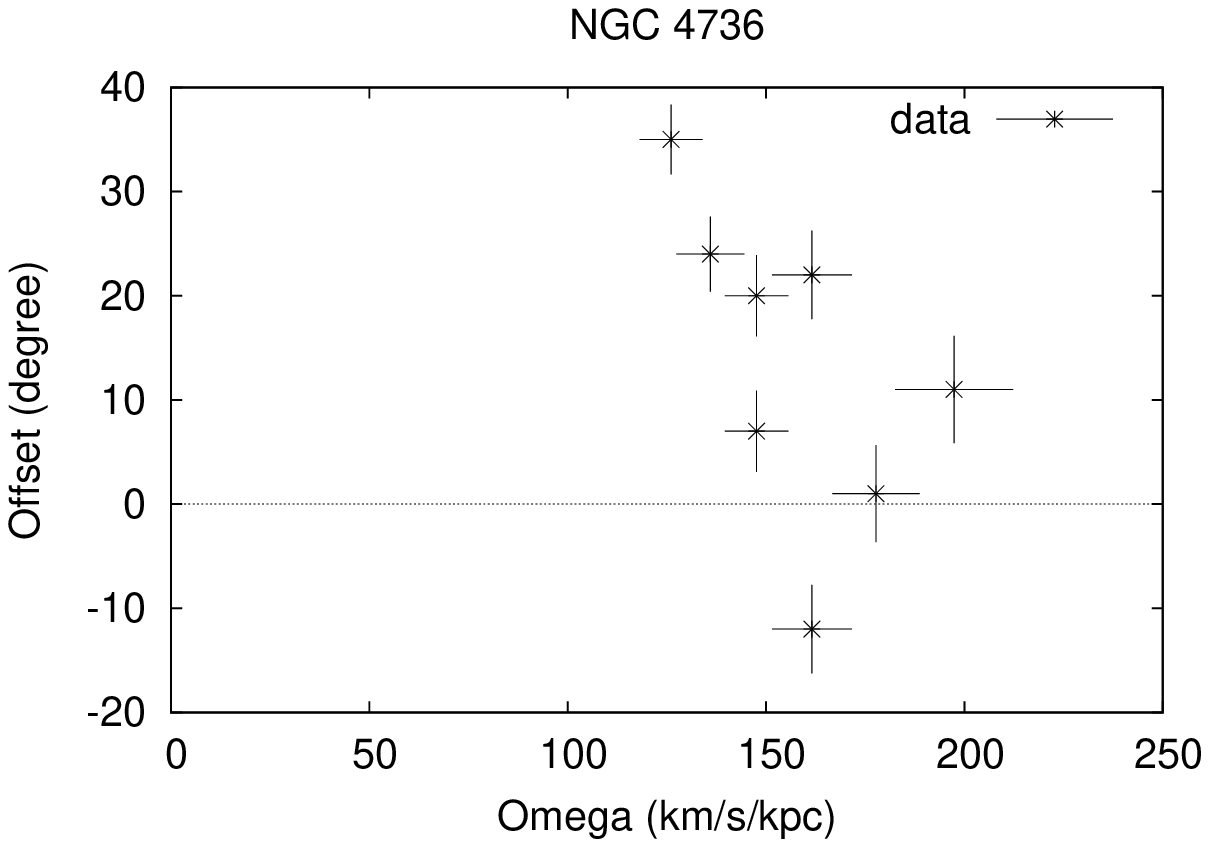}
\end{center}
\end{minipage}
\hspace{0.03\linewidth}
\begin{minipage}{0.3\linewidth}
\begin{center} 
\includegraphics[width=\linewidth]{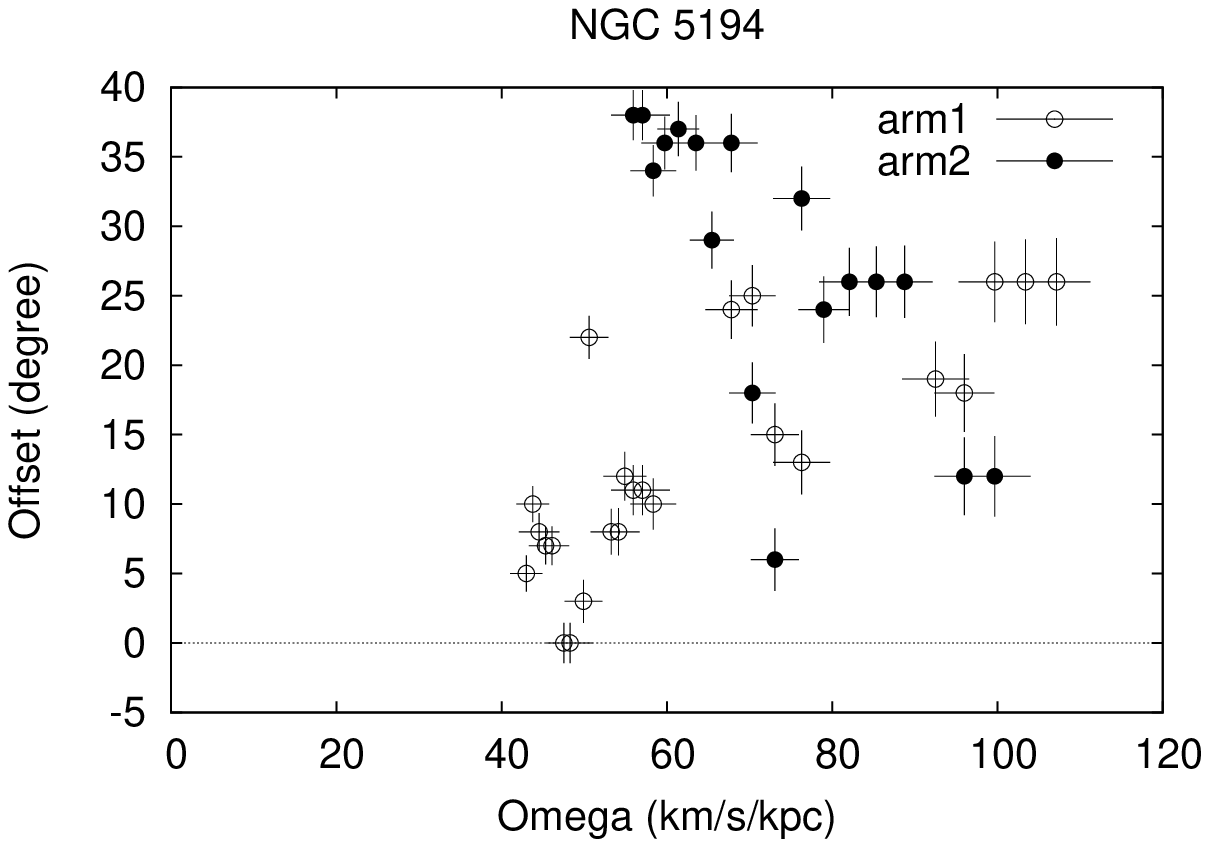}
\end{center}
\end{minipage}
\hspace{0.03\linewidth}
\begin{minipage}{0.3\linewidth}
\begin{center} 
\includegraphics[width=\linewidth]{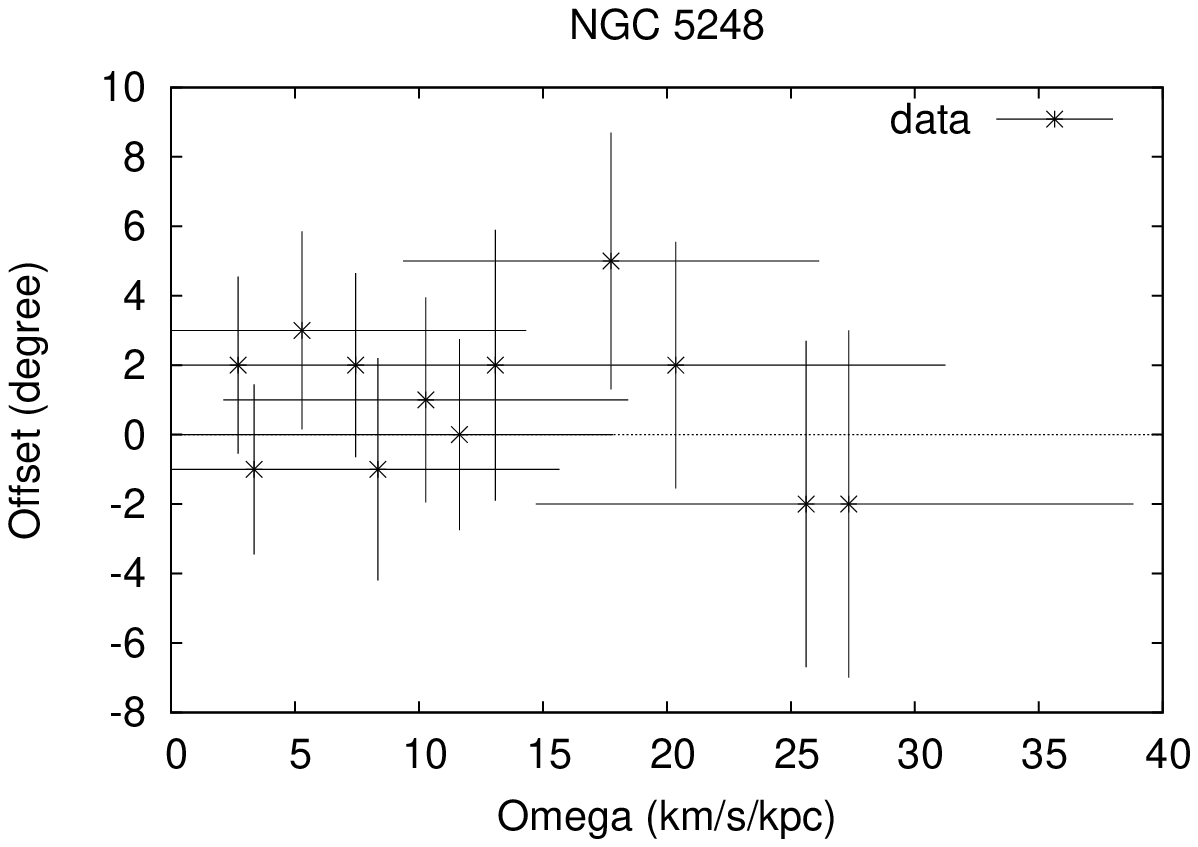}
\end{center}
\end{minipage}\\
\vspace{12pt}
\begin{minipage}{0.3\linewidth}
\begin{center} 
\includegraphics[width=\linewidth]{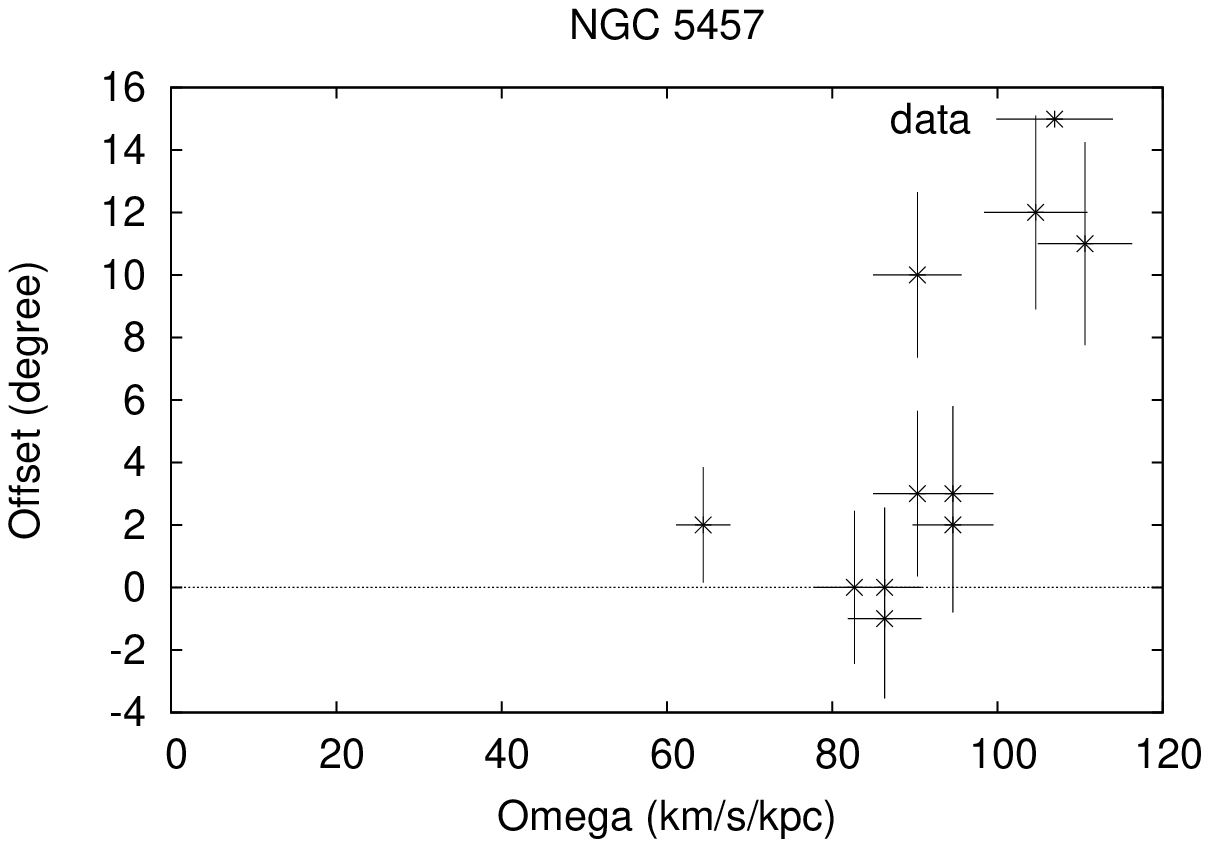}
\end{center}
\end{minipage}
\hspace{0.03\linewidth}
\begin{minipage}{0.3\linewidth}
\begin{center} 
\includegraphics[width=\linewidth]{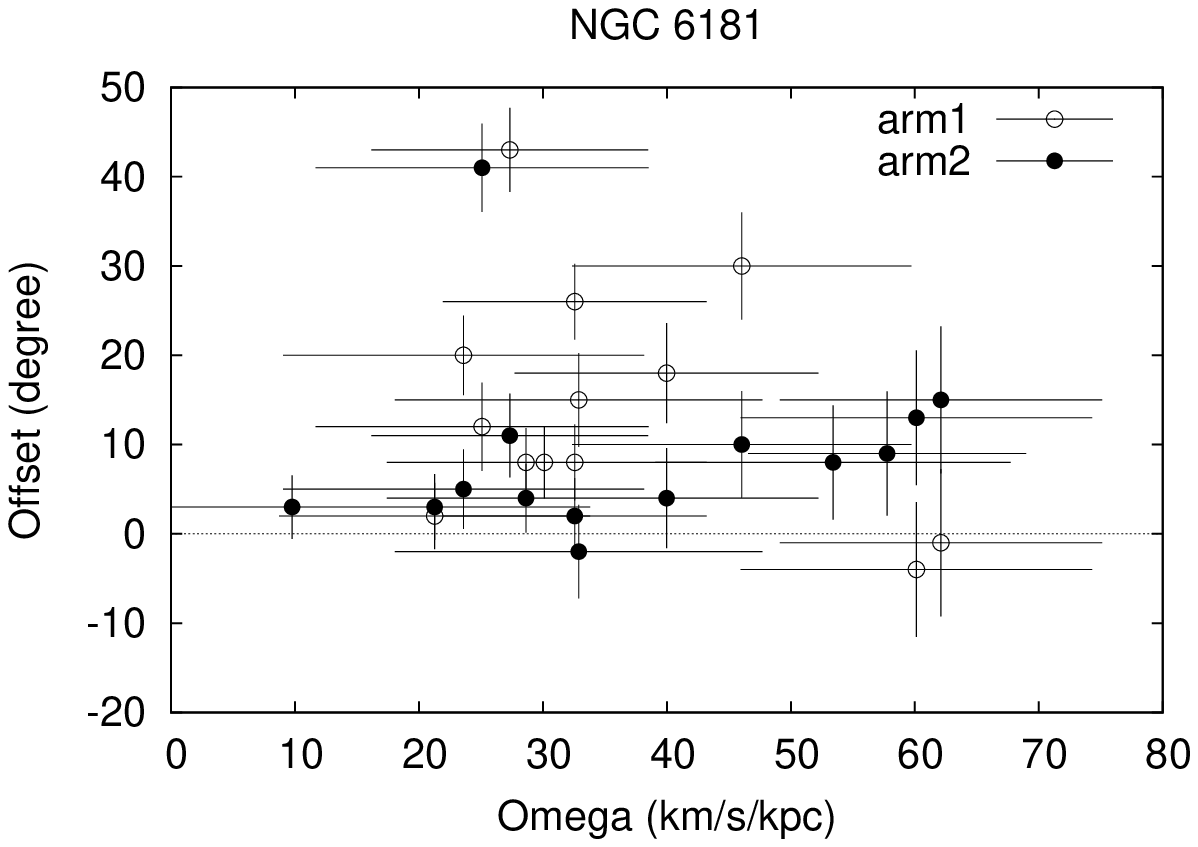}
\end{center}
\end{minipage}
\hspace{0.03\linewidth}
\begin{minipage}{0.3\linewidth}
\begin{center} 
\includegraphics[width=\linewidth]{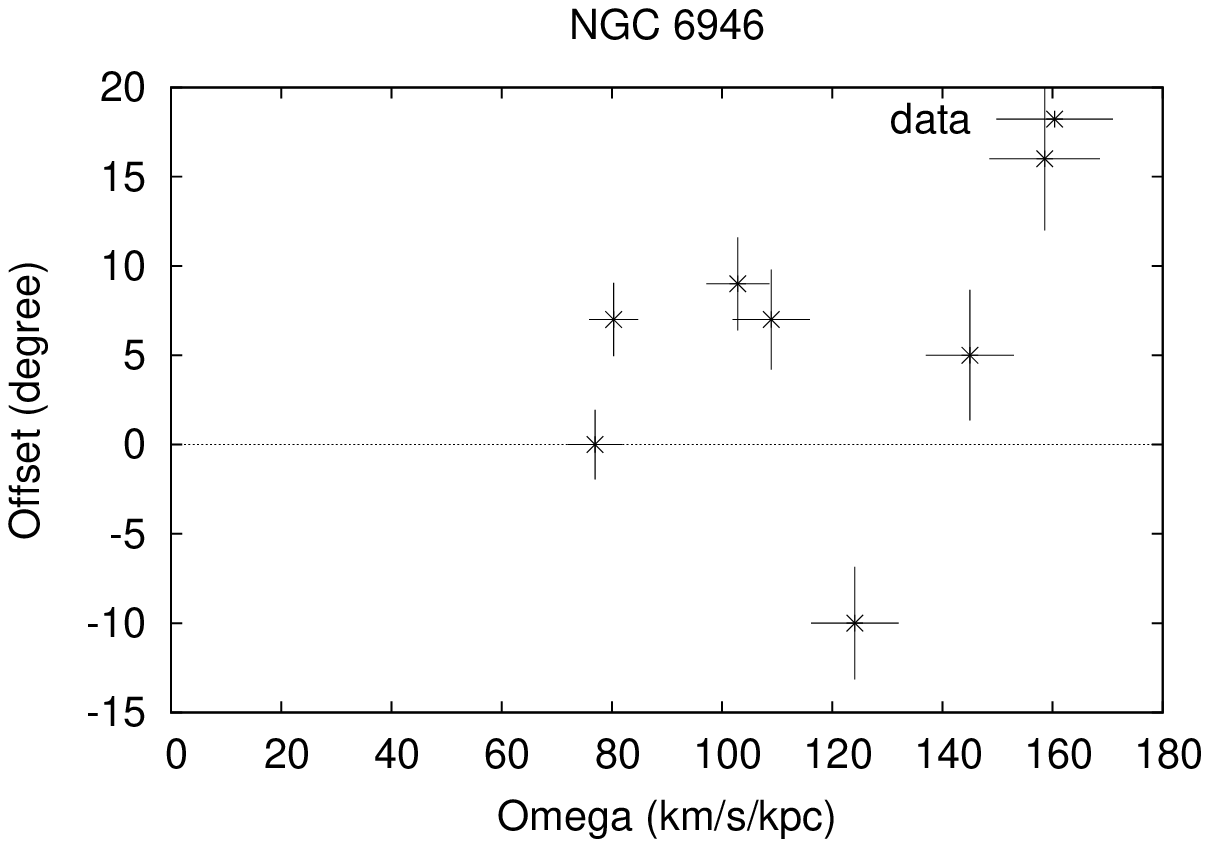}
\end{center}
\end{minipage}
\caption{Plot of $\theta$ against $\Omega$. 
Horizontal lines indicate $\theta=0$.
For several galaxies, two symbols (open and filled circles) are used to show 
difference between arms.}
\label{ofs.fig}
\end{center}
\end{figure*}

The measured offsets ($\theta$) were plotted against $\Omega$ in Figure \ref{ofs.fig}.
 The error bar in $\theta$ corresponds to the spatial resolution of CO data, which is the largest factor 
of uncertainty in $\theta$.  The uncertainty 
in $\Omega$ is calculated from errors of the RCs.
 Since errors of RCs from \citet{RC99} were not available, 
we assumed an error of $\pm 20$ \kms.
 The least $\chi^2$ method was used to fit the $\Omega-\theta$ plot and to derive 
\OP ~and \tsf ~with errors.
 The inclusion of the uncertainty in $\Omega$ and the use of the least $\chi^2$ method 
for the fitting are major revisions from what we have presented in the previous paper.
 Results from the fitting procedure are presented in \S \ref{sec.dis_indivi}.

\section{Result\label{sec.dis_indivi}}
 As described in the previous section, we show a plot of 
$\theta$ against $\Omega$ for 12 sample galaxies in Figure \ref{ofs.fig}. 
 For several galaxies with a substantial number (about 20 or more) of offsets, offset distributions for each arm 
are shown by different symbols. 
 We found that the $\Omega-\theta$ distribution varies with galaxies, and that 
for some galaxies, $\theta$ can even show a negative dependence on $\Omega$, 
which is not explained by our picture or Figure \ref{fig.draw}.
 From these plots and the CO and \Ha ~images, we categorized the sample galaxies into 3 types as 
\begin{description}
\item [C:] galaxies with clear offsets, which show clear spiral structures in CO and \Ha, 
and the fit to their $\Omega-\theta$ plot results in $t_{\rm SF} > 0$ and 
$\Delta\Omega_{\rm P}/\Omega_{\rm P}\lesssim 1$,
\item [N:] galaxies with no offsets, which show clear spiral structures, but their offsets in arm regions are almost zero, 
\item [A:] galaxies with ambiguous offsets, which cannot be categorized into the above two, 
mostly because their $\Omega-\theta$ plot show a negative correlation or large dispersion 
(i.e., $t_{\rm SF} < 0$ or $\Delta\Omega_{\rm P}/\Omega_{\rm P}\gtrsim 1$).
\end{description}

 For our sample of 12 galaxies, we have 5 in C, 2 in N, and 5 in A.
 For the 5 galaxies in category C, we derived meaningful values of \OP ~and \tsf ~by the $\chi^2$ fitting.
 Results of the fitting and the offset category are listed in Table \ref{oc.tb}.

 Results and discussions for each galaxy are presented in \S \ref{sec.indivi}, and 
characteristics and differences in physical and observational parameters of the 3 categories 
are discussed in \S \ref{sec.dis_cat}.
 Discussions on the derived \OP ~and \tsf ~are presented in 
\S \ref{sec.dis_Op} and \ref{sec.dis_tsf}, respectively.

\subsection{Description for individual galaxies\label{sec.indivi}}
\subsubsection{Galaxies with Clear Offsets\label{sec.dis_indiviC}}
\begin{figure*} 
\begin{center}
\includegraphics[width=0.45\linewidth]{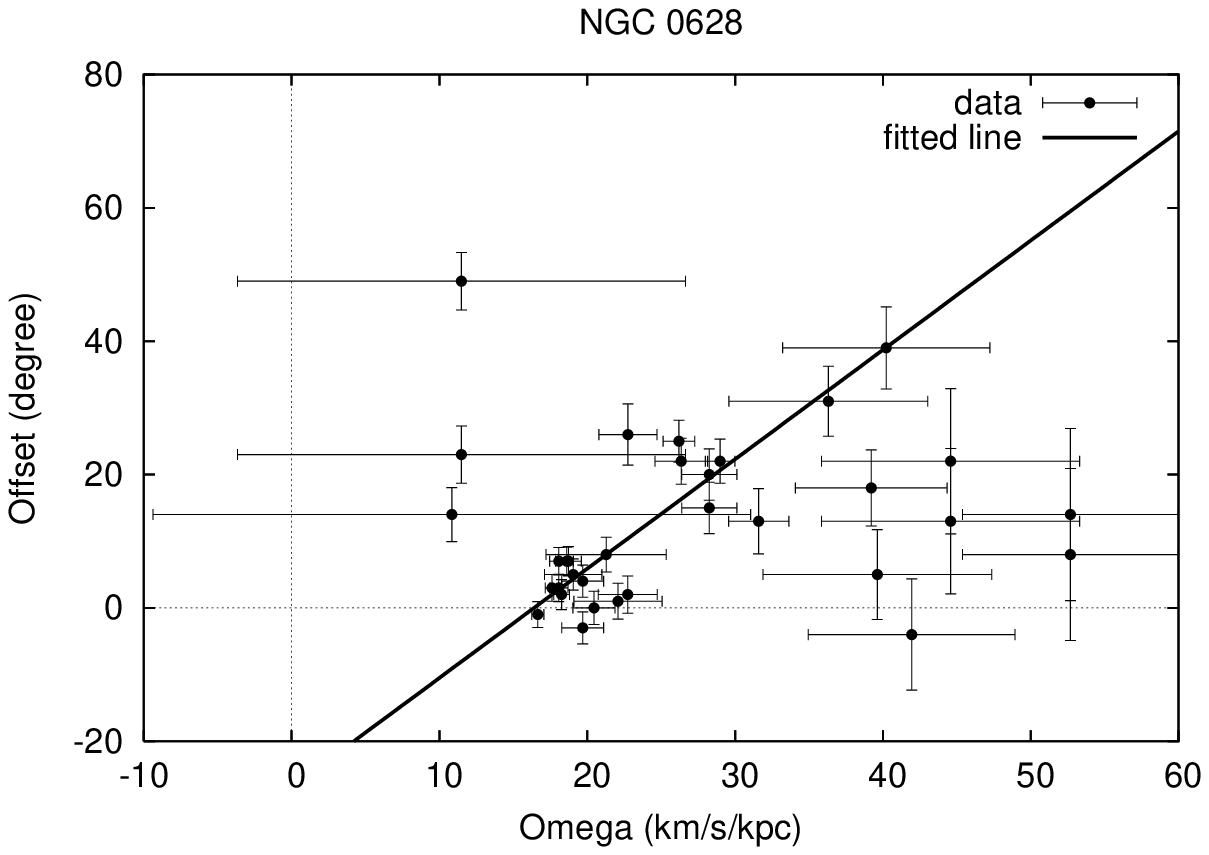}
\hspace{0.05\linewidth}
\includegraphics[width=0.45\linewidth]{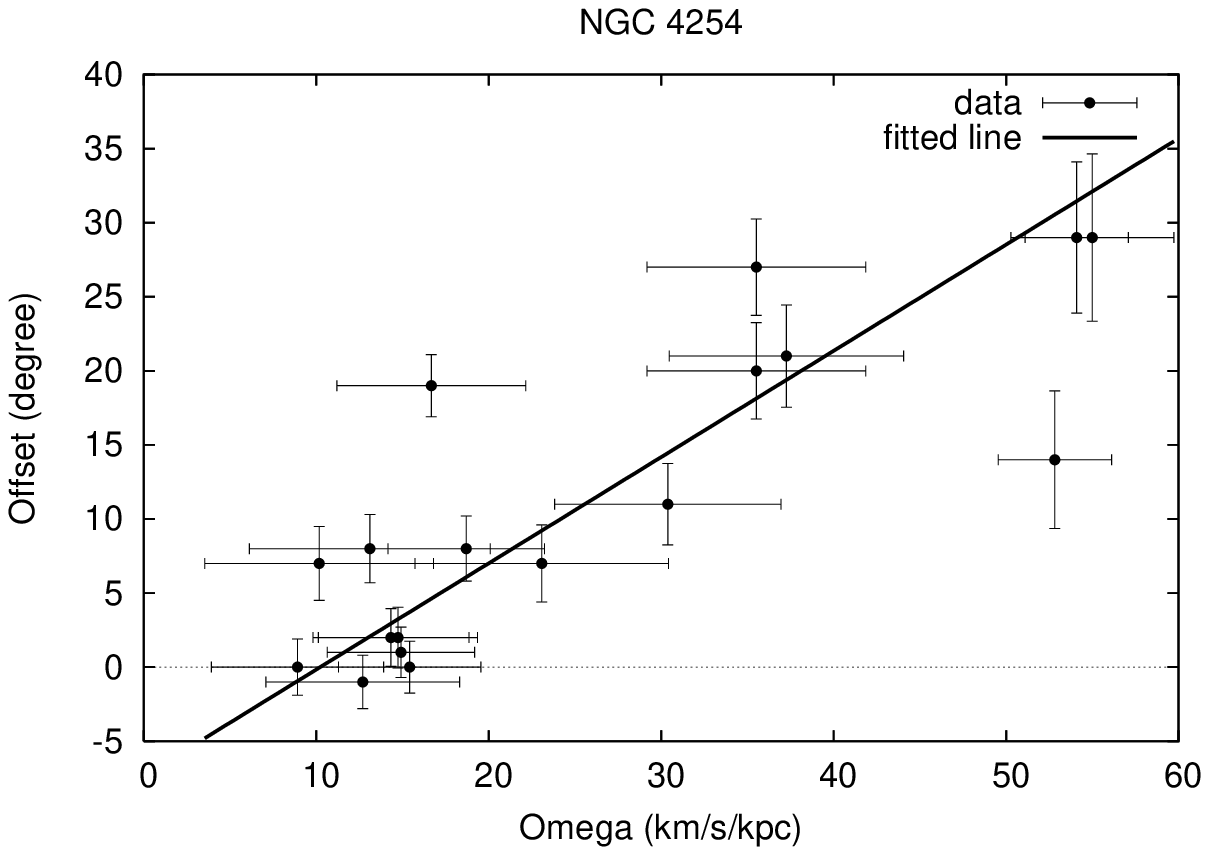}\\
\vspace{12pt}
\includegraphics[width=0.45\linewidth]{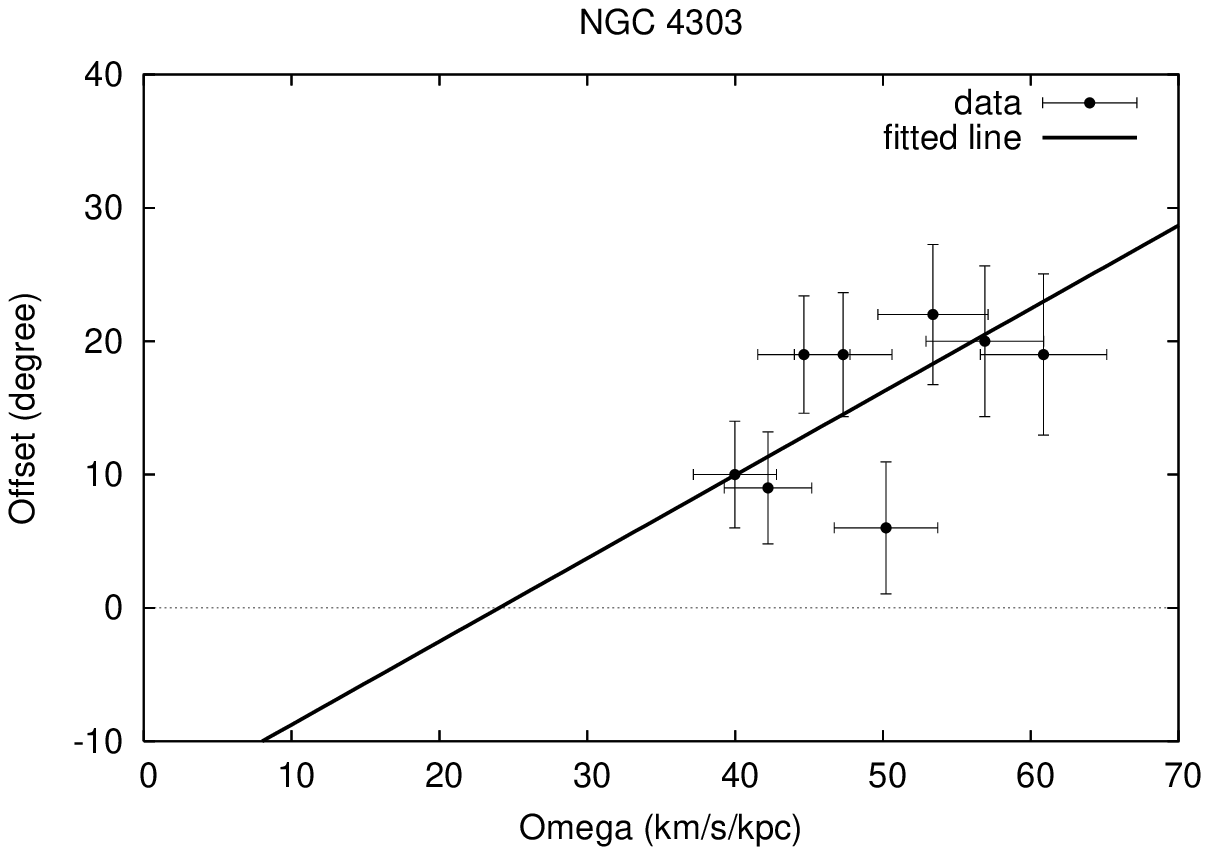}
\hspace{0.05\linewidth}
\includegraphics[width=0.45\linewidth]{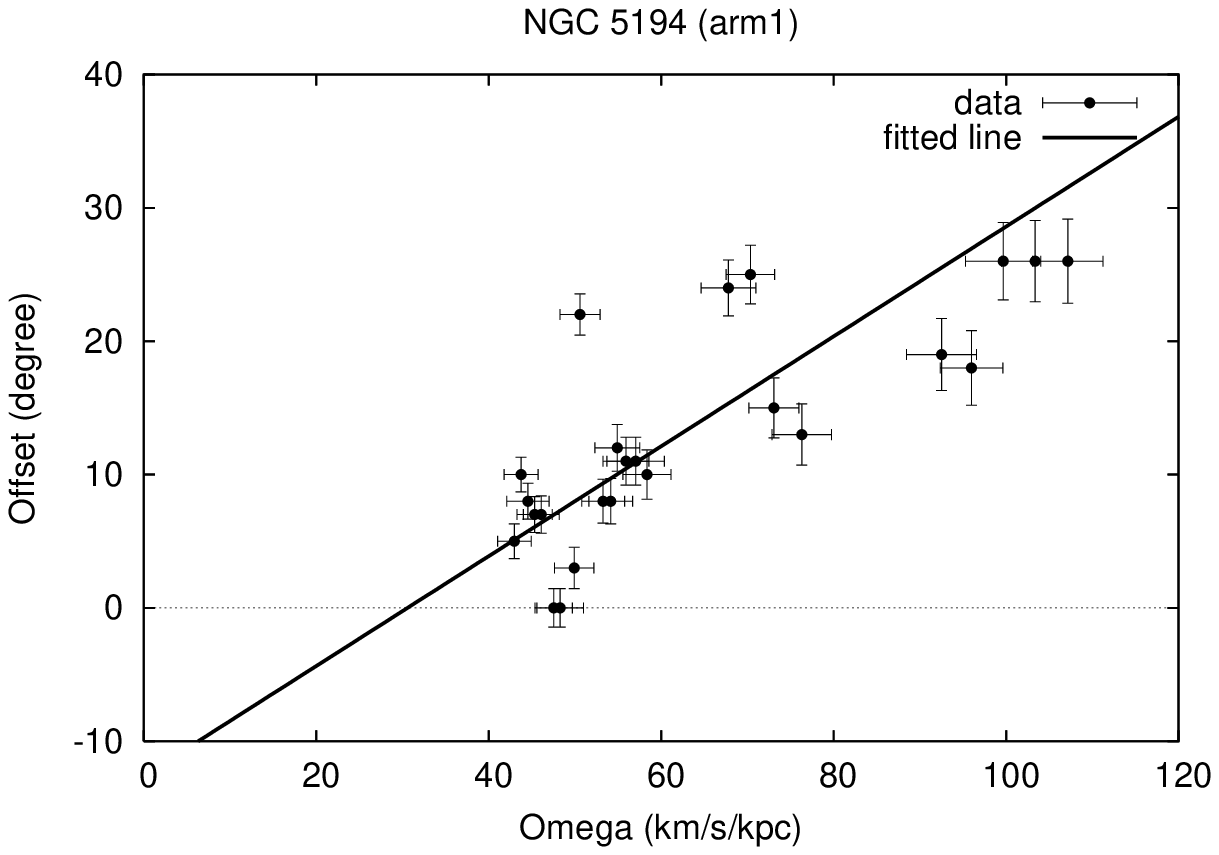}\\
\vspace{12pt}
\includegraphics[width=0.45\linewidth]{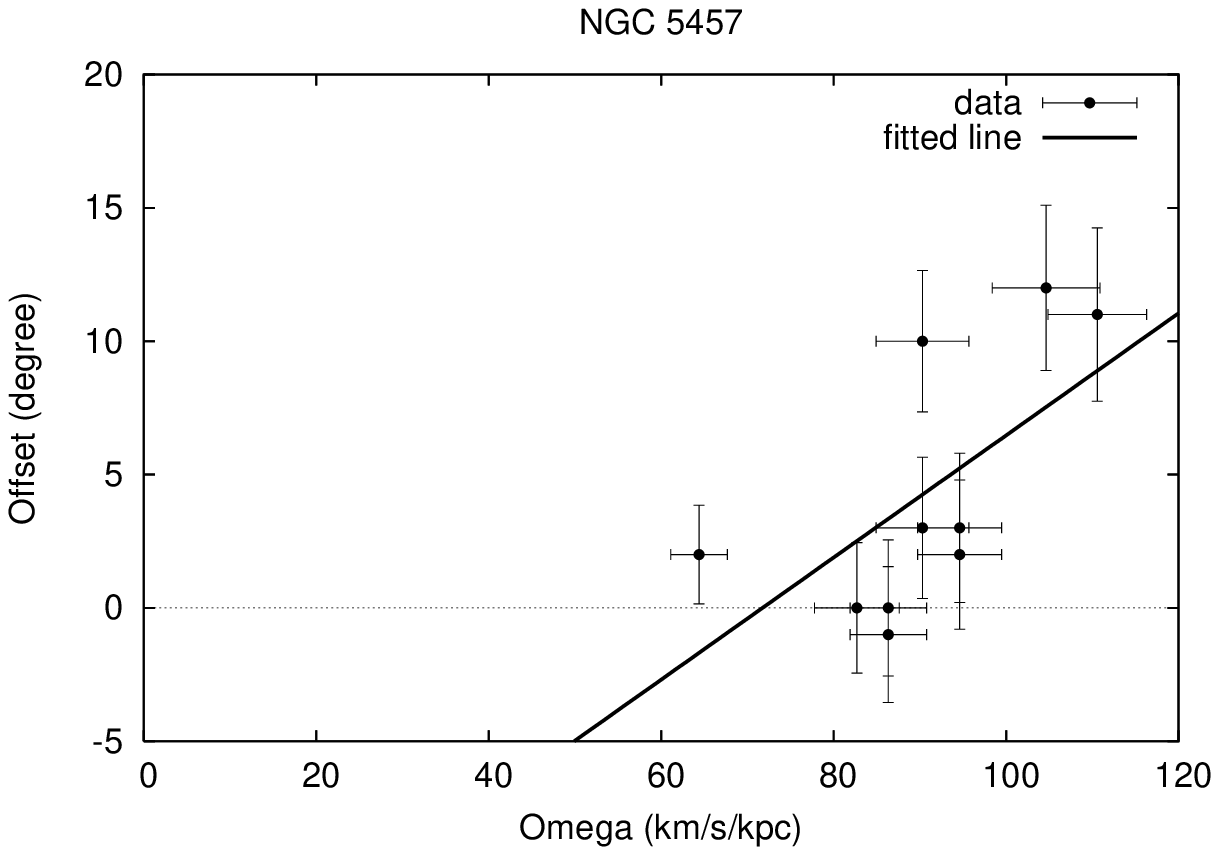}
\caption{$\Omega-\theta$ plot with the fitted line for each galaxy in the C category.
The gradient and horizontal-axis intercept of the fitted line correspond to \tsf ~and \OP, respectively.}
\label{fit.fig}
\end{center}
\end{figure*}

\begin{figure*}[h!]
\begin{center}
\includegraphics[width=0.45\linewidth]{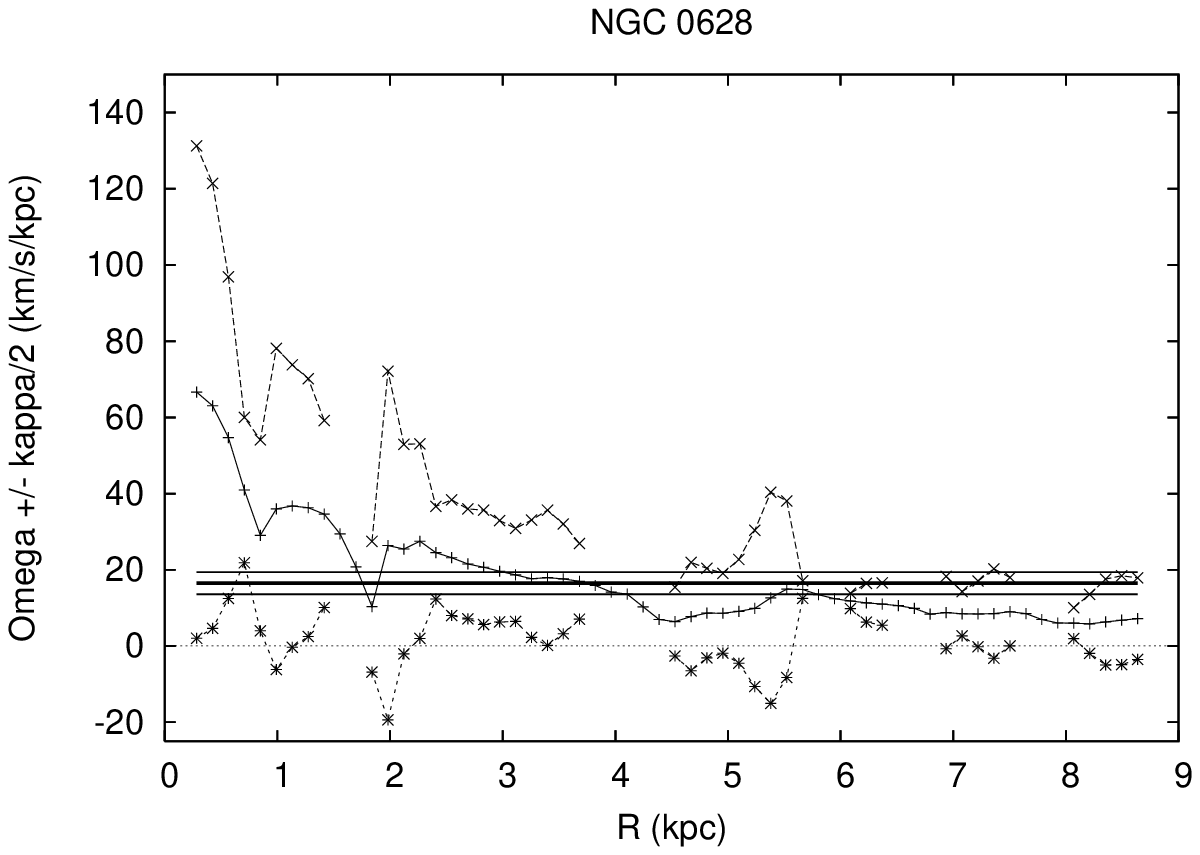}
\hspace{0.05\linewidth}
\includegraphics[width=0.45\linewidth]{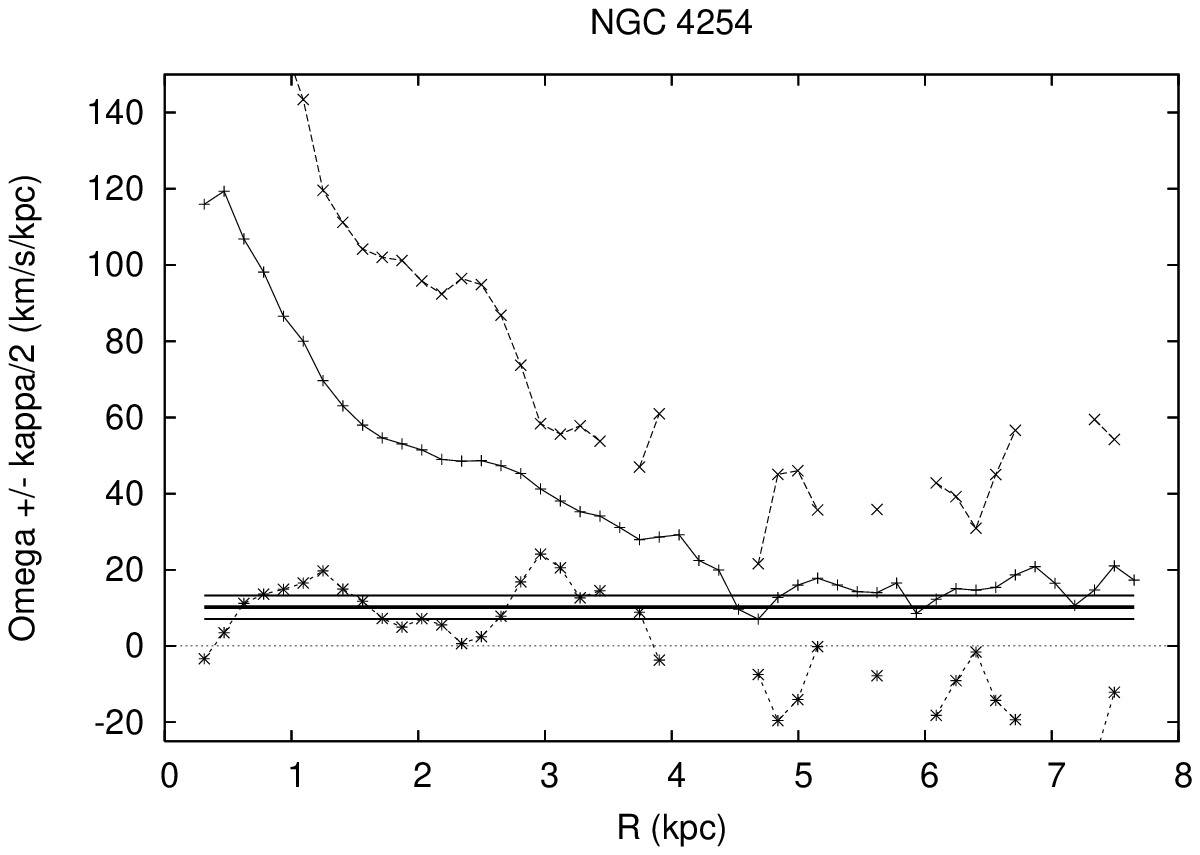}
\hspace{0.05\linewidth}
\includegraphics[width=0.45\linewidth]{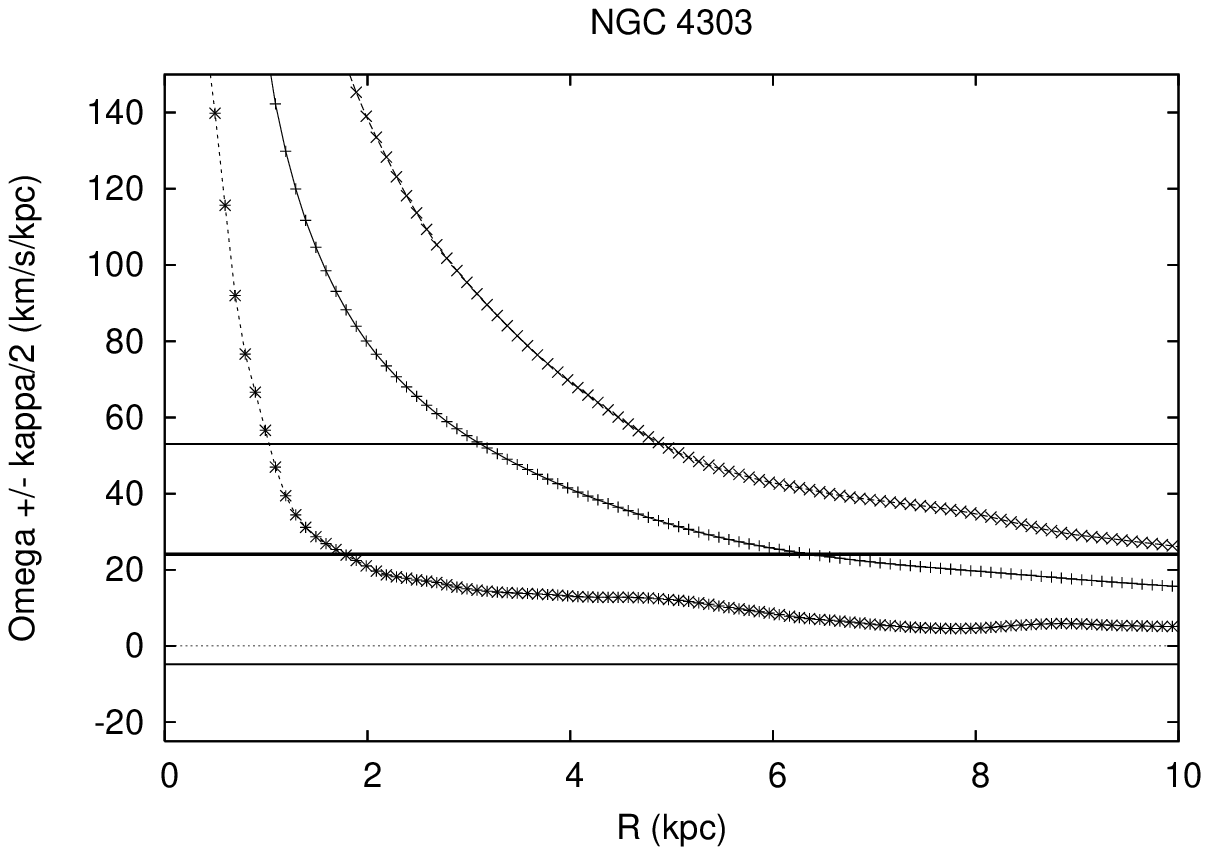}
\hspace{0.05\linewidth}
\includegraphics[width=0.45\linewidth]{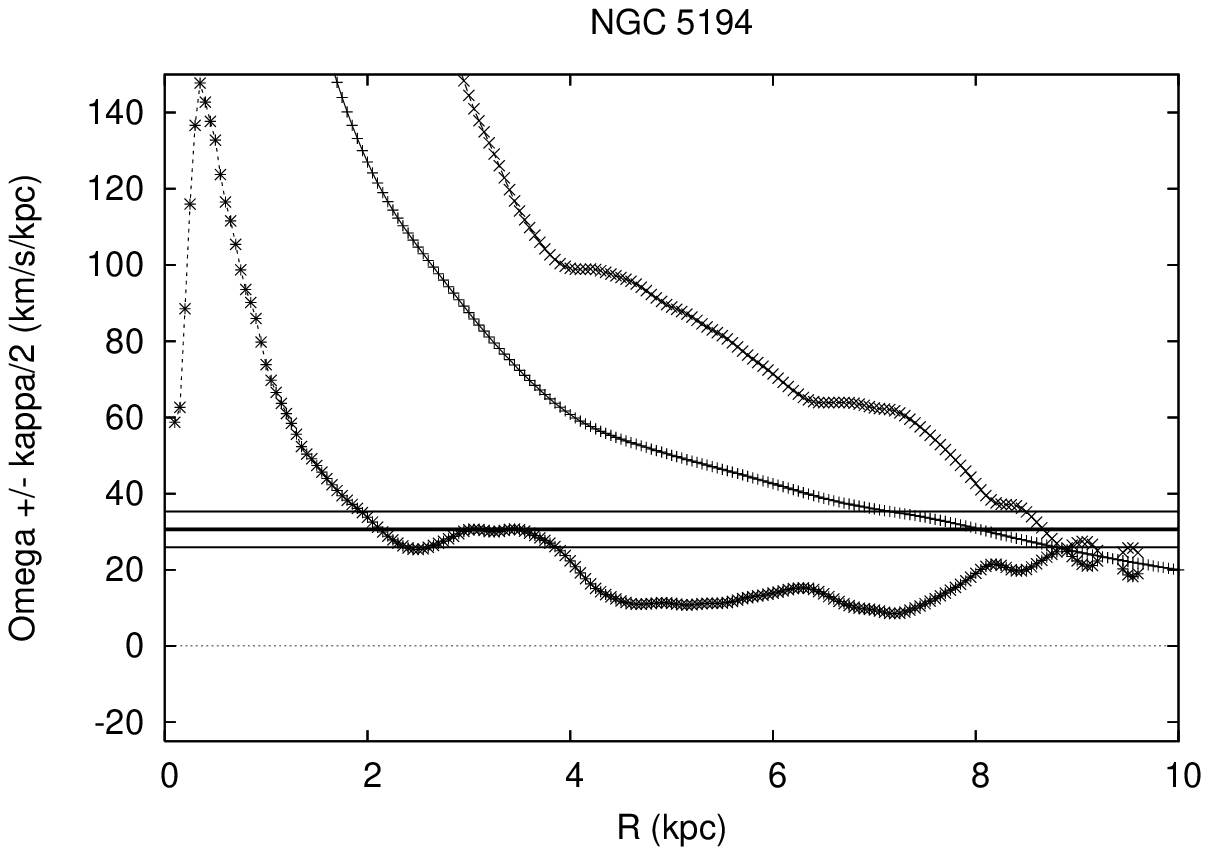}
\hspace{0.05\linewidth}
\includegraphics[width=0.45\linewidth]{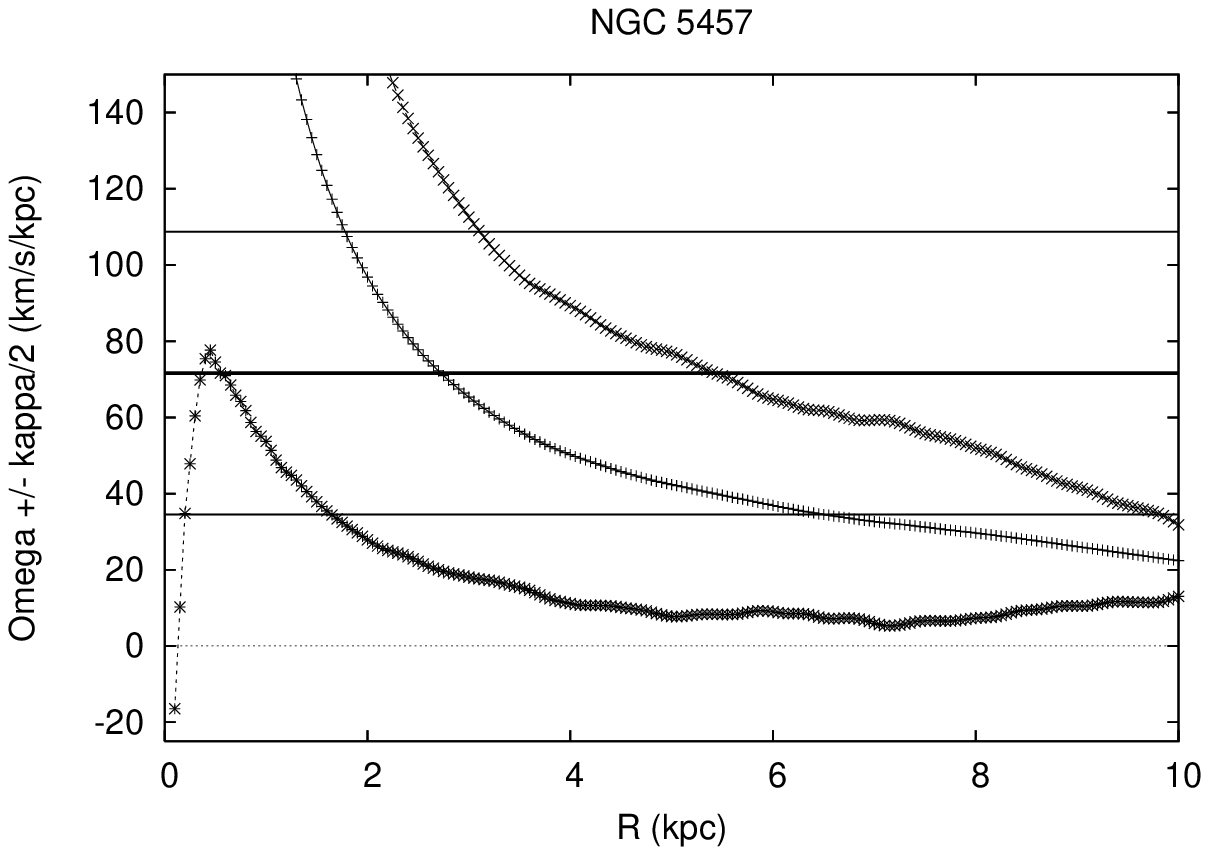}
\caption{Plot of $\Omega$ (solid line with '+'), $\Omega + \kappa/2$ (dashed line with '$\times$'), 
and $\Omega - \kappa/2$ (dashed line with '*') as a function of radius. 
Derived \OP ~is presented as a thick horizontal line with thin solid lines indicating the uncertainty. 
The corotation and Lindblad resonances are 
derived to be where $\Omega_{\rm P}=\Omega$ and $\Omega_{\rm P}=\Omega\pm \kappa/2$, 
respectively.}
\label{resonance.fig}
\end{center}
\end{figure*}

 Here, we describe our results for ``C'' galaxies.
 The $\Omega-\theta$ plot with the fitted line is shown in Figure \ref{fit.fig}. 
 In Figure \ref{resonance.fig}, we plot $\Omega$ and $\Omega\pm \kappa/2$ as a function of radius 
to locate the kinetic resonances.
 We could derive the lower and upper limit of \Rcr ~for 4 out of 5 ``C'' galaxies, 
while only the lower limit was derived 
for the remaining 1 galaxy, NGC 4303, whose uncertainty in \OP ~is as large as \OP ~itself.
 On the other hand, the insufficient quality of RC for some galaxies or at some radii 
hampers reliable derivation of $\kappa$, which makes it difficult to derive the Lindblad resonances.
 We could thus locate the ILR for NGC 5194 only.

 Comparison with previous studies is also presented in this section.
 Values of \OP ~and \Rcr ~from this and other studies are listed in Table \ref{OP_C.tb}.

\paragraph{NGC 0628 (M 74)} 
 We were able to measure 33 offsets between CO and \Ha ~at $r=0.5-3.7$ kpc ($0'.2-1'.7$), 
while $\Omega$ was calculated from the RC of \citet{SINGSHa} with a correction for 
the adopted inclination angle of 24 degree. 
 We should note that even though the $\Omega-\theta$ plot shows a large dispersion, 
the fitted line has a positive gradient since the outlying data with large errorbars 
have lower weights in the $\chi^2$-fitting.
 Removing such data points with $\Delta \Omega/\Omega > 1$ did not substantially change 
the fitted result.
 As offset distributions for the 2 arms appear to be consistent, 
we derive only one value for \tsf ~and \OP.

 From the derived value of \OP ~and the RC, the corotation radius 
is calculated to be $R_{\rm CR}\simeq 2-4$ kpc, corresponding to $0'.9-1'.9$.
 With $R_{\rm CR}\simeq 5$ kpc or $1'.8$ located by \citet{CB90} from the arm/interarm ratio of SFE, 
\citet{Fath07a} derived $\Omega_{\rm P}=31^{+5}_{-3}$ \kmskpc ~with the same RC which we used. 
 We should mention that the discrepancy in \OP ~is due to the difference in the assumed distance and inclination angle, 
and that the location of the corotation resonance is consistent within errors.
 However, \citet{EEM92} examined the $B$-band morphology of this galaxy and derived $R_{\rm CR}\sim 2'.4$, 
which is larger than our result.

\paragraph{NGC 4254 (M 99)}
 For this 3-armed spiral galaxy, we were able to measure 18 offsets between CO and \Ha ~for the arm1, 
extending from east to south, 
and arm2, extending from west to north, at $r= 2-6$ kpc ($0'.4-1'.3$).
 The difference in the $\Omega-\theta$ distribution for the 2 arms is not clear, 
since the number of offsets is small for arm2 (see Figure \ref{ofs.fig}).
 We could not measure any offsets for arm3, originating from the same position as the arm1 
but with much smaller pitch angle extending to north, since the \HII ~regions were not 
well aligned where the CO emission delineates the arm.
 
 This galaxy has an asymmetric structure, as one spiral arm (arm1) is 
much prominent than the other two, which could be explained by a 
ram pressure effect from intra-cluster matter in the Virgo cluster \citep{Hida02, ViCS3}. 
 In addition, a faint but large \HI ~structure, VIRGOHI 21, which might be interacting 
with this galaxy, has recently been found \citep{HGK07, Min07} and is also thought to be responsible for the asymmetry.
 This global asymmetry could explain why only a few offsets could be measured in arm2, and none in arm3.

 From the derived value of \OP ~and the RC, the corotation radius 
was calculated to be $R_{\rm CR}\simeq 4.5-6.0$ kpc, corresponding to $1'.0$--$1'.3$.
 Our result is consistent with \citet{Kranz01}, 
who used results from hydrodynamic simulations to locate the \Rcr.

 In our previous work, \OP ~and \tsf ~were derived to be $28^{+10}_{-6}$ \kmskpc ~and 
$4.8\pm 1.2$ Myr, respectively, while 
this work gives $10\pm 3$ \kmskpc ~and $12.4\pm 1.3$ Myr. 
 The difference between the previous and current work is the fitting scheme, 
CO data (both in map and RC), and the adopted inclination angle.
We have included the uncertainty in $\Omega$ in the fitting procedure, and furthermore,
 the number of data and the radial range of RC have been increased 
by our new CO observations with 3 pointings with NMA.
 These two aspects should have improved the reliability of results.
 The assumed inclination angle is changed from $34^\circ$ to $52.4^\circ$.
 However, since $\sin (34^\circ)/\sin (52.4^\circ)\sim 0.7$, this is not enough to explain 
the discrepancy in \OP ~and \tsf ~(See \S \ref{sec.valid}).
 This implies that the uncertainties from the fitting could be underestimated.

\paragraph{NGC 4303 (M 61)}
 We were able to measure only 8 offsets between CO and \Ha ~at $r= 2.5-4$ kpc ($0'.5-0'.9$), 
so that the uncertainty in \OP ~is too large to derive \Rcr ~for this galaxy.
 We only give the lower limit of $R_{\rm CR} \gtrsim 3$ kpc or $0'.6$.

 As NGC 4303 has a double bar and an active nuclei, studies of this galaxy have concentrated on the 
dynamics and morphology in the central region.
 \citet{CW00} performed multiphase hydrodynamical simulations and derived 
$\Omega_{\rm P}\sim 0.5$ Myr$^{-1} \sim 500$ \kmskpc ~for the nuclear bar. 
 They locate the CR at the end of the nuclear bar, which is about $2''$ in radius. 
 On the other hand, \citet{RSL05} derived 
$R_{\rm CR}=1'.5\pm 0'.1$ for the outer bar with the size of $1'.5$.
 Although their \Rcr ~gives \OP ~close to our mean value of 24 \kmskpc, 
it is still unclear whether the bar and spiral arms have the same pattern speed.
 \citet{Koda06} calculated orbits in the bar from CO data, taken as a part of the Virgo High-Resolution 
CO Survey \citep{ViCS1}, and estimated ages of young clusters apparent in \Ha ~to be $\lesssim 10$ Myr, 
which is consistent to our result of $t_{\rm SF}\simeq 10$ Myr.

\paragraph{NGC 5194 (M 51)}
 We were able to measure 41 offsets between CO and \Ha ~at $r= 2.5-6$ kpc ($0'.9-2'.2$), 
and found that their dependence on $\Omega$ is different for the 2 spiral arms. 
 The offsets of arm2, directly connected to the companion galaxy NGC 5195, 
showed negative dependence on $\Omega$.  We have therefore derived 
\OP ~and \tsf ~by the $\chi^2$-fitting to the offsets of arm1 only.
The difference in 
the $\Omega-\theta$ distribution for the 2 arms, may be attributed to the tidal interaction
with the companion.
This is because the interaction can cause asymmetry in the disk, and the outer part of 
the spiral arm connected to the companion could be a material arm.

 From the derived \OP ~and the RC, we calculate 
$R_{\rm CR}=8.0^{+0.8}_{-1.0}$ kpc, or $2'.9^{+0.2}_{-0.4}$, and 
the ILR to be at $r\sim 2-4$ kpc, which is slightly outside of the inner edge of spiral arms.
 The derived values of \Rcr ~in Table \ref{OP_C.tb} span a range more than a factor of 2, 
indicating the difficulty of its determination.
 \citet{Zim04} derived \OP ~to be 
about 40 \kmskpc, from the TW method using the 45m CO data \citep{Naka94}, while 
\citet{Mei08} applied the radial TW method to other CO data \citep{Shet07} and derived 
two pattern speeds of about 90 and 50 \kmskpc ~for $r\lesssim 2$ kpc and $2\lesssim r \lesssim 4$ kpc, 
respectively, with a possible third pattern of $\Omega_{\rm P}\sim 20$ \kmskpc ~for 
$4\lesssim r \lesssim 5$ kpc.
 As the results from \citet{Zim04} and this work are both between 50 and 20, 
derived by \citet{Mei08}, these three works are in good agreement.
 On the other hand, \citet{Knap92} calculated the arm/interarm ratio of SFE along the two arms, 
and showed that positions of ILR, CR (at $r\sim 2'$ or 6 kpc), and OLR from \citet{EES} were in good 
agreement with the minimum of this ratio.
 They also found that the pattern of this ratio along the position in arms is 
quite similar for the two arms, and claimed that some global triggering mechanisms for star formation should be at work.
This pattern they found is not consistent with our result showing different offset 
distributions for the two arms, as well as the location of the CR.

 \citet{Bast05} estimated ages of \HII ~regions in this galaxy as $\lesssim 10$ Myr based on 
population synthesis models, and the derived value of \tsf ~is consistent to their ages.
 Recently published extinction-corrected distribution of star forming regions \citep{Kenn07} 
and the latest CO observations with the 45m and CARMA \citep{Koda09} of this galaxy will be able to improve 
the current results in terms of reliability of location of \HII ~regions, the spatial resolution 
of CO map, and the quality of the rotation curve.

\paragraph{NGC 5457 (M 101)}
 We measured 10 offsets at $r=1.7-3.1$ kpc ($0'.8-1'.5$). 
 Although the data range in $\Omega$ is rather narrow, we were able to derive \tsf ~and \OP.
 We should note that the region where we measured offsets is only the central part of 
the whole disk, whose $R_{25}$ is $11'.9$.

 \citet{CC02} derived the \Ha ~equivalent width from observations with narrow band filters, 
and estimated a mean age of \HII ~regions at $r\lesssim 5'$ to be about 1.6--4 Myr, 
supporting our result of $t_{\rm SF}=4.0\pm 1.3$ Myr.

 The fitted \OP ~of $72\pm 37$ \kmskpc ~locates the corotation at $R_{\rm CR}=2.7^{+3.7}_{-0.9}$ kpc, 
or $1.3^{+1.8}_{-0.4}$ arcmin.
 \citet{Wall97} derived $\Omega_{\rm P}=19\pm 5$ \kmskpc ~using offsets between 
CO and FUV at $r\sim 1'.5$, assuming the timescale from CO to FUV 
to be 3 Myr, locating $R_{\rm CR}=5'.5\pm 1'.5$.
 \citet{EEM92} also located the corotation at $5'.5$ from the optical appearance.
 This discrepancy implies a radial variance of the pattern speed or the inclination angle, which could be 
caused by interactions with companions, NGC 5474 and NGC 5477.
 If the inner disk where we measured offsets, is more inclined than the adopted inclination angle of $18^\circ$, 
the resultant \OP ~would be smaller or closer to other results mentioned above than the value presented here.
 In terms of the inclination, the adopted value for this galaxy is one of the smallest in our sample, 
implying a difficulty in deriving RC, which could be also inferred by the poor consistency in 
available RCs (see Figure \ref{RC3.fig}).
 The uncertainties derived by the fitting, thus, could be underestimated.

\subsubsection{Galaxies with No Offsets\label{sec.dis_indiviN}}
 Here, we describe our findings for ``N'' galaxies in comparison with previous researches, 
and discuss why they do not show offsets.
 Values for \OP ~from the literature are listed in Table \ref{OP_N.tb}.

\paragraph{NGC 4321 (M 100)}
 We measured 34 offsets at $r= 1.6-6.6$ kpc ($0'.3-1'.4$), 
and found that the offsets in the arm region ($r\sim 5-6.5$ kpc) are almost zero, while 
offsets close to the bar region ($r\lesssim 4.5$ kpc) are about 10 degrees, 
which had already been found by \citet{Sheth02}.
 There are several possible reasons for no offsets found in the arms: (1) material arms, 
(2) corotation at where offsets were measured, and 
(3) elliptical orbits nearly parallel to the spiral arms.
 Since there are offsets close to the bar region, where orbits should have higher ellipticity than in the arm region, 
(3) seems to be less plausible at least for this galaxy.

 \citet{RW04} derived the pattern speed of this galaxy 
to be $28^{+4}_{-5}$ \kmskpc ~with the TW method applied to the CO data.
 With the rotation curve from \citet{RC99}, this value locates the CR
at about 10 kpc ($\sim 2'$) in radius.
 Previous studies listed in Table \ref{OP_N.tb} also suggest $R_{\rm CR}\gtrsim 2'$.
 These measurements are well beyond where we measured offsets, so cannot explain 
the lack of offsets.
 
 The northern arm is weaker than the southern arm both in CO and \Ha.
 This asymmetry is thought to be due to the existence of the companion galaxy NGC 4322, 
the ram pressure from intragalactic matter in the Virgo cluster, or the central bar. 
 \citet{Knap96} investigated the arm/interarm ratio of the SFE, and found no symmetric patterns 
though this ratio was larger than unity for most part of the two arms.
 They concluded that this was because
their observed region did not include any resonances, 
although star formation on arms was actually enhanced and it 
might be triggered by spiral density waves.
 Thus, we deduce that the spiral density wave might be rather unstable or localized at least in the region we see, 
and that (1) would be the most plausible among the three.
 If this grand-design galaxy is confirmed to be without density waves, 
it will give a new picture of spiral galaxy formation.

\paragraph{NGC 5248}
 We were able to measure 13 offsets at $r=4.2-9$ kpc ($0'.6-1'.4$), which were found to be almost zero, 
while 2 spiral arms were clearly traced.
 Three possible reasons for no observed offsets in NGC 4321 are also applicable to this galaxy.

 In addition to a small bar with size of $0'.4$, \citet{Jog02a} found a large-scale 
stellar bar with a semimajor axis of $1'.6$, indicating that the spiral structure seen in 
CO and \Ha ~could be inside the bar.
 They also estimated $R_{\rm CR}\sim 1'.9$ from the assumed ratio of the corotation radius 
to the bar semimajor axis (1.2), while \citet{EEM92} derived a comparable value.
 Following the concept of the larger bar, 
\citet{YY06} developed a nonlinear model for the spiral density waves excited by a bar potential, 
and succeeded in explaining the observational features of this galaxy.
 Their orbits show strong inward streaming motions along the spiral arm at $r\lesssim 0'.8$, 
or 5.8 kpc using our distance.
 As we found that offsets of up to $1'.4$ in radius were almost zero, 
this non-circular orbit can partially explain this feature.

\subsubsection{Galaxies with Ambiguous Offsets\label{sec.dis_indiviA}} 
 In this section, we describe our findings for ``A'' galaxies in comparison with previous studies, 
and discuss possible reasons for their ambiguity.
 These reasons are not the same for all the galaxies, and we expect that if data with 
higher quality become available, some of them will be recategorized as C.

\paragraph{NGC 3184}
 We measured 19 offsets at $r= 1.6-3.6$ kpc ($0'.6-1'.4$), 
and found a negative correlation between $\theta$ and $\Omega$, 
with large dispersion.

\begin{figure*}
\begin{center}
\includegraphics[width=0.45\linewidth]{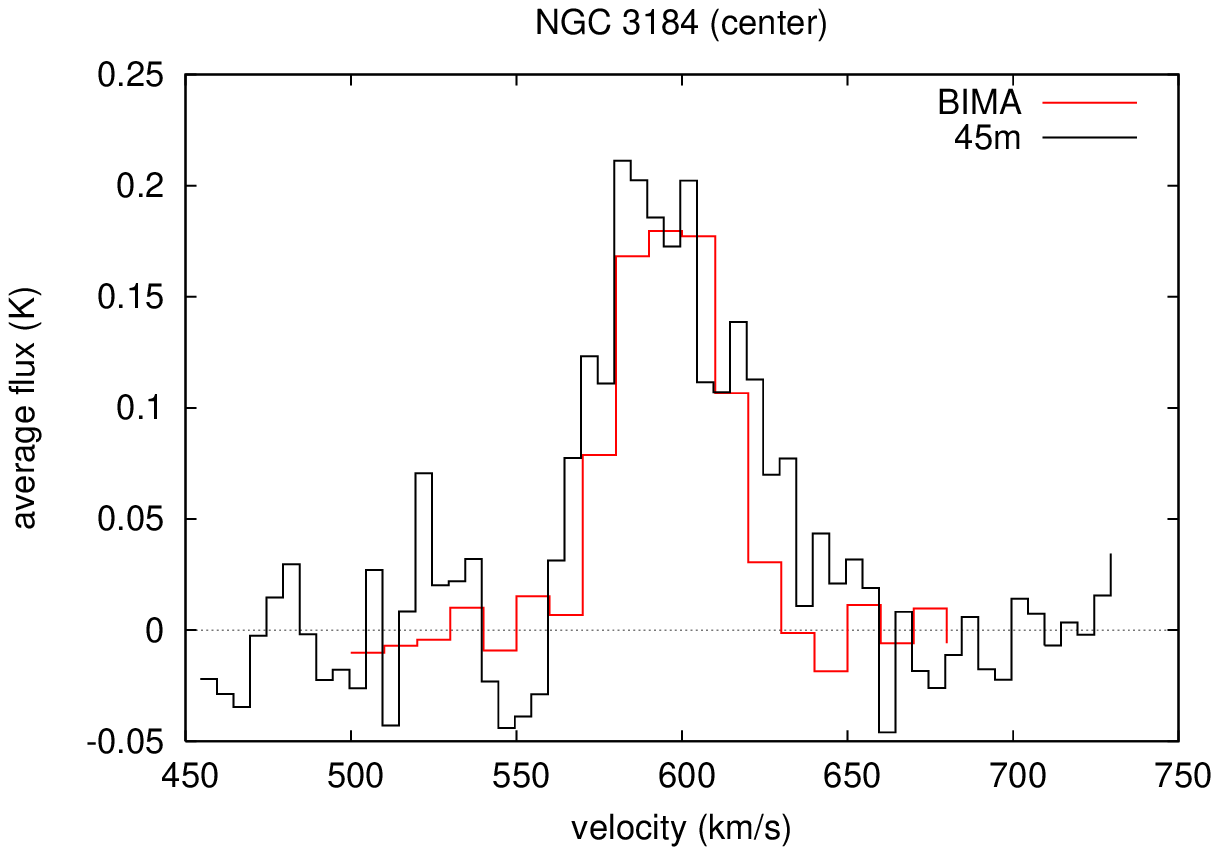}
\hspace{0.05\linewidth}
\includegraphics[width=0.45\linewidth]{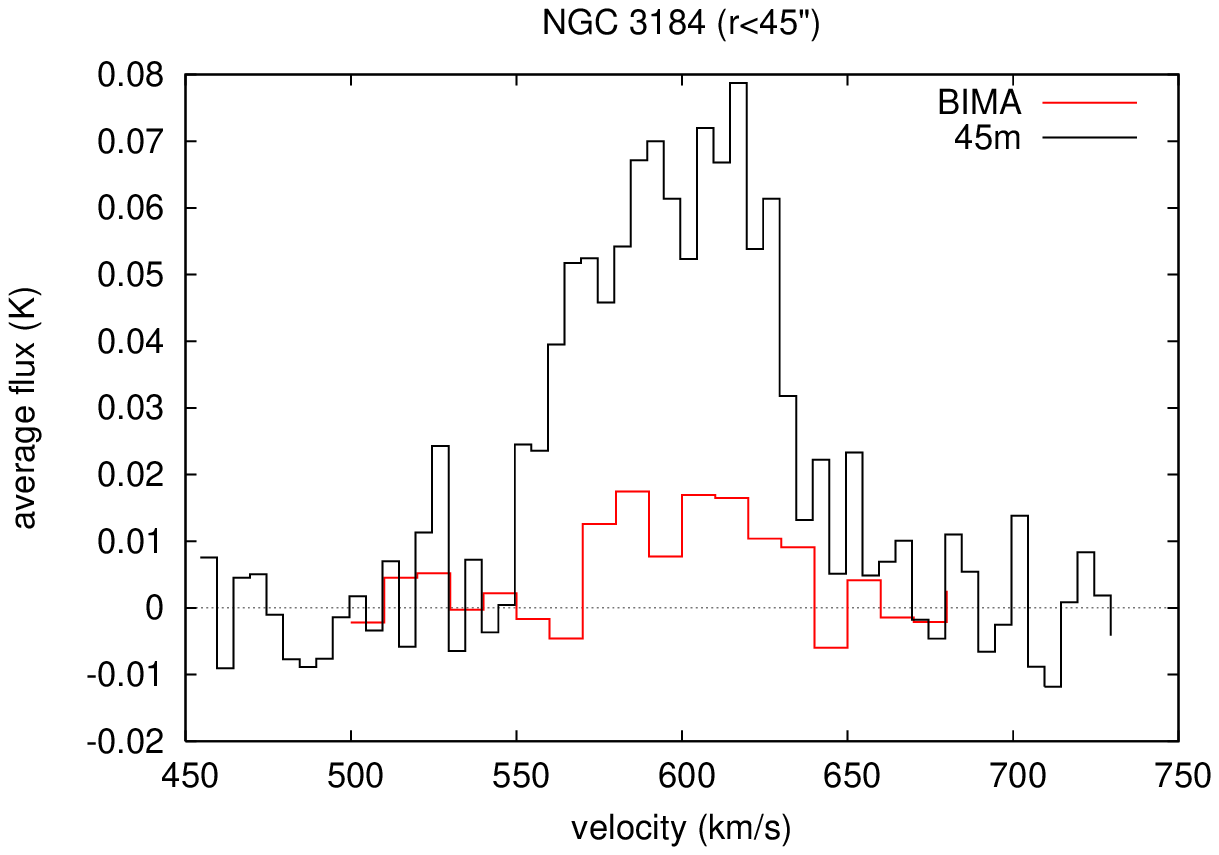}
\caption{Spectra from BIMA data (dotted) and 45m data (solid). 
Flux is averaged for the central 21$''$ region (left) and for the central 90$''$ region (right). 
The missing flux derived by ``1 - BIMA/45m" is larger when averaged for most of the disk than 
for the center.}
\label{n3184ispec.fig}
\end{center}
\end{figure*}

 Since CO data from BIMA SONG do not include data from single-dish observations, 
they are subject to the missing flux problem. 
 In order to quantify how much flux is missed in the BIMA observations, 
we compared its spectra to those from the 45m observations \citep{Kuno07} in Figure \ref{n3184ispec.fig}.
 While about 90\% of the total flux was detected in the central 21$''$ region, 
about 70\% was missed if the flux was averaged at $r\leq 45''$.

 We do not, however, attribute the ambiguity in the offset to the missing flux, since 
spiral arms are thought to be comprised of smaller molecular clouds, 
which should be selectively detected in the interferometric observations.
 Since the spatial resolution of the CO data ($\sim 250$ pc) is small enough to resolve 
typical molecular arms and offsets, the field of view and/or sensitivity should be insufficient.
 As spiral arms can be traced in the \Ha ~to larger radii, 
we expect that the $\Omega-\theta$ distribution will become clearer with smaller dispersion, 
if CO data with a larger field of view and higher sensitivity become available.

\paragraph{NGC 3938}
 We measured 9 offsets at $r= 3-4.6$ kpc ($0'.6-0'.9$), 
and found a negative correlation between $\theta$ and $\Omega$, 
with a large dispersion.
 This large dispersion is partly because of the large errors and unreliability of the CO rotation 
curve, which was used to calculate $\Omega$.
 The reason we did not use the RC from \citet{SINGSHa} is that it is systematically different 
from ours, suggesting that the inclination angle mentioned in their paper is not correct.
 In addition, the CO image and resolution around 500 pc indicates 
that molecular arms are not sufficiently resolved and detected, 
while spiral arms are clear in the \Ha ~image.
 Thus we need further CO observations to discuss the spiral structure of this galaxy 
based on the $\Omega-\theta$ distribution.

\paragraph{NGC 4736 (M 94)}
 Though we were able to measure 8 offsets at $r= 1-1.5$ kpc ($0'.7-1'.0$), 
which are clear in the CO and \Ha ~images, they show a negative dependence on $\Omega$.
 We derived the rotational velocity from CO data, but it was not consistent with the RC from \citet{RC99} 
 where we measured offsets.
 Given that the RC from \citet{RC99} is flat, which is typical of spiral galaxies, we use this to 
calculate $\Omega$.
 However, even if we used the RC from CO data, the dependence of offsets on $\Omega$ would 
still be negative.

 NGC 4736 is the only galaxy in our sample to be classified as flocculent (AC=3).
 A star forming ring or short spiral arms are apparent at $r\sim 1-1.5$ kpc in the \Ha ~image, 
while molecular arms are longer and apparent at $r\sim 1-2$ kpc.
 This difference in morphology between CO and \Ha, 
which has been also noted by \citet{WB00}, 
could be responsible for the negative correlation between $\theta$ and $\Omega$.
 Therefore, we deduce that there are other triggers of star formation than the spiral density wave in this galaxy.
 
 One possible trigger is an expanding poststarburst ring, proposed by \citet{vdK74} and \citet{SB76}. 
 They explained that the ring resulted from a central starburst which occurred about 10 Myr ago, from 
spectroscopic observations and hydrodynamical calculations.
 \citet{Maoz95} found two compact UV sources separated by $2''.5$ in the nucleus 
with diffuse emission centered on the one source, presumably corresponding to the optical nucleus.
 They interpreted their results to be the final stage of a merger, 
which is consistent with the poststarburst scenario.
 \citet{MT04} adopted another possibility of  
this ring to be located at the OLR of the central bar as well as the ILR of the outer disk.
 The pattern speed for each structure should be 85 and 27 \kmskpc, respectively, 
and the latter could explain the location of outer faint ring at $r\sim 5'$.
 They also noted that the FUV ring is located slightly outward from the \Ha ~ring, 
implying an inward-propagating star formation, which is totally opposite with the poststarburst scenario.

 \citet{RW04} applied the TW method to BIMA CO data, and derived 
$\Omega_{\rm P}=152\pm 28$ \kmskpc, corrected for our distance. 
 This value is much larger than that derived by \citet{MT04}, and 
locates the corotation at $r=0'.8-1'.2$, 
which is almost same as where we measure the offsets.

\paragraph{NGC 6181}
 We were able to measure 27 offsets for 2 spiral arms at $r=2-5$ kpc ($0'.2-0'.5$).
 While the offsets of arm1 show a negative dependence on $\Omega$, 
those of arm2 show a positive dependence (Figure \ref{ofs.fig}).
 We thus fit the $\Omega-\theta$ plot for arm2 only, but the uncertainties of the resultant values for 
\tsf ~and \OP ~are too large to give a quantitative conclusion.
 As the large uncertainty is mainly due to large errors in the RC we used, 
further observations will be able to give more reliable results at least for arm2.

\paragraph{NGC 6946}
 At the region where CO emission was detected, the spiral structure is not clear, 
so we were able to measure only 7 offsets at $r=1.1-2.4$ kpc ($0'.7-1'.5$).
 This small number of data resulted in the large uncertainty of \OP, 
so we categorized this galaxy as A.
 The positive slope in the $\Omega-\theta$ plot, however, gives the fitted value of 
$t_{\rm SF}=1.1\pm 0.8$ Myr.
 This is slightly smaller than that for C galaxies, but still has the same order of magnitude.

 \citet{Zim04} applied the TW method to the IRAM 30m CO data \citep{Wals02}, and derived 
$\Omega_{\rm P}=39\pm 8$ \kmskpc, consistent with our mean value of 38 \kmskpc.
 \citet{Fath07b} also applied the TW method using \Ha ~data from \citet{SINGSHa}, 
whose spatial resolution was higher than that of CO data, and obtained two pattern speeds: 
$\Omega_{\rm P}=47^{+3}_{-2}$ \kmskpc ~for the inner structure ($r\lesssim 1'$) and 
$\Omega_{\rm P}=22^{+4}_{-1}$ \kmskpc ~for the outer structure ($r\gtrsim 1'$).
 They also claimed that spatially smoothed \Ha ~data gave a comparable value 
to \OP ~from \citet{Zim04}.
 As we measured offsets around the border of the inner and outer structure, 
our \OP, which is in between the two values, is not inconsistent to their results despite the large uncertainty.

 Since the spiral structure is more conspicuous and 
 stronger CO emission is detected in the outer regions \citep{Wals02}, 
further CO observations pointed to such regions will give understanding of the property of 
the spiral arms and star formation in this galaxy.
 We anticipate that this galaxy will be recategorized as C if we can measure a larger number of offsets 
and thus derive \OP ~with smaller uncertainty.

\subsection{Property of Galaxies in Each Category\label{sec.dis_cat}}
In order to understand relationships between spiral structures and other physical parameters, 
we have examined several properties of galaxies according to their category.
 Similar analyses are performed for all ``C" galaxies, and their results are shown in \S \ref{sec.dis_tsf}.

\subsubsection{Molecular Gas Property}
 As we found that the missing flux in CO data of NGC 3184 data could be as large as about 70\%, 
we excluded this galaxy from the following quantitative analysis of molecular gas.
NGC 6181, whose CO data are also from interferometric observations only, 
is included in the analysis, since we find almost no flux is missed in the data by comparison with spectra 
from \citet{FCRAO}.

\begin{figure*}
\begin{center}
\includegraphics[width=0.7\linewidth]{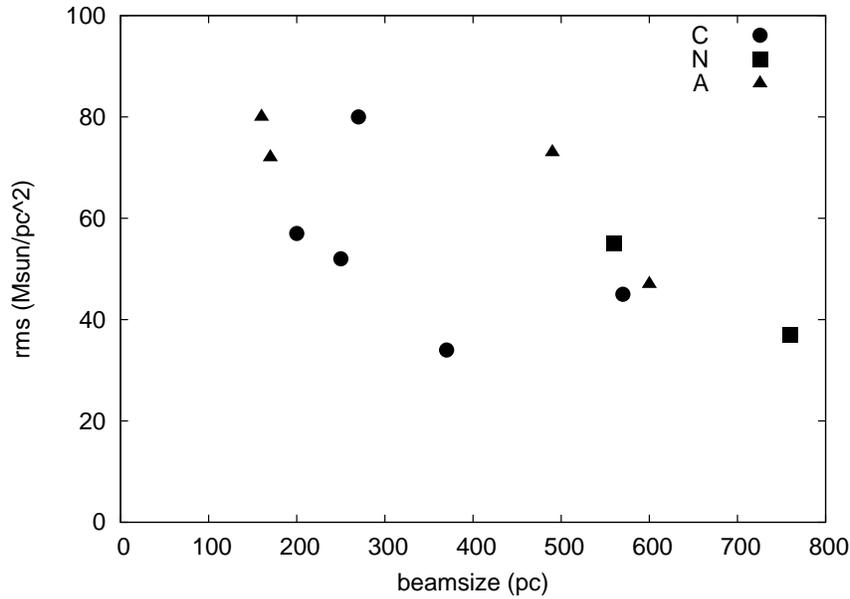}
\caption{Plot of sensitivity (rms) against beam major axis (bmaj) of the CO images 
listed in Table \ref{bmaj+rms.tb} for each category.}
\label{bmaj+rms_oc.fig}
\end{center}
\end{figure*}

 First, we reassess the quality of CO data, which can cause biases to the categorization.
 We list the beamsizes (bmaj) and the noise levels (rms) of the CO images together with the offset category in Table \ref{bmaj+rms.tb}, 
and show a plot of rms against bmaj in Figure \ref{bmaj+rms_oc.fig}.
 From this figure, we find no systematic differences or biases to the category.

\begin{figure*}
\begin{center}
\includegraphics[width=0.45\linewidth]{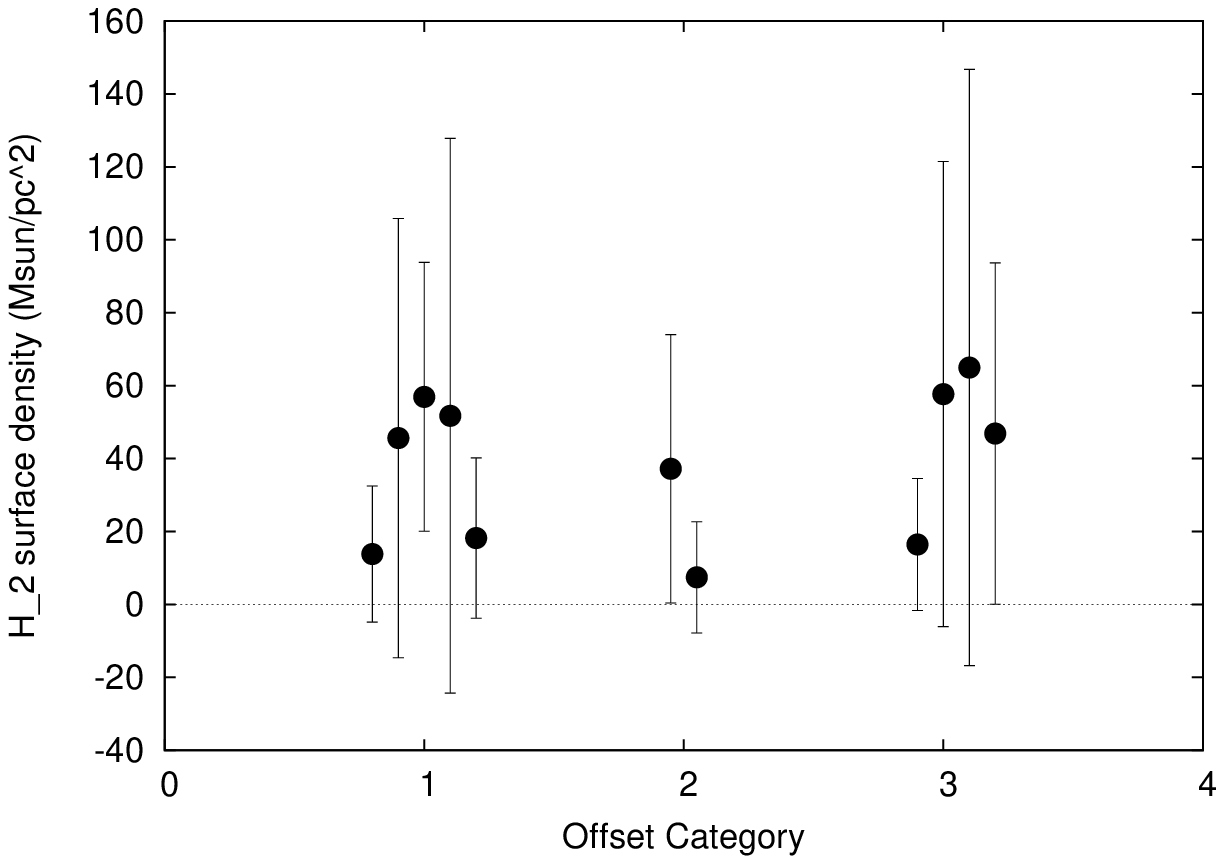}
\hspace{0.05\linewidth}
\includegraphics[width=0.45\linewidth]{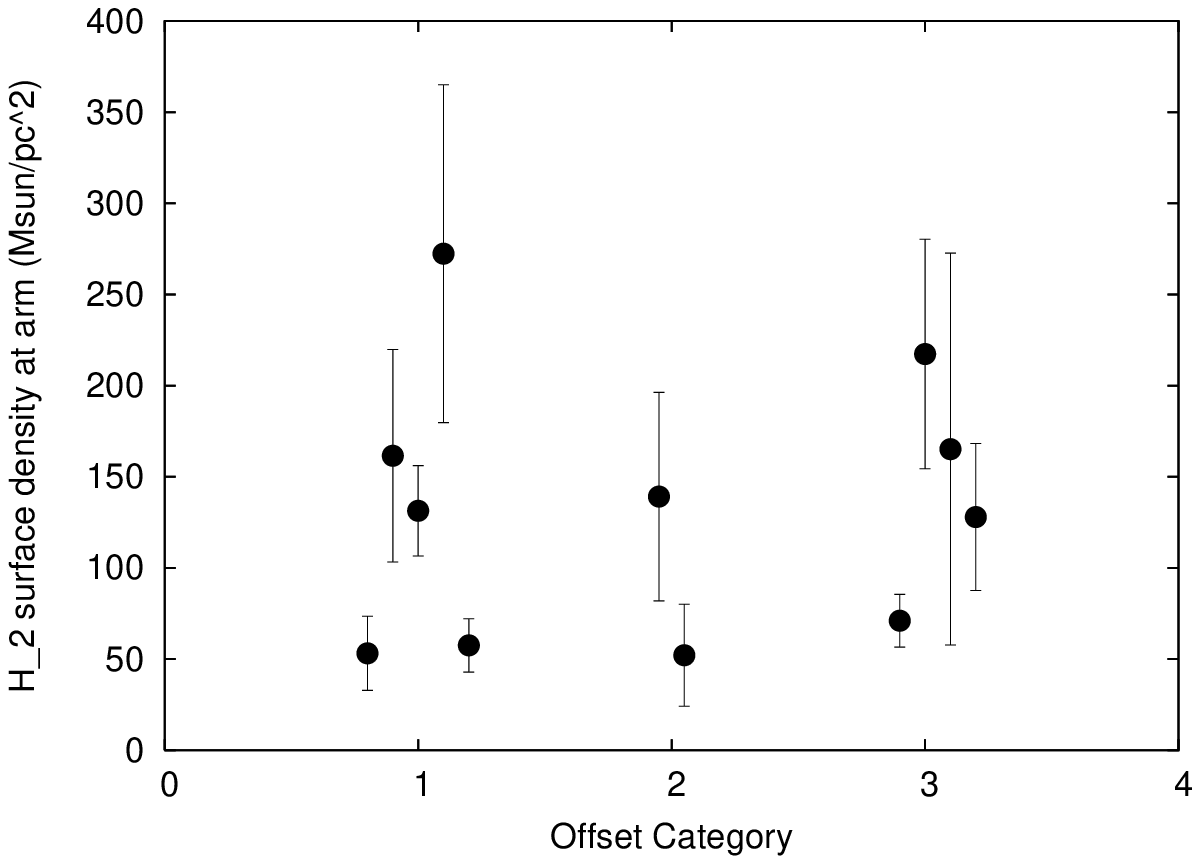}
\caption{Mean H$_2$ surface density against the offset category (1=C, 2=N, 3=A).
Values in the left panel are derived by averaging azimuthally, while those in the right 
are derived by averaging only the arm regions.}
\label{oc-mol.fig}
\end{center}
\end{figure*}

 Next, we examine global properties for each category.
 We calculate the mean H$_2$ surface density $\Sigma$ using a conversion factor of 
$X_{\rm CO}=3.0\times 10^{20}$ cm$^{-2}$ (K \kms )$^{-1}$ at the radial region where we measured offsets.
 In addition, we derive $\Sigma$ only for the arm regions, centered on CO peaks, which were used 
to derive offsets, with a width of 500 pc.
 From Figure \ref{oc-mol.fig}, where disk-averaged and arm-only $\Sigma$ are plotted for each category, 
we find no clear dependence of both values on the offset category.
 This indicates that larger ($>500$ pc) scale properties of gas are not 
closely relevant to the spiral structure.

\subsubsection{Morphological Property\label{sec.mor-oc}}
\begin{figure*}
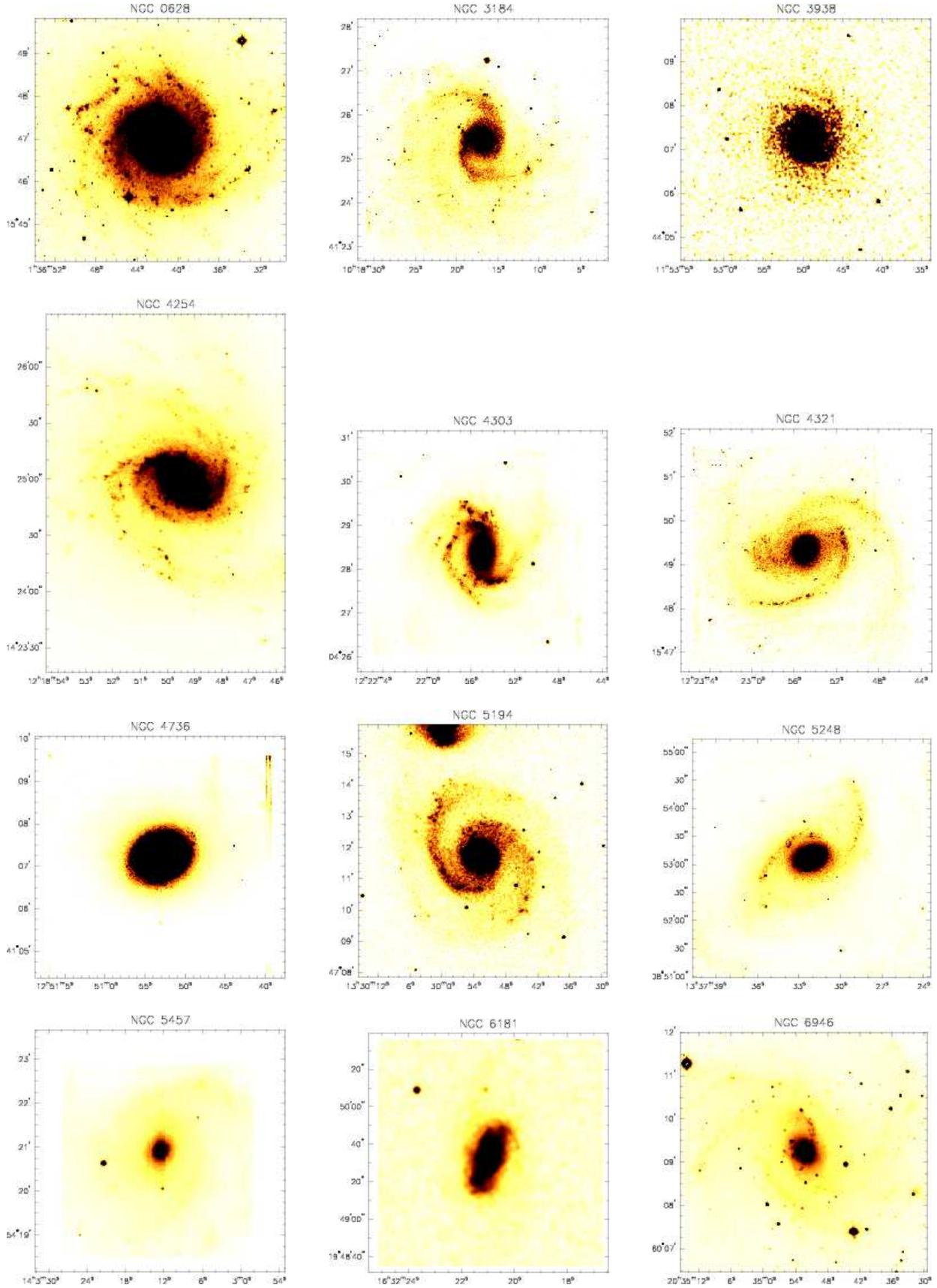

\begin{center}
\includegraphics[width=0.3\linewidth]{f11a.eps2}
\hspace{0.03\linewidth}
\includegraphics[width=0.3\linewidth]{f11b.eps2}
\hspace{0.03\linewidth}
\includegraphics[width=0.3\linewidth]{f11c.eps2}\\
\vspace{12pt}
\includegraphics[width=0.3\linewidth]{f11d.eps2}
\hspace{0.03\linewidth}
\includegraphics[width=0.3\linewidth]{f11e.eps2}
\hspace{0.03\linewidth}
\includegraphics[width=0.3\linewidth]{f11f.eps2}\\
\vspace{12pt}
\includegraphics[width=0.3\linewidth]{f11g.eps2}
\hspace{0.03\linewidth}
\includegraphics[width=0.3\linewidth]{f11h.eps2}
\hspace{0.03\linewidth}
\includegraphics[width=0.3\linewidth]{f11i.eps2}\\
\vspace{12pt}
\includegraphics[width=0.3\linewidth]{f11j.eps2}
\hspace{0.03\linewidth}
\includegraphics[width=0.3\linewidth]{f11k.eps2}
\hspace{0.03\linewidth}
\includegraphics[width=0.3\linewidth]{f11l.eps2}\\
\caption{$K$-band images of sample galaxies.}
\label{k.fig}
\end{center}
\end{figure*}

\begin{figure*} 
\begin{center}
\includegraphics[height=0.3\textheight]{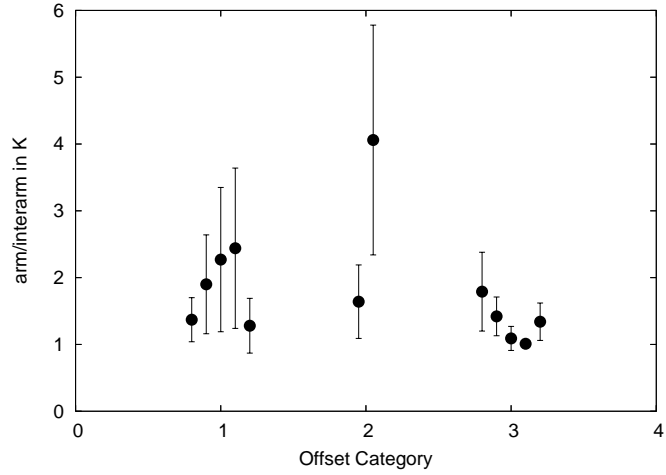}
\caption{Plot of the arm/interarm ratio in the $K$-band image 
against the offset category (1=C, 2=N, 3=A).
Points are slightly shifted in the horizontal direction so as not to overlap.}
\label{oc-ai.fig}
\end{center}
\end{figure*}

 In order to quantify the morphological properties of the underlying gravitational potential of the spiral disk, 
we used the arm/interarm ratio in $K$-band images as an indicator of 
the strength of the galactic shock or spiral arms, since the luminosity in $K$ is less sensitive 
to extinction and thought to be proportional 
to the total mass of low-mass stars, which dominate total stellar mass, assuming a constant mass-to-luminosity ratio.
 The $K$-band images are taken from \citet{KJS03} and 2MASS, and shown in Figure \ref{k.fig}.
 We defined the arm region as a region with 500 pc width centered on the H$\alpha$ peak which was measured in order to 
derive offsets, while the interarm region was set to the same width centered at the midpoint 
of the two arm regions, or 90 degrees away from an arm region if only one arm region was defined 
at that radius.
 The arm/interarm $K$ flux ratio was calculated at each radius where H$\alpha$ peaks (i.e., offsets) 
were measured and averaged over the disk.
 In Figure \ref{oc-ai.fig}, the averaged ratio is plotted for each offset category.
 Although there is no strong correlation, ``A'' galaxies tend to have a smaller ratio.

\begin{figure*} 
\begin{center}
\includegraphics[height=0.3\textheight]{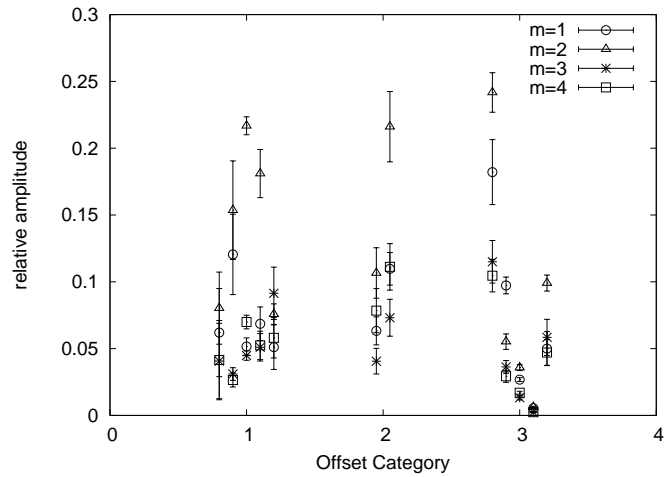}
\end{center}
\caption{Plot of $A(r,m)/A(r,m=0)$ for $m=1-4$ against the offset category (1=C, 2=N, 3=A).
Points are slightly shifted in the horizontal direction so as not to overlap.}
\label{oc-amp.fig}
\end{figure*}

 As an alternative way of estimating the arm strength, 
we performed an one dimensional discrete Fourier transformation, 
\begin{equation}
A(r,m)=\sum_{\phi=0}^{360-1} F(r,\phi) \exp (-2\pi im\phi/360), 
\label{eq.DFT}
\end{equation}
where $F(r,\phi)$ is the $K$-band flux at the polar coordinate $(r,\phi)$, 
in order to know which mode $m$ is dominant within a galaxy and 
 obtain the typical strength of asymmetric components.
 We calculate the relative amplitude $|A(r,m)|/|A(r,m=0)|$ for $m=1-4$, and 
the mean of these components is shown against the offset category in Figure \ref{oc-amp.fig}.
 The $m=2$ component is the strongest for most of the galaxies, regardless of category.
We find no clear dependence of the amplitude on the category, similar to that in the $K$ flux ratio.
 The smallest amplitudes, however, belong to ``A'' galaxies.

 This tendency between the arm strengths and the offset category implies that 
stronger stellar arms or a deeper potential generates clearer offsets.
 As the potential depth should be tightly correlated with the strength of spiral density waves, 
the ambiguity in offsets for ``A" galaxies at least partially 
results from weaker density waves.

 In Table \ref{bar-oc.tb}, we list the number of SA, SAB, and SB galaxies in each category.
 While our sample is not large enough for a robust statistical study, 
``N" galaxies seem to have stronger bar than others.
 Thus, bars could be responsible for no offsets as discussed in \S \ref{sec.dis_indiviN}.

\section{Discussion\label{sec.dis_param}}
 For 5 galaxies in the ``C'' category, we were able to derive their \OP ~and \tsf ~by the $\chi^2$-fitting. 
 In this section, we discuss the dependency of the derived values on other physical parameters.
 As we have only 5 galaxies, the statements here are still tentative and require a larger 
sample to confirm them statistically.

\subsection{Pattern Speed\label{sec.dis_Op}}
\begin{figure*}[h!]
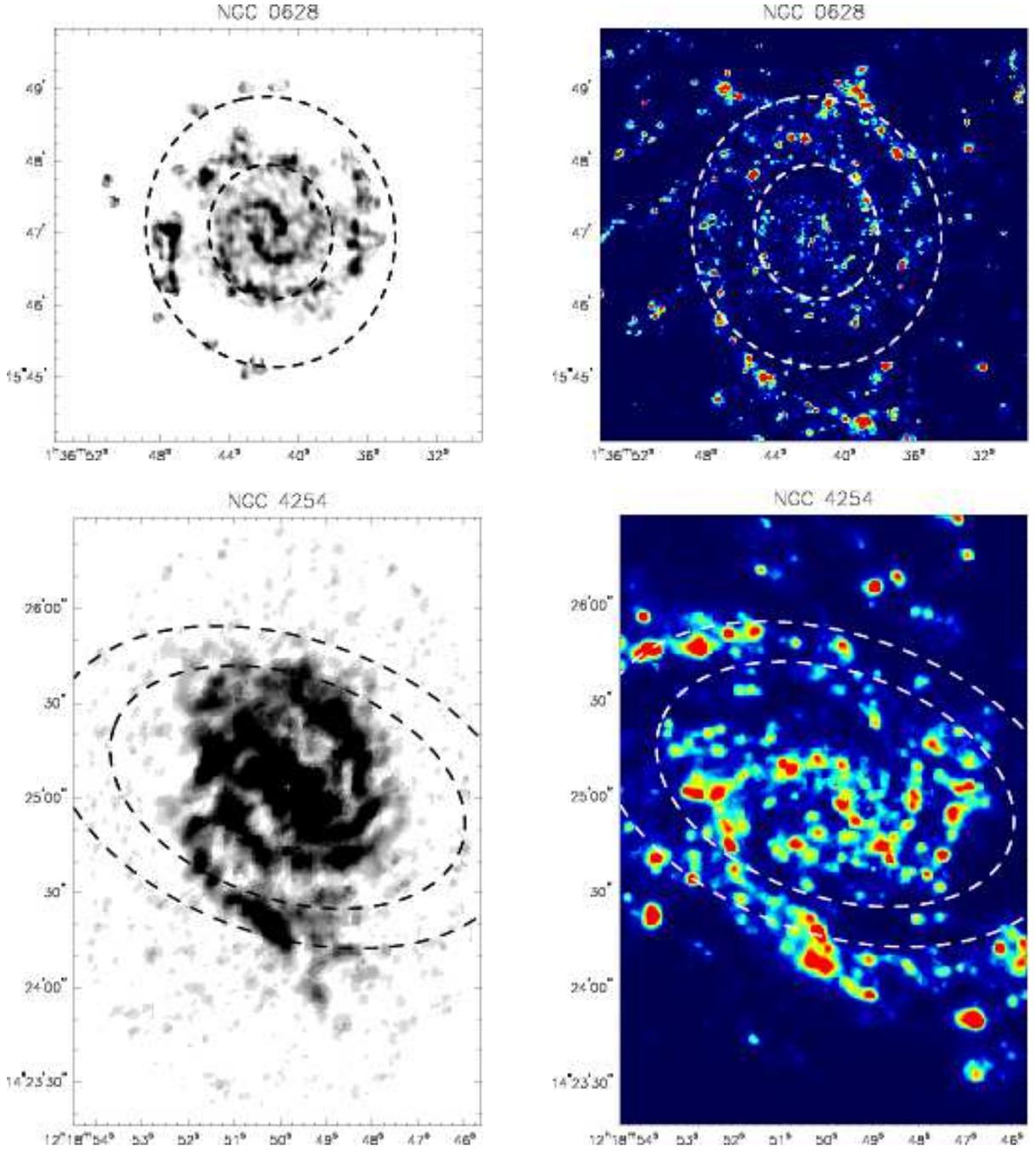

\begin{center}
\includegraphics[width=0.45\linewidth]{f14a.eps2}
\hspace{0.05\linewidth}
\includegraphics[width=0.45\linewidth]{f14b.eps2}\\
\vspace{12pt}
\includegraphics[width=0.45\linewidth]{f14c.eps2}
\hspace{0.05\linewidth}
\includegraphics[width=0.45\linewidth]{f14d.eps2}\\
\caption{Location of the corotation on CO (left) and \Ha ~(right) images. 
Coordinates are (RA, Dec) in J2000.
Solid line indicates the best-fitted corotation location and dashed lines 
indicate the range of uncertainty. For NGC 0628 and NGC 4254, only dashed 
lines are drawn due to the large radial variation of their rotation curve.}
\label{Rcr.fig}
\end{center}
\end{figure*}

 In Figure \ref{Rcr.fig}, we plot ellipses on CO and \Ha ~maps to show the location of the derived corotation, 
except for NGC 4303, whose uncertainty in \OP ~is too large to derive a reliable \Rcr.
We find that \Rcr ~is nearly 
consistent with the edge of CO data, while \Ha ~arms are more extended.
 This is a natural result from the fact that we did not find large negative offsets, 
but we cannot exclude the possibility that the small CO field of view has caused 
a bias to this result. 
 For a more robust determination, larger CO maps are needed.
 In Table \ref{Rcr.tb}, the radius of the optical disk $R_{25}$ from Nearby Galaxies Catalog \citep{NBG} 
and the derived \Rcr ~are listed.
 From the ratio $R_{\rm CR}/R_{25}$, we find that the corotation radius is about half of $R_{25}$,
except for NGC 5457. 
 We should note that this result is fully independent of the assumed distance, and less sensitive 
to the assumed inclination angle.


\subsection{Star Formation Timescale\label{sec.dis_tsf}}
 As the derived values of \tsf ~are about $5-30$ Myr and the free-fall time ($\sim 1/\sqrt{G\rho}$) of 
a typical molecular cloud ($\rho \sim 100$ cm$^{-3}$) is about 6 Myr, we deduce that 
the gravitational instability should be a dominant trigger for star formation in the spiral arms.
 Here, we examine how the gravitational instability and other factors 
affect the star formation timescale.
 Similar analyses are performed for each offset category and their results are shown in \S \ref{sec.dis_cat}.

 We should bare in mind that \tsf ~is linearly dependent on the assumed distance, and 
also dependent on the assumed inclination, where its error is 
substantial for galaxies with low inclination angles.
 Its dependency is discussed in \S \ref{sec.valid}.

\subsubsection{Gravitational Instability}
\begin{figure*}
\begin{center}
\includegraphics[width=0.4\linewidth]{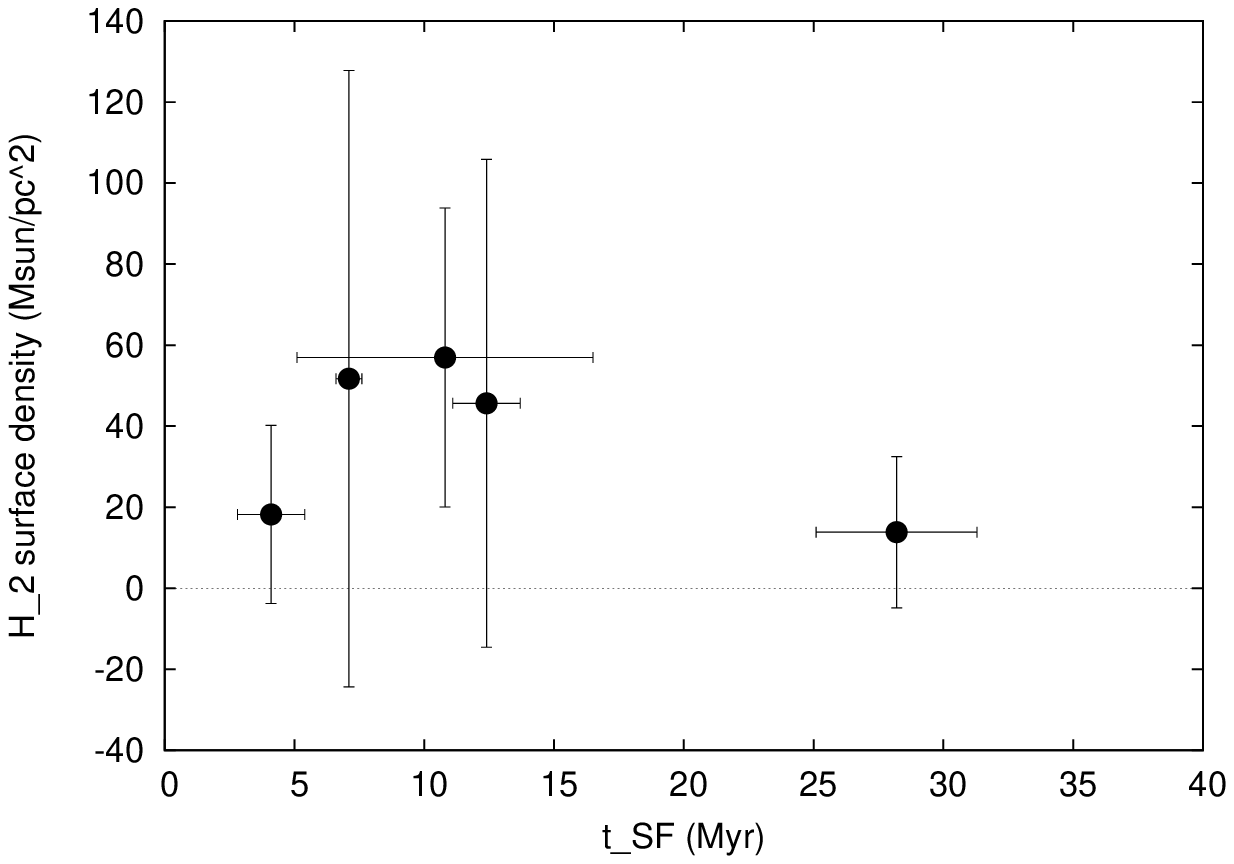}
\hspace{12pt}
\includegraphics[width=0.4\linewidth]{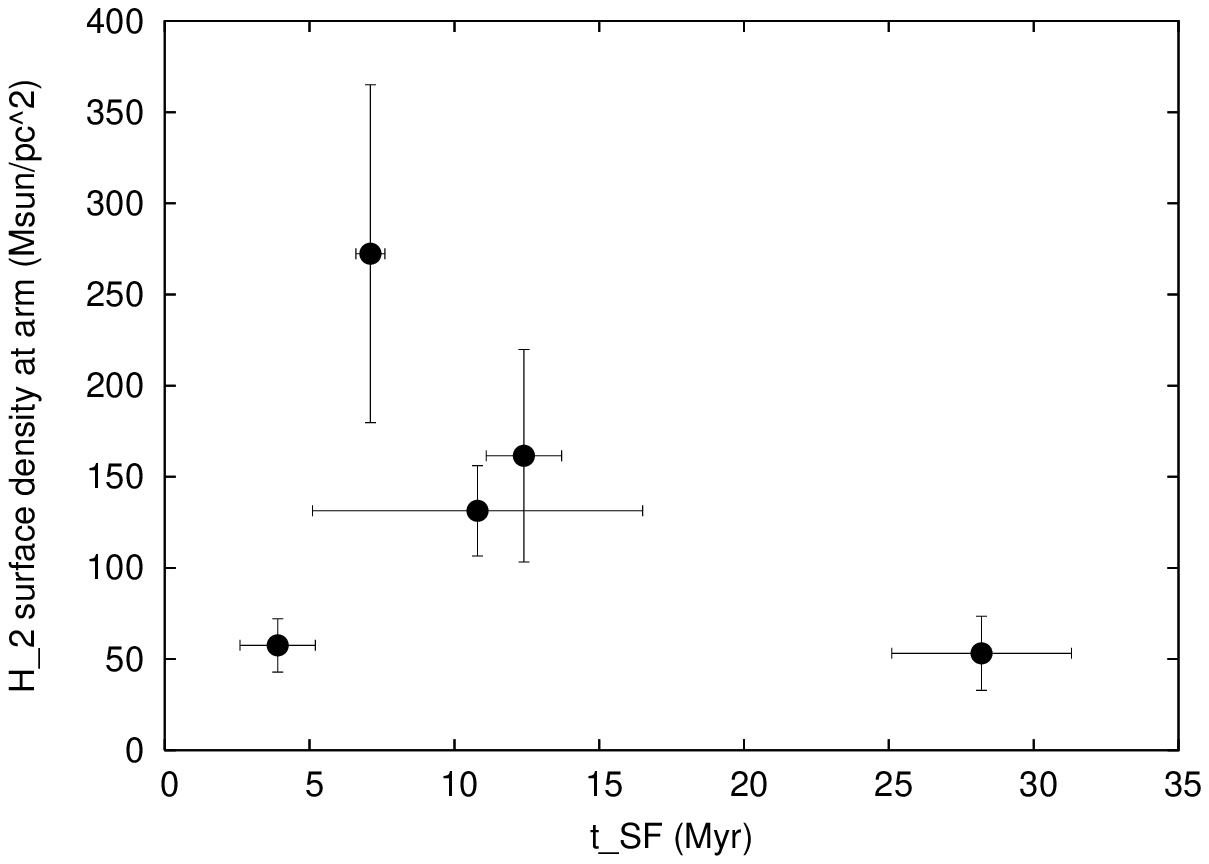}
\caption{Plot of mean H$_2$ surface density $\Sigma$ 
against 
\tsf. 
Values in the left panel are derived by averaging azimuthally, while those in the right  
are derived by averaging only the arm regions.}
\label{sigH2.fig}
\end{center}
\end{figure*}

 In Figure \ref{sigH2.fig}, the mean H$_2$ surface density $\Sigma$ derived from CO data is plotted against \tsf.
 Values in the left panel are derived by averaging the data azimuthally, while those in the right panel 
are derived only from arm regions.
 As we obtained \tsf ~consistent with the gravitational collapse timescale, 
we expect larger density with smaller \tsf.
 However, there is no clear correlation between \tsf ~and $\Sigma$, suggesting 
that molecular properties at scale $> 500$ pc (adopted width of arm regions) are not directly relevant to 
star formation activities.
 Since the typical size of a GMC is about 50 pc, which is much smaller than 
the spatial resolution of CO data we used, we need CO data with higher resolution 
as well as larger sample of spiral galaxies for further study.

\subsubsection{Metallicity}
\begin{figure*}[h!]
\begin{center}
\includegraphics[width=0.45\linewidth]{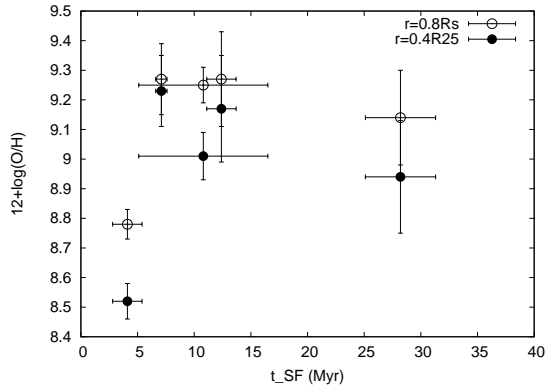}
\caption{Plot of metallicity $12+\log({\rm O/H})$ versus \tsf.
Each galaxy has 2 metallicity values measured at $r=0.4R_{\rm 25}$ (filled circle) and 
$r=0.8R_{\rm s}$ (open circle), where $R_{\rm 25}$ is a radius of optical disk and 
$R_{\rm s}$ is a scale length.}
\label{metal.fig}
\end{center}
\end{figure*}

 We plot the heavy element abundance in the form of $12+\log({\rm O/H})$ 
from \citet{Zari94} against \tsf ~in Figure \ref{metal.fig}.
 Each galaxy has 2 metallicity values measured at $r=0.4R_{\rm 25}$ and 
$r=0.8R_{\rm s}$, where $R_{\rm s}$ is the scale length.
 We measured offsets around $0.8R_{\rm s}$ with the exception of NGC 5457, 
for which we used the region of $r<1'.5$ while $0.8R_{\rm s}=1'.7$.
 Since heavy elements can help the cooling and collapse of molecular clouds by 
their transitional lines, we expect smaller \tsf ~with higher metallicity.
 Such a correlation is, however, not apparent here, implying that 
these effects by heavy elements are not critical to star formation processes.

\subsubsection{Spiral Strength}
\begin{figure*}[h!]
\begin{center}
\includegraphics[height=0.3\textheight]{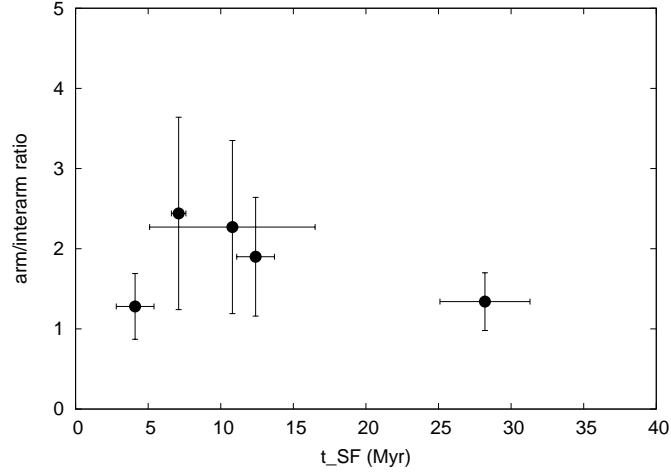}
\caption{Plot of the arm/interarm ratio in the $K$-band image against the derived \tsf}
\label{Kratio.fig}
\end{center}
\end{figure*}

\begin{figure*}[h!]
\begin{center}
\includegraphics[height=0.3\textheight]{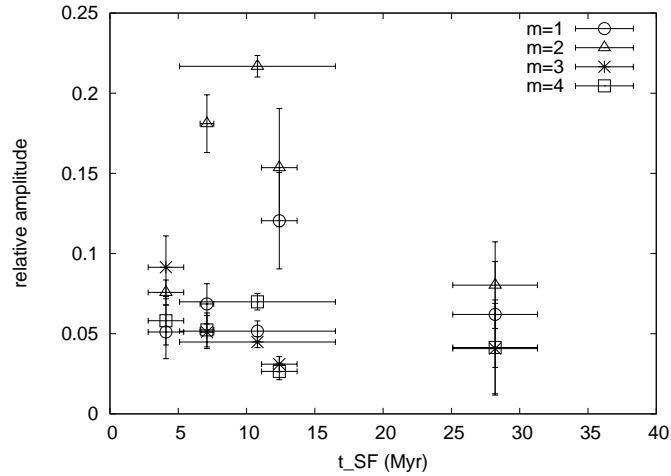}
\caption{A mean of the relative amplitude $A(r,m)/A(r,m=0)$ against the derived \tsf. 
Four relative amplitudes of $m=1-4$ are plotted for each galaxy.
Larger values of amplitude indicate that the corresponding asymmetric component 
is larger, presumably due to the spiral density waves in that galaxy.}
\label{tsf-amp.fig}
\end{center}
\end{figure*}

 In Figure \ref{Kratio.fig} and \ref{tsf-amp.fig}, we plot the arm/interarm ratio and 
the Fourier amplitudes $|A(r,m)|/|A(r,m=0)|$ for $m=1-4$ from 
$K$-band images against \tsf.
 Neither plots show strong dependence on \tsf.
 If we can confirm that the spiral arm strength has no correlation with \tsf, 
the spiral shock or potential should have no direct effect on star formation and 
its role could be just to agglomerate small pre-existing molecular clouds to form 
massive molecular clouds, in which star formation eventually occurs, 
as suggested by \citet{Koda06}.
This is important for understanding the relationship 
between galactic dynamics and star formation activities, and 
we need a larger sample of galaxies for further study.

\subsection{Validity of Offset Method\label{sec.valid}}
 Out of 12 nearby spiral galaxies, we have succeeded in deriving \OP ~and \tsf ~for 5 galaxies, which 
are referred as ``C" galaxies.
 The rest of the sample, comprised of 7 galaxies, are categorized as ``N" or ``A" according to their offset distributions.
 
We have examined the surface mass density $\Sigma$ of the molecular disk, 
and strength of spiral arms in \S \ref{sec.dis_cat}, to find any differences in physical properties between the categories.
 While no clear difference is found in the molecular properties, there is an indication that 
galaxies which show ambiguous offsets tend to have a smaller stellar density contrast 
between arm and interarm regions.
 Thus, the $K$-band morphology can be used as an additional selection criterion, 
when a larger sample is needed for further study.

 The validity of the assumptions we have used in the offset method (i.e., circular rotation, constant \tsf ~and 
\OP) can be estimated from the $\Omega-\theta$ plot. 
 Several features found in this work, such as the lack of offsets and different distributions for individual arms, 
indicate that these assumptions or even the density wave theory are not applicable to all the spiral galaxies.
 A presence of a central bar \citep[e.g.,][]{Atha80,Rome07} and an interaction with 
companion galaxies \citep[e.g.,][]{TT72} can drive material or short-lived spiral structures 
or enhance pre-existing density waves.
 Possible explanations for individual galaxies are mentioned in \S \ref{sec.indivi}, 
but a full consideration of these effects is beyond the scope of this paper.

 Here we consider the effects of the adopted parameters on the results.
 The parameter we must care most is the inclination angle $i$, as it changes both \tsf ~and \OP. 
 Since $\Omega$ is derived as $V_{\rm obs}/\sin i/r$, 
larger \tsf ~and smaller \OP ~will be derived, if larger $i$ is adopted.
 In addition, different $i$ can change the shape of spiral arms, 
since $r=r_{\rm obs}/\sin i$ for the minor axis direction. 
 This change is difficult to quantify, because it does not change anything regarding the major axis direction.
 These effects are especially significant for the most face-on (small $i$) galaxies.
 For example, if $i=12^\circ$ is adopted for NGC 0628, which is half of our value, 
the resultant \OP ~and \tsf ~will be changed by a factor of 2 from our result, since 
$\sin(12^\circ)/\sin(24^\circ)\sim 0.51$.
 While this can directly change the result on \tsf, 
the location of the corotation is almost independent of $i$, 
since $\Omega$ is also inversely dependent on $\sin i$ as well as \OP.

 Another parameter we must keep in mind is the distance $D$.
 As $r$ is proportional to $D$, $\Omega$ is inversely proportional to $D$.
 Though $D$ affects \OP ~and \tsf  ~in the same way as $\sin i$, 
its uncertainty is about 20\%, so that the effect is not as large as $\sin i$ for small $i$.
 We should note that the location of corotation is fully independent on $D$.
 
 In addition, the poor determination of the RC for some galaxies and 
the discrepancy between this work and previous work for NGC 4254, both indicate that 
the uncertainty of the fitted \OP ~and \tsf ~could be underestimated.
 The latter and several other works also suggest a radial variation of \OP.
 In order to solve these problems, an extensive CO survey of nearby spiral galaxies 
with both interferometers and single dishes is critical.

\section{Conclusion\label{sec.end}}
 We have presented a revised method to determine both the pattern speed (\OP)
and star formation timescale (\tsf) of spiral galaxies simultaneously, proposed by \citet{egu04}.
 This method utilizes offsets between molecular and young-stellar arms, 
and we refer to it as the ``Offset Method".

 We applied the offset method to CO and \Ha ~data of 13 nearby galaxies, and 
derived \OP ~and \tsf ~for 5 galaxies.
 Since their offsets are clear, we categorize them as ``C" galaxies.
 From the results of these 5 galaxies, we find the followings.
\begin{itemize}
\item The corotation radius calculated by the derived \OP ~is near the edge of CO data, 
and is about half of the optical radius for 3 galaxies.
\item The derived \tsf ~is roughly consistent with the typical free-fall timescale of molecular clouds, 
which indicates that gravitational instability is a dominant mechanism 
triggering star formations in spiral arms.
\item The surface density of molecular gas calculated from CO data, heavy elements abundance, 
and spiral arm strengths evaluated from $K$-band images 
do not show clear dependence on \tsf.
\end{itemize}

 We also find that 2 out of the remaining 8 galaxies show no offsets between CO and \Ha ~and 
categorized them as ``N" galaxies.
 As their arms are clearly traced, the spiral density wave is usually thought to be at work in these galaxies. 
 There are several possible reasons to explain the lack of offset in these galaxies: (1) material arms, 
(2) corotation where arms are seen, and 
(3) elliptical orbits nearly parallel to spiral arms. 
 Although the current data are not sufficient to confirm which case is at work, 
we have gained an insight that a central bar could account for this feature, 
since these 2 galaxies are both barred.
 
 With 1 galaxy excluded from our analysis due to its poor rotation curve, 
the remaining 5 galaxies have ambiguous offsets, 
whose dependence on the rotational frequency is not clear or different from expectation. 
 We categorize them as ``A" galaxies.
 The major reasons for this ambiguity are  
(1) the density wave is weaker, and/or (2) observational resolution and sensitivity are not sufficient to detect arms 
and their offsets clearly.
 The former is supported by our finding that the arm strengths of ``A" galaxies 
are slightly weaker than those of ``C" galaxies.

 In addition to the results for individual galaxies, properties of molecular gas and morphology were examined 
according to the category.
 We list the results in the following.
\begin{itemize}
\item The disk-averaged and arm-only value of molecular surface density do not appear to correlate 
with the category, indicating that larger ($>500$ pc) scale properties of gas do not vary according to a spiral structure.
\item Analyses of the $K$-band images show that the amplitude of stellar spiral arms is 
slightly smaller for ``A" galaxies.
\end{itemize}

 From the results for individual galaxies and offset categories as mentioned above, 
we summarize our findings as  
\begin{enumerate}
\item Star formation in spiral arms is dominantly triggered by gravitational instability of molecular clouds.
\item Stellar spiral arms could affect the appearance of offsets between molecular gas and young stars.
\end{enumerate}

\acknowledgments
 We are grateful to Dr.\ Rebecca Koopmann, Dr.\ Crystal Martin, Dr.\ Johan Knapen, and Mr.\ Olivier Daigle 
for kindly providing their data and advices on how to use them.
 This work has made use of 
the NASA/IPAC Extragalactic Database (NED) and 2MASS Image Services 
via the NASA/IPAC Infrared Science Archive, which is operated by 
the Jet Propulsion Laboratory, California Institute of Technology, under contract with 
the National Aeronautics and Space Administration; 
the VizieR catalogue access tool and the Aladin Sky Atlas, CDS, Strasbourg, France.
 F.~E.\ and S.~K.\ are thankful to the financial support by 
the Japan Society for the Promotion of Science during this research. 
 A part of this study was also financially supported
by the MEXT Grant-in-Aid for Scientific Research on
Priority Areas No.\ 15071202.




\begin{deluxetable}{cccccccc}
\tabletypesize{\footnotesize}
\tablewidth{0pt}
\tablecaption{General Property of Sample Galaxies\label{sample.tb}}
\tablehead{\colhead{NGC} & \colhead{R.A.~(J2000)} & \colhead{Dec.~(J2000)} & 
\colhead{Velocity} & \colhead{Distance} & \colhead{Morphology} & \colhead{T$^\dagger$} & \colhead{AC$^\ddagger$}\\
 & (h m s) & (d m s) & (\kms) & (Mpc) &}
\startdata
0628 & 01  36  41.77 & +15 47 00.5  &
 657\tablenotemark{a} & 7.3\tablenotemark{a} & SA(s)c & 5 & 9\\
3184 & 10  18  16.98 & +41 25 27.8  &
 592\tablenotemark{a} & 8.7\tablenotemark{a} & SAB(rs)cd & 6 & 9\\
3938 & 11  52  49.45 & +44 07 14.6  &
 809\tablenotemark{a} & 17\tablenotemark{a} & SA(s)c & 5 & 9\\
4254 & 12  18  49.63 & +14 24 59.4  &
 2407\tablenotemark{c} & 16.1\tablenotemark{c} & SA(s)c & 5 & 9\\
4303 & 12  21  54.90 & +04 28 25.1  &
 1566\tablenotemark{a} & 16.1\tablenotemark{c} & SAB(rs)bc & 4 & 9\\
4321 & 12  22  54.90 & +15 49 20.6  &
 1571\tablenotemark{a} & 16.1\tablenotemark{c} & SAB(s)bc & 4 & 12\\
4535 & 12  34  20.31 & +08 11 51.9  &
 1961\tablenotemark{a} &16.1\tablenotemark{c} & SAB(s)c & 5 & 9\\
4736 & 12  50  53.06 & +41 07 13.7  &
 308\tablenotemark{a} & 5.1\tablenotemark{b} & SA(r)ab & 2 & 3\\
5194 & 13  29  52.71 & +47 11 42.6  &
 463\tablenotemark{a} & 9.6\tablenotemark{b} & SA(s)bc & 4 & 12\\
5248 & 13  37  32.07 & +08 53 06.2  &
 1153\tablenotemark{a} & 22.7\tablenotemark{a} & SB(rs)bc & 4 & 12\\
5457 & 14  03  12.59 & +54 20 56.7  &
 241\tablenotemark{a} & 7.2\tablenotemark{b} & SAB(rs)cd & 6 & 9\\
6181 & 16  32  20.96 & +19 49 35.6  &
 2375\tablenotemark{d} & 36.7\tablenotemark{d} & SA(rs)c & 5 & 12\\
6946 & 20  34  52.34 & +60 09 14.2  &
 48\tablenotemark{a} & 5.5\tablenotemark{b} & SAB(rs)cd & 6 & 9\\
\enddata
\tablecomments{(RA, Dec) and morphology are from NED. 
$\dagger$: T is a numeric index corresponding to the morphology, and is called the Hubble T type.
$\ddagger$: AC is the arm class from \citet{EE87}}
\tablerefs{(a) \citet{BIMA2}, (b) \citet{RC99}, (c) \citet{ViCS1}, (d) \citet{NBG}}
\end{deluxetable}

\begin{deluxetable}{cccccccccc}
\tabletypesize{\footnotesize}
\tablewidth{0pt}
\tablecaption{Property of Data for Sample Galaxies\label{sample.data.tb}}
\tablehead{\colhead{NGC} & 
\multicolumn{4}{c}{CO} & &
\multicolumn{3}{c}{\Ha} & \colhead{RC$^*$}\\
\cline{2-5}\cline{7-9}
& \colhead{bmaj} & \colhead{bmin} & \colhead{rms$^\dagger$} & \colhead{tel.} & &
\colhead{seeing} & \colhead{tel.} & \colhead{ref}\\
& ($''$) & ($''$) & (mJy beam$^{-1}$) & & & ($''$) & }
\tablecolumns{10}
\startdata
0628 & 7.2 & 5.3 & 51 & BIMA+12m & & 0.43 & CTIO & 1 & {\bf 6}, CO\\
3184 & 5.9 & 5.4 & 50 & BIMA & & 1.7 & JKT & 2 & {\bf 6}, CO\\
3938 & 5.9 & 5.4 & 59 & BIMA+12m & & 0.3 & KP2 & 1 & 6, {\bf CO}\\
4254 & 4.8 & 3.6 & 15 & NMA+45m & & 2.1 & KP9 & 3 & {\bf CO}\\
4303 & 7.3 & 5.5 & 47 & BIMA+12m & & 1.4 & JKT & 2 & {\bf 7}, CO\\
4321 & 7.2 & 4.9 & 50 & BIMA+12m & & \nodata & S90 & 4 & 6, {\bf 7}, CO\\
4535 & 7.3 & 5.7 & 63 & BIMA & & 1.9 & JKT & 2 & CO\\
4736 & 6.9 & 5.0 & 64 & BIMA+12m & & 1.4 & JKT & 2 & {\bf 7}, CO\\
5194 & 5.8 & 5.1 & 61 & BIMA+12m & & \nodata & BS & 4 & 6, {\bf 7}, CO\\
5248 & 6.9 & 5.8 & 38 & BIMA+12m & & 1.4 & JKT & 2 & {\bf CO}\\
5457 & 5.7 & 5.4 & 45 & BIMA+12m & & 2.4 & INT & 2 & {\bf 7}, CO\\
6181 & 3.4 & 2.8 & 11.5 & NMA & & 4.7 & P60 & 5 & {\bf CO}\\
6946 & 6.0 & 5.0 & 61 & BIMA+12m & & 1.4 & JKT & 2 & 6, {\bf 7}, CO\\
\enddata
\tablecomments{
$\dagger$: The rms of CO data was calculated for a velocity width of 10 \kms.
$*$: ``CO" means a rotation curve from the velocity field of CO data, 
and data in bold font were used for the application. (See \S \ref{sec.RC} for detail.)\\
Abbreviations for telescopes: BIMA (Berkeley-Illinois-Maryland Association), 12m (NRAO 12m telescope), 
NMA (Nobeyama Millimeter Array), 45m (Nobeyama 45m Telescope), 
CTIO (Cerro Tololo Inter-American Observatory 1.5m), 
JKT (1m Jacobus Kapteyn Telescope), KP2 (Kitt Peak 2.1m), KP9 (Kitt Peak 0.9m), S90 (Steward 90inch), 
BS (Kitt Peak Burrell-Schmidt), INT (2.5m Isaac Newton Telescope), P60 (Palomar 60inch)}
\tablerefs{(1) \citet{Kenn03}, (2) \citet{Knap04}, (3) \citet{Koop01}, (4) \citet{MK01}, 
(5) Koopmann et al., private communication, (6) \citet{SINGSHa}, (7) \citet{RC99}}
\end{deluxetable}

\begin{deluxetable}{ccccccccc}
\tabletypesize{\footnotesize}
\tablewidth{0pt}
\tablecaption{Parameters Derived from CO Velocity Field\label{param.tb}}
\tablehead{ & \multicolumn{3}{c}{Dynamical Center} & & & &\multicolumn{2}{c}{Adopted Value}\\
\cline{2-4}\cline{8-9}
\colhead{NGC} & \colhead{R.A.~(J2000)} & \colhead{Dec.~(J2000)} & \colhead{error}
 & \colhead{Velocity} & \colhead{P.A.$^\dagger$} & \colhead{$i^\ddagger$} & \colhead{P.A.} & \colhead{$i$}\\
& (h m s) & (d m s) & & (\kms) & ($^\circ$) & ($^\circ$) & ($^\circ$) & ($^\circ$)}
\tablecolumns{9}
\startdata
0628 & 01  36  41.589 & +15 47 01.30  & 3.3$''$ & $650.2\pm 1.5$ 
& $11.8\pm 1.1$ & \nodata & 11.8 & 24\tablenotemark{a}\\
3184 & \nodata & \nodata & \nodata & \nodata 
& \nodata & \nodata & 176.4\tablenotemark{c} & 16.7\tablenotemark{c}\\
3938 & 11  52  49.502 & +44 07 20.50  & 9.0$''$ & $812.1\pm 3.7$ 
& $202.2\pm 5.1$ & \nodata & 202.2 & 24\tablenotemark{a}\\
4254 & 12  18  49.799 & +14 25 03.90  & 3.2$''$ & $2393.7\pm 8.5$ 
& $70.9\pm 4.1$ & $52.4\pm 2.3$ & 70.9 & 52.4\\
4303 & 12  21  54.929 & +04 28 21.60  & 6.1$''$ & $1557.8\pm 5.4$ 
& $304.8\pm 9.1$ & $27.7\pm 9.5$ & 318\tablenotemark{b} & 27\tablenotemark{b}\\
4321 & 12  22  55.192 & +15 49 20.10  & 5.7$''$ & $1575.2\pm 7.8$ 
& $161.7\pm 9.5$ & $34.5\pm 23.3$ & 146\tablenotemark{b} & 27\tablenotemark{b}\\
4535 & 12  34  20.427 & +08 11 54.10  & 3.2$''$ & $1943.8\pm 9.8$ 
& \nodata & \nodata & \nodata & \nodata\\
4736 & \nodata & \nodata & \nodata & \nodata 
& \nodata & \nodata & 108\tablenotemark{b} & 35\tablenotemark{b}\\
5194 & 13  29  52.565 & +47 11 42.20  & 4.2$''$ & $469.2\pm 2.4$
 & \nodata & \nodata & 22\tablenotemark{b} & 20\tablenotemark{b}\\
5248 & 13  37  32.140 & +08 53 03.90  & 7.1$''$ & $1165.0\pm 10.9$ 
& $109.6\pm 0.2$ & \nodata & 109.6 & 43\tablenotemark{a}\\
5457 & 14  03  12.330 & +54 20 56.99  & 4.3$''$ & $258.1\pm 2.3$ 
& \nodata & \nodata & 38\tablenotemark{b} & 18\tablenotemark{b}\\
6181 & 16  32  20.916 & +19 49 35.23  & 0.6$''$ & $2395.7\pm 4.0$ 
& $332.8\pm 0.7$ & $66.8\pm 3.9$ & 332.8 & 66.8\\
6946 & 20  34  52.703 & +60 09 13.39  & 4.3$''$ & $51.3\pm 9.0$ 
& $244.8\pm 15.7$ & \nodata & 64\tablenotemark{b} & 30\tablenotemark{b}\\
\enddata
\tablecomments{
$\dagger$: P.A.~is the position angle of the disk major axis, which is defined as $0^\circ$ at north and 
increases counterclockwise.
$\ddagger$: $i$ is the inclination angle of the disk, which is defined as $0^\circ$ when the disk is face-on, 
and $90^\circ$ when edge-on. 
Data fields shown as ``\nodata '' indicate that meaningful values are not derived by the analysis 
described in \S \ref{sec.RC}}
\tablerefs{(a) \citet{BIMA2}, (b) \citet{RC99}, (c) \citet{SINGSHa}}
\end{deluxetable}

\begin{deluxetable}{cccccc}
\tabletypesize{\footnotesize}
\tablewidth{0pt}
\tablecaption{Result of Fitting with Offset Category\label{oc.tb}}
\tablehead{\colhead{NGC} & \colhead{\tsf ~(Myr)} & \colhead{\OP ~(\kmskpc)} 
& \colhead{\# of data} & \colhead{$\chi^2$} & \colhead{Category$^\dagger$}}
\startdata
0628 & $28.2\pm 3.1$ & $16\pm 3$ & 33 & 91.78 & C\\
3184 & $-34.1\pm 10.9$ & $51\pm 24$ & 19 &  26.93 & A\\
3938 & $-6.4\pm 6.5$ & $20\pm 28$ & 9 & 5.2 & A\\
4254 & $12.4\pm 1.3$ & $10\pm 3$ & 18 & 29.51 & C\\
4303 & $10.8\pm 5.7$ & $24\pm 29$ & 8 & 7.09 & C\\
4321 & $2.3\pm 0.8$ & $31\pm 20$ & 34 & 86.80 & N\\
4736 & $-17.2\pm 4.8$ & $166\pm 66$ & 8 & 15.67 & A\\
5194 & $13.8\pm 0.7$ & $40\pm 4$ & 41 & 811.54 & C\\
(arm1 & $7.1\pm 0.5$ & $31\pm 5$ & 24 & 157.32 & C)\\
5248 & $-0.1\pm 2.6$ & $270\pm 10860$ & 13 & 4.16 & N\\
5457 & $4.0\pm 1.3$ & $72\pm 37$ & 10 & 16.09 & C\\
6181 & $-85.7\pm 102.7$ & $38\pm 150$ & 27 & 31.4 & A\\
(arm2 & $43.4\pm 54.5$ & $33\pm 119$ & 14 & 19.52 & A)\\
6946 & $1.1\pm 0.8$ & $36\pm 79$ & 7 & 38.07 & A\\
\enddata
\tablecomments{
$\dagger$: ``C" galaxies are categorized if their \tsf ~and \OP ~can be 
determined by the $\Omega-\theta$ plot. ``N" galaxies show no offsets 
between CO and \Ha ~arms. ``A" galaxies are those which do not fall into the
 above two categories. See \S \ref{sec.dis_indivi} for details.}
\end{deluxetable}

\begin{deluxetable}{llccccccc}
\tabletypesize{\scriptsize}
\tablewidth{0pt}
\tablecaption{\OP ~Values for ``C'' Galaxies\label{OP_C.tb}}
\tablehead{\colhead{NGC} & \colhead{Authority} & \colhead{$D$} & 
\colhead{P.A.} & \colhead{$i$} & \colhead{\OP} & \colhead{$R_{\rm CR}$} & \colhead{Method} & \colhead{Data}\\
& & \colhead{(Mpc)} & \colhead{($^\circ$)} & \colhead{($^\circ$)} & \colhead{(\kmskpc)} & \colhead{(arcmin)} & }
\tablecolumns{9}
\startdata
0628 &
\citet{EEM92}
 & \nodata & \nodata & 21\tablenotemark{a} & \nodata & $2.4$ & morphology & $B$\\
& \citet{Fath07a}
 & 9.7 & 25 & 6.5 & $31^{+5}_{-3}$ & 1.8\tablenotemark{b} & SFE & \HI, \Ha\\
& This work (2009)
 & 7.3 & 11.8 & 24 & $16\pm 3$ & $0.9$ --$1.9$ & offset & CO, \Ha\\
\tableline
4254 &
\citet{EEM92}
 & \nodata & \nodata & 27\tablenotemark{a} & \nodata & 1.5 & morphology & $B$\\
& \citet{Kranz01}
 & 20 & 67.5 & 41.2 & 20 & $1.3\pm 0.2$ & simulation & $K$, \Ha\\
& This work (2009)
 & 16.1 & 70.9 & 52.4 & $10\pm 3$ & $1.0$ --$1.3$ & offset & CO, \Ha\\
\tableline
4303 & 
\citet{CW00} 
& 16.1 & 40 & \nodata & 500 & $2''$ & simulation & \\
& \citet{RSL05}
& \nodata & \nodata & \nodata & \nodata & 1.5 & simulation & $H$ \\
& This work (2009)
 & 16.1 & 318 & 27 & $24\pm 29$ & $\gtrsim 0.6$ & offset & CO, \Ha\\
\tableline
5194 &
\citet{Tul74}
 & 4 & -10 & 20 & 90 & 2.2 & morphology & \Ha \\
& \citet{EEM92}
 & \nodata & \nodata & 45\tablenotemark{a} & \nodata & 2.2 & morphology & $B$\\
& \citet{GB93_2}
 & 9.6 & 170 & 20 & 27 & 2.7 & simulation & CO\\
& \citet{Vogel93}
 & \nodata & -10 & 20 & \nodata & 2.1 & streaming motion & \Ha \\
& \citet{Kuno95}
 & 9.6 & -10 & 20 & 14 & 5.0 & morphology & optical\\
& \citet{Oey03}
 & 9.6 & \nodata & \nodata & \nodata & 5.0 & isochrones & \Ha\\
& \citet{Zim04}
 & 9.5 & -10 & 20 & $38\pm 7$ & \nodata & TW & CO\\
& This work (2009)
 & 9.6 & 22 & 20 & $31\pm 5$ & $2.9^{+0.2}_{-0.4}$ & offset & CO, \Ha\\
\tableline
5457 & 
\citet{EEM92}
 & \nodata & \nodata & 12\tablenotemark{a} & \nodata & $5.5$ & morphology & $B$\\
& \citet{Wall97}
 & 7.4 & 39 & 18 & $19\pm 5$ & $5.5\pm 1.5$ & offset & CO, FUV\\
& This work (2009)
 & 7.2 & 38 & 18 & $72\pm 37$ & $1.3^{+1.8}_{-0.4}$ & offset & CO, \Ha\\
\enddata
\tablenotetext{a}{Inclination is calculated from the ratio of major axis to minor axis, from \citet{RC2}.}
\tablenotetext{b}{\Rcr ~is from \citet{CB90} but \OP ~is derived from their new rotation curve.}
\end{deluxetable}

\begin{deluxetable}{llccccccc}
\tabletypesize{\scriptsize}
\tablewidth{0pt}
\tablecaption{\OP ~Values for ``N'' Galaxies\label{OP_N.tb}}
\tablehead{\colhead{NGC} & \colhead{Authority} & \colhead{$D$} & 
\colhead{P.A.} & \colhead{$i$} & \colhead{\OP} & \colhead{$R_{\rm CR}$} & \colhead{Method} & \colhead{Data}\\
& & \colhead{(Mpc)} & \colhead{($^\circ$)} & \colhead{($^\circ$)} & \colhead{(\kmskpc)} & \colhead{(arcmin)} & }
\tablecolumns{9}
\startdata
4321 &
\citet{EEM92}
 & \nodata & \nodata & 27\tablenotemark{a} & \nodata & 2.0 & morphology & $B$\\
& \citet{Semp95}
 & 20 & 155 & 32 & $20$ & 1.8 & Canzian, simulation & \HI\\
& \citet{Canz97}
 & 17.1 & 153 & 28 & 31 & $1.6\pm 0.2$ & Canzian & \Ha\\
& \citet{Wada98}
 & 17.1 & 155 & 30 & 65 (bar) & $0.2$ (bar) & simulation & CO\\
& \citet{Oey03}
 & 16.1 & \nodata & \nodata & \nodata & 2.6 & isochrones & \Ha\\
& \citet{RW04}
 & 16.1 & 153 & 27 & $28^{+4}_{-5}$ & \nodata & TW & CO\\
\tableline
5248 &
\citet{EEM92}
 & \nodata & \nodata & 41\tablenotemark{a} & \nodata & 1.7 & morphology & $B$\\
& \citet{Jog02a}
 & 15 & 135 & \nodata & \nodata & 1.9 & bar length & $K$ \\
\enddata
\tablenotetext{a}{Inclination is calculated from the ratio of major axis to minor axis from \citet{RC2}.}
\end{deluxetable}

\begin{deluxetable}{cccc}
\tabletypesize{\footnotesize}
\tablewidth{0pt}
\tablecaption{Spatial Resolution and Sensitivity of CO data\label{bmaj+rms.tb}}
\tablehead{\colhead{NGC} & \colhead{bmaj\tablenotemark{\dag} (pc)} & 
\colhead{rms\tablenotemark{\ddag} ($M_{\odot}$ pc$^{-2}$)} 
& \colhead{Category}}
\startdata
0628 & 250 & 52 & C\\
3938 & 490 & 73 & A\\
4254 & 370 & 34 & C\\
4303 & 570 & 45 & C\\
4321 & 560 & 55 & N\\
4736 & 170 & 72 & A\\
5194 & 270 & 80 & C\\
5248 & 760 & 37 & N\\
5457 & 200 & 57 & C\\
6181 & 600 & 47 & A\\
6946 & 160 & 80 & A\\
\enddata
\tablenotetext{\dag}{Major axis of the synthesized beam from the CO data.}
\tablenotetext{\ddag}{Sensitivity of CO channel maps (10 \kms ~per channel) measured 
as the root mean square of emission-free regions.}
\end{deluxetable}

\begin{deluxetable}{ccccc}
\tabletypesize{\footnotesize}
\tablewidth{0pt}
\tablecaption{Bar Classification and Offset Category\label{bar-oc.tb}}
\tablehead{ & \colhead{C} & \colhead{N} & \colhead{A} & \colhead{total}}
\startdata
SA & 3 & 0 & 3 & 6\\
SAB & 2 & 1 & 2 & 5\\
SB & 0 & 1 & 0 & 1\\
total & 5 & 2 & 5 & 12\\
\enddata
\end{deluxetable}


\begin{deluxetable}{cccc}
\tabletypesize{\footnotesize}
\tablewidth{0pt}
\tablecaption{$R_{25}$ and \Rcr \label{Rcr.tb}}
\tablehead{\colhead{NGC} & \colhead{$R_{25}$ (arcmin)} 
& \colhead{\Rcr ~(arcmin)} & \colhead{$R_{\rm CR}/R_{25}$}}
\startdata
0628 & 3.5 & $\simeq 0.9-1.9$ & $0.3-0.5$\\
4254 & 2.5 & $\simeq 1.0-1.3$ & $0.4-0.5$\\
5194 & 6.8 & $2.9^{+0.2}_{-0.4}$ & $0.4-0.5$\\
5457 & 11.9 & $1.3^{+1.8}_{-0.4}$ & $0.1-0.3$\\
\enddata
\end{deluxetable}

\begin{figure*}[h!]
\begin{center}
\begin{minipage}{0.3\linewidth}
\begin{center} 
\includegraphics[width=\linewidth]{f2g.eps2}
\end{center}
\end{minipage}
\hspace{0.03\linewidth}
\begin{minipage}{0.3\linewidth}
\begin{center} 
\includegraphics[width=\linewidth]{f2h.eps2}
\end{center}
\end{minipage}
\hspace{0.03\linewidth}
\begin{minipage}{0.3\linewidth}
\begin{center} 
\includegraphics[width=\linewidth]{f2i.eps2}
\end{center}
\end{minipage}\\
\vspace{12pt}
\begin{minipage}{0.3\linewidth}
\begin{center} 
\includegraphics[width=\linewidth]{f2j.eps2}
\end{center}
\end{minipage}
\hspace{0.03\linewidth}
\begin{minipage}{0.3\linewidth}
\begin{center} 
\includegraphics[width=\linewidth]{f2k.eps2}
\end{center}
\end{minipage}
\hspace{0.03\linewidth}
\begin{minipage}{0.3\linewidth}
\begin{center} 
\includegraphics[width=\linewidth]{f2l.eps2}
\end{center}
\end{minipage}\\
\vspace{12pt}
\begin{minipage}{0.3\linewidth}
\begin{center} 
\includegraphics[width=\linewidth]{f2m.eps2}
\end{center}
\end{minipage}\\
\vspace{12pt}
Figure \ref{COHA.fig}--continued
\end{center}
\end{figure*}

\clearpage

\clearpage

\begin{figure*}
\begin{center}
\includegraphics[width=0.45\linewidth]{f4g.eps2}
\hspace{0.05\linewidth}
\includegraphics[width=0.45\linewidth]{f4h.eps2}\\
\vspace{0.05\textheight}
\includegraphics[width=0.45\linewidth]{f4i.eps2}
\hspace{0.05\linewidth}
\includegraphics[width=0.45\linewidth]{f4j.eps2}\\
\vspace{0.05\textheight}
\includegraphics[width=0.45\linewidth]{f4k.eps2}
\hspace{0.05\linewidth}
\includegraphics[width=0.45\linewidth]{f4l.eps2}\\
\vspace{0.05\textheight}
Figure \ref{phase2.fig}--continued
\end{center}
\end{figure*}

\clearpage

\begin{figure*}[h!]
\begin{center}
\includegraphics[width=0.45\linewidth]{f14e.eps2}
\hspace{0.05\linewidth}
\includegraphics[width=0.45\linewidth]{f14f.eps2}\\
\vspace{12pt}
\includegraphics[width=0.45\linewidth]{f14g.eps2}
\hspace{0.05\linewidth}
\includegraphics[width=0.45\linewidth]{f14h.eps2}\\
\vspace{12pt}
Figure \ref{Rcr.fig}--continued
\end{center}
\end{figure*}

\clearpage

\end{document}